 \documentclass[final,1p,times]{elsarticle}

\usepackage{amssymb}
\usepackage{amsmath}
\usepackage{float}

\graphicspath{{./figures/}}

\usepackage{tikz}
\usepackage{pgfplots}
    \pgfdeclarelayer{background}
    \pgfdeclarelayer{foreground}
    \pgfsetlayers{background,main,foreground}
\usepackage{subfig}
\usetikzlibrary{arrows.meta, graphs, positioning, quotes}
\usepackage{siunitx}
\usepackage{comment}

\usepackage{booktabs}

\newif\ifmargincomments 
\margincommentstrue
\ifmargincomments

\else

\fi

\def\tsc#1{\csdef{#1}{\textsc{\lowercase{#1}}\xspace}}
\tsc{WGM}
\tsc{QE}

\pgfplotsset{compat=1.18} 

\journal{Acta Materialia}

\begin{document}

\begin{frontmatter}


\title{Double-tough Ceramics: Optimization-supported Multiscale Computational Design}


\author{Jian Zhang}
\author{Francesco Aiello}

\affiliation[]{organization={Eindhoven University of Technology},
            addressline={PO Box 513}, 
            city={Eindhoven},
            postcode={5600 MB}, 
            state={North Brabant},
            country={Netherlands}}

\author{Mauro Salazar}
\ead{m.r.u.salazar@tue.nl}
\author{Diletta Giuntini}
\ead{d.giuntini@tue.nl}
\cortext[1]{Corresponding author}

\begin{abstract}
To overcome the brittleness limitation of ceramics, various toughening mechanisms have been proposed. Some of the most remarkable, especially for oxides, include the tetragonal-to-monoclinic phase transformation leading to crack shielding in zirconia, and bioinspired brick-and-mortar microstructures fostering crack deflection. It has, however, proven challenging to incorporate both these mechanisms into a single all-ceramic material.
In this work, we propose a computational methodology for the design of a material that combines these two toughening strategies, using a multiscale modeling approach that captures both their individual contributions and the overall fracture performance. This is achieved by developing an all-ceramic composite with a brick-and-mortar microstructure, in which the nanocrystalline mortar is transformation-toughened. Key factors influencing phase transformation, such as grain boundary properties, grain orientations, and kinetic coefficients, are analyzed, and the resulting transformation stress-strain behavior is incorporated into the microscale mortar constitutive model.
We demonstrate that the synergistic effect of the two toughening mechanisms is achievable, and that it is an extremely effective strategy to boost fracture performance. The influence of brick size, mortar thickness, and properties of the constituent materials is then systematically investigated. Finally, a gradient-free optimization algorithm is employed to identify optimal geometric and material parameters, revealing that longer, thinner bricks with minimal mortar thickness provide the best fracture resistance. Optimal combinations of material properties are identified for given brick sizes and mortar thicknesses.
\end{abstract}

%


\end{frontmatter}



\section{Introduction}\label{sec:Intro}


Ceramics and all-ceramic composites possess a unique combination of properties: high strength, thermal stability, chemical inertness, biocompatibility, and superior corrosion resistance~\cite{Lai2013}.
As such they are considered as a promising replacement of metallic alloys for many structural applications, and as key enablers for technologies requiring mechanical performance in challenging environments.
However, even though they have been increasingly used in various fields, such as thermal barrier coatings~\cite{Cao2004, Darolia2013}, biomedical implants~\cite{Piconi1999, Chevalier2007, Denry2008}, and automobile brake pads~\cite{Bhakuni2023}, their widespread application remains limited due to their intrinsic brittleness. 
It is therefore critical to develop toughening strategies against this backdrop.
Several approaches have been adopted over the last decades. Two remarkable ones are crack shielding enabled by the tetragonal-to-monoclinic (T$\rightarrow$M) phase transformation in zirconia, already broadly adopted by industry, and more recently crack deflection via bio-inspired brick-and-mortar structures. 
Interestingly, an effective combination of these two mechanisms is yet to be achieved. Since they operate at different length scales and have specific microstructural requirements, it is indeed very challenging to concurrently introduce them into a single all-ceramic material.

Remarkably, a recent experimental study has developed proof-of-concept ‘double-tough' ceramics that incorporate both these toughening mechanisms~\cite{Francesco2024}. Building on these advancements, this study presents the first numerical analysis of such double-tough ceramics, with the T$\rightarrow$M phase transformation implemented in the mortar of a brick-and-mortar architecture. The aim is to identify the optimal material design to boost the fracture performance across multiple length scales.
Instead of relying on largely empirical considerations, the proposed computational platform can efficiently guide the material design, thus significantly reducing experimental iterations towards the desired solution.

The transformation toughening behavior in zirconia, first discovered nearly five decades ago~\cite{GARVIE1975}, has been extensively studied through experimental observations~\cite{Ruhle1988, Becher1992, Rauchs2002, Liens2020}. The T$\rightarrow$M phase transformation, involving a shear deformation of $\sim$0.16 and volume expansion of $\sim$0.04, alters the stress field near crack tips, enhancing fracture toughness by promoting crack closure and preventing propagation. Theoretical studies complement these findings, exploring the formation of a transformed ‘process' zone at growing cracks~\cite{Porter1979, EVANS1980, Marshall1983}. Two primary methodologies have been implemented to capture this toughening phenomenon: linear elastic fracture mechanics, focusing on stress shielding at crack fronts~\cite{McMEEKING1982}, and fracture energy estimation during crack propagation~\cite{Budiansky1983}. Comprehensive reviews detail these models~\cite{Hannink2000, Kelly2002, Basu2005, Chevalier2009}.

Numerical simulations play a crucial role in predicting the T$\rightarrow$M phase transformation and demonstrating its toughening effects, complementing experimental and theoretical studies. Techniques like finite element (FE) analyses~\cite{Wang2010, Platt2014}, molecular dynamics (MD)~\cite{Wang2015, Deng2020}, peridynamics (PD)~\cite{Platt2017}, and phase field (PF) methods~\cite{Mamivand2013Review, Abubakar2015} have been employed for this purpose. The PF method is widely used to solve evolving boundary problems, such as crack propagation~\cite{Abdollahi2012, Clayton2015, HansenDorr2019}, solidification~\cite{Boettinger2002, Provatas2005}, grain growth~\cite{Chen2002}, dislocation dymanics~\cite{Levitas2012, Javanbakht2016}, and solid-state phase transformations~\cite{Steinbach2006, Lv2022}, and it has gained popularity for simulating the T$\rightarrow$M phase transformation. Here, the governing equation is represented by the time-dependent Ginzburg–Landau kinetic equation~\cite{Falk1982} and the PF micro-elasticity theory~\cite{Khachaturyan1969, AArtemev2001}. The former controls the temporal and spatial evolution of the field variable used to describe the transformation process, and the latter links the field variable to the strain evaluation used to calculate the elastic energy. 
Mamivand et al. developed PF models for the T$\rightarrow$M phase transformation in 2D and 3D single crystals~\cite{Mamivand2013, Mamivand2015}, later extending these to polycrystals with inhomogeneous and anisotropic elasticity~\cite{Mamivand2014Poly}. They demonstrated how external loading and nucleation locations influence transformation behaviour, including stress-induced crack closure~\cite{Mahmood2014}. Cissé and Zaeem further incorporated plasticity to examine grain orientation and boundary effects in yttria-stabilized zirconia~\cite{Cisse2020}.

A large number of studies effectively simulate the T$\rightarrow$M phase transformation and its toughening effects, but often do not include fracture analysis. To address this gap, recent efforts focus on investigating the toughening effect on real-time crack propagation.
The numerical techniques to deal with fracture problems can be classified into two categories, namely discrete and continuum methodologies~\cite{Adrian2019}.
Discrete methods, such as the finite element method (FEM)~\cite{Tracey1971, Barsoum1976}, eXtended/Generalized FEM~\cite{Oden1998, Moes1999, Fries2010}, and discontinuity-enriched FEM~\cite{Aragon2017, Zhang2019}, treat cracks as topological discontinuities requiring explicit or implicit representations~\cite{Zhang2022}. These methods often require additional criteria to model complex crack paths and capture crack tip singularities~\cite{Park2009}.
In contrast, continuum methods avoid discontinuities by diffusing cracks into a scalar field, enabling smoother transitions and simpler modeling of fracture evolution. Among these, phase field (PF) methods have gained prominence over the past two decades, particularly through the variational approach introduced by Francfort and Marigo~\cite{Francfort1998}, where brittle fracture is treated as an energy minimization problem. 
Bourdin et al. later regularized this approach using a length scale parameter~\cite{Bourdin2008}. Comprehensive reviews detail PF methods for fracture analysis, highlighting their advantages and applications~\cite{Ambati2015, Wu2020}.

The PF framework is then able to effectively couple the T$\rightarrow$M phase transformation with the fracture behavior, enabling a realistic analysis of the toughening effect. Zhao et al. developed a PF model for single-crystalline tetragonal zirconia, highlighting the impact of lattice orientations on transformation toughening~\cite{Zhao2016}. This was later extended to polycrystalline zirconia to study the effects of grain boundary properties, crystal orientations, and external loading on crack paths~\cite{Zhu2017, Zhu2020}. Similarly, Moshkelgoshaa and Mamivand explored crystal orientation effects on toughening and crack patterns in 2D and 3D~\cite{Moshkelgosha2020, Moshkelgosha20213D}. They validated their multiphysics PF model against experimental data, concluding that an optimal grain size balances transformation toughening and compression from grain refinement for maximum fracture toughness~\cite{Moshkelgosha2021}.

As for toughening via brick-and-mortar structures, inspiration has come from natural materials consisting of brittle components and yet featuring exceptional fracture properties. The prime example is nacre, consisting of 95 vol.\% hard aragonite-based (a polymorph of $\text{CaCO}_{3}$) platelets (bricks) bonded together with a thin layer of 5 vol.\% soft organic material (mortar)~\cite{Currey1977}. Despite its high mineral content, the fracture toughness of nacre is up to three orders of magnitude greater than monolithic aragonite~\cite{Wegst2004}. 
Decades of research have revealed its complex hierarchical structure, spanning nanoscale to macroscale, with various toughening mechanisms active across these scales~\cite{Jackson1988, Wang2001, Meyers2008b, Barthelat2010, Yourdkhani2011, Song2003, Meyers2008, Evans2001, Katti2004, Katti2005, Wang1995, Lin2009}. The brick-and-mortar arrangement at the microscale~\cite{Sarikaya1989} has been identified as the core contributor to the outstanding mechanical properties of nacre~\cite{Mayer2005}. Theoretical models have been developed to explore the structure-property relationships and toughening mechanisms in brick-and-mortar structures, often using simple representative volume elements (RVEs) based on tension-shear chain or shear-lag models~\cite{Gao2006, Yan2022, Kotha2001, Baron2013, Okumura2001, Shao2012, Nie2023}. However, these analytical models typically focus on individual deformation mechanisms, making it challenging to derive unified formulas incorporating all physical processes. Furthermore, many models use 2D RVEs with only a few bricks, limiting their ability to capture detailed structural responses, and focus on softer, usually polymeric mortars, while it is of high interest to work with ceramic mortars not to compromise on the key features of this material class.

Numerical simulations offer a promising alternative for to study larger models with complex deformation mechanisms. Techniques such as MD \cite{Zhang2016}, discrete lattice models \cite{Anup2015}, particle-spring models~\cite{Sen2011, Dimas2013}, FEM \cite{Katti2001, Genet2014, Djumas2016}, and discrete element methods \cite{Abid2018, Radi2020} provide deeper insights into the structure-property relationships of these complex systems. FEM-based numerical simulations are also widely used in this context. Xie and Yao demonstrated that crack deflection depends on the fracture strength of bricks, while flaw tolerance is ensured by the mortar's ductility~\cite{Xie2014}. Niebel et al. modeled actual composite microstructures to analyze how constituent properties influence macroscopic performance~\cite{Niebel2016}. Mirkhalaf and Barthelat examined the effect of brick concentration on material behavior using RVEs with varying brick content and orientation~\cite{Mirkhalaf2016}. Askarinejad and Rahbar used beam elements for bricks and link elements for organic matrices in their FE models, aligning well with experimental data and showing that all components significantly impact performance~\cite{Askarinejad2015}. Optimization techniques have also been integrated into FE simulations. Park et al. used multi-objective Bayesian optimization to design nacre-inspired composites, identifying key design parameters like brick volume, overlap, and aspect ratio to balance strength and toughness~\cite{Park2023}. 

In this work, we develop a new all-ceramic material design able to leverage both the T$\rightarrow$M phase transformation in zirconia and the brick-and-mortar structure to boost fracture toughness. We set the brick-and-mortar layout as the microscale structure, and simultaneously exploit the phase transformation in zirconia at the nanoscale within the mortar phase. Since these two mechanisms function at different length scales, we propose a multiscale modeling approach to capture their coupling effects on the macroscopic material behavior, including fracture. For the nanoscale model, FE analyses and PF method are combined to simulate the T$\rightarrow$M phase transformation in zirconia, and the corresponding stress-strain curve is adopted as the constitutive law of the mortar phase. At this scale, we investigate the impact of grain orientations and grain boundary properties on the phase transformation process. At the microscale, we demonstrate the toughening effect caused by crack deflection in the brick-and-mortar structure. We then prove that the coupled nano- and microscale toughening mechanisms effectively lead to increased energy dissipation until material failure. Furthermore, we investigate the effect of key material design parameters - brick dimensions and aspect ratio, mortar thickness, and mechanical properties of the single constituents - on the overall toughening. We finally adopt the particle swarm optimization (PSO) algorithm to maximize fracture toughness by updating these parameters, leading to the development of optimal material designs for double-tough ceramics.

\section{Methods and formulations}\label{sec:Formulations}
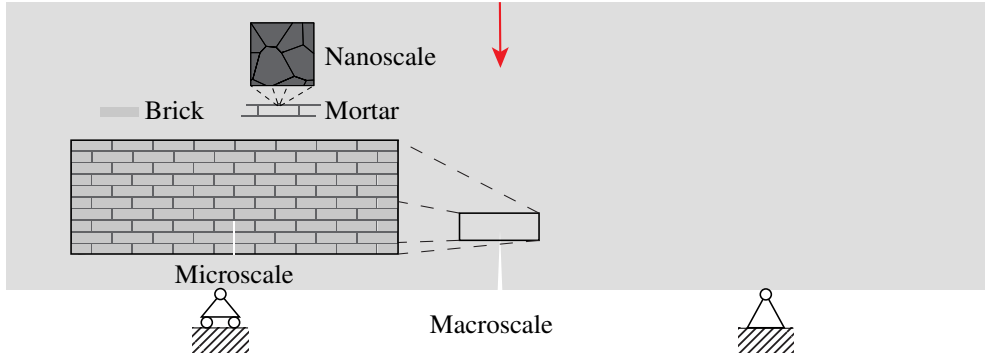
\begin{figure}[!ht]
\centering
\def\svgwidth{0.975\textwidth}
\begingroup%
  \makeatletter%
  \providecommand\color[2][]{%
    \errmessage{(Inkscape) Color is used for the text in Inkscape, but the package 'color.sty' is not loaded}%
    \renewcommand\color[2][]{}%
  }%
  \providecommand\transparent[1]{%
    \errmessage{(Inkscape) Transparency is used (non-zero) for the text in Inkscape, but the package 'transparent.sty' is not loaded}%
    \renewcommand\transparent[1]{}%
  }%
  \providecommand\rotatebox[2]{#2}%
  \newcommand*\fsize{\dimexpr\f@size pt\relax}%
  \newcommand*\lineheight[1]{\fontsize{\fsize}{#1\fsize}\selectfont}%
  \ifx\svgwidth\undefined%
    \setlength{\unitlength}{1899.99993896bp}%
    \ifx\svgscale\undefined%
      \relax%
    \else%
      \setlength{\unitlength}{\unitlength * \real{\svgscale}}%
    \fi%
  \else%
    \setlength{\unitlength}{\svgwidth}%
  \fi%
  \global\let\svgwidth\undefined%
  \global\let\svgscale\undefined%
  \makeatother%
  \begin{picture}(1,0.35)%
    \lineheight{1}%
    \setlength\tabcolsep{0pt}%
    \put(0,0){\includegraphics[width=\unitlength,page=1]{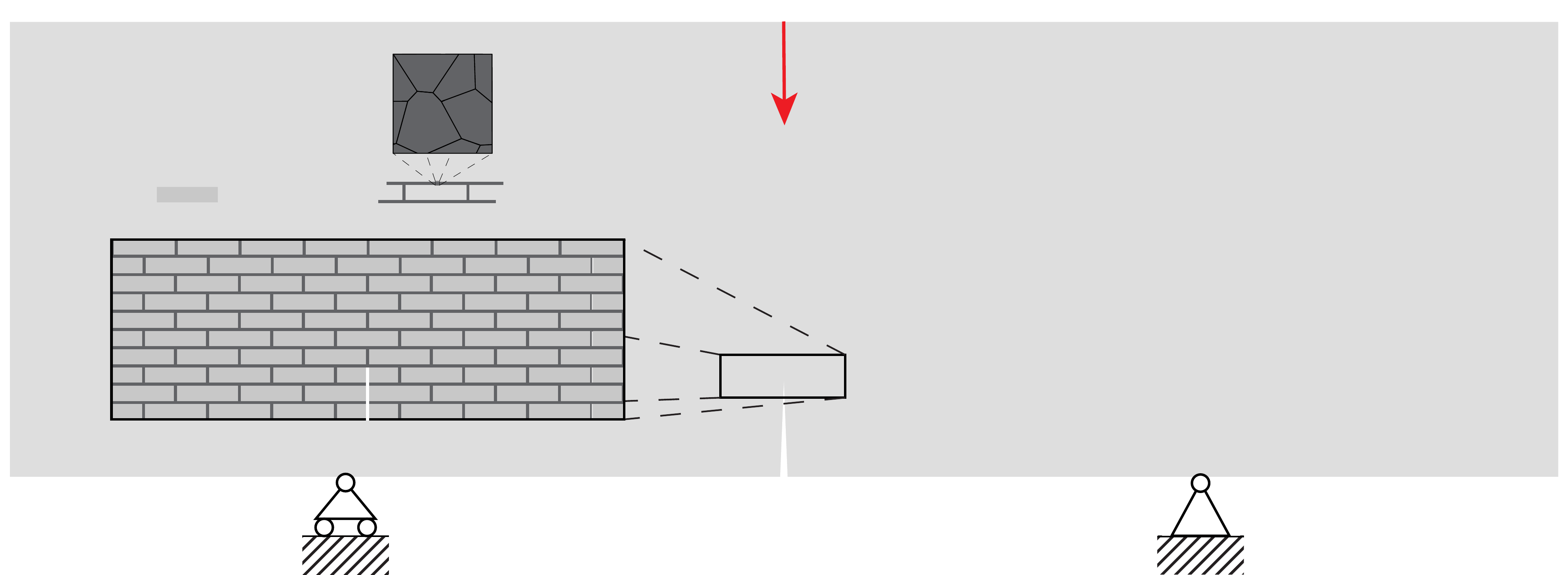}}%
    \put(0.145,0.235){\color[rgb]{0,0,0}\makebox(0,0)[lt]{\lineheight{1.25}\smash{\begin{tabular}[t]{l}Brick\end{tabular}}}}%
    \put(0.325,0.235){\color[rgb]{0,0,0}\makebox(0,0)[lt]{\lineheight{1.25}\smash{\begin{tabular}[t]{l}Mortar\end{tabular}}}}%
    \put(0.175,0.07){\color[rgb]{0,0,0}\makebox(0,0)[lt]{\lineheight{1.25}\smash{\begin{tabular}[t]{l}Microscale\end{tabular}}}}%
    \put(0.43,0.02){\color[rgb]{0,0,0}\makebox(0,0)[lt]{\lineheight{1.25}\smash{\begin{tabular}[t]{l} Macroscale \end{tabular}}}}%
    \put(0.325,0.29){\color[rgb]{0,0,0}\makebox(0,0)[lt]{\lineheight{1.25}\smash{\begin{tabular}[t]{l}Nanoscale\end{tabular}}}}%
  \end{picture}%
\endgroup%
	
\caption{Homogenized macroscale model featuring a centrally located notch and subjected to three-point bending loading conditions. The model has a brick-and-mortar structure at the microscale, with alumina bricks and ceria-stabilized zirconia mortar, the behavior of which is captured at the nanoscale.}
\label{fig:Multiscale_model}
\end{figure}

We aim to design nacre-like ceramics composed of high-strength alumina bricks (platelets) with an average particle size in the micrometer range, interconnected by a sub-micrometer ceria-stabilized zirconia mortar. Ceria-stabilized zirconia is selected for its high toughness and resistance against low-temperature degradation~\cite{Francesco2024}. To explore the toughening mechanisms across different scales, we develop a multiscale model encompassing the macro-, micro-, and nano-scales (see Figure~\ref{fig:Multiscale_model}).
The three-point bending boundary condition (BC) is imposed on the macroscale model, and the fracture analysis is performed on the microscale model with the brick-and-mortar structure, where the corresponding governing equation under the phase field (PF) model is given in~\S\ref{sec:Crack_propagation}. Together with the crack propagation, the tetragonal-to-monoclinic (T$\rightarrow$M) phase transformation happens in the mortar at the nanoscale. \S\ref{sec:Phase_transformation} describes the formulations representing the phase transformation in zirconia.

\begin{figure}[!ht]
\centering
\hspace*{\fill}
     \subfloat[]{
        \centering
        \def\svgwidth{0.3\textwidth}
\begingroup%
  \makeatletter%
  \providecommand\color[2][]{%
    \errmessage{(Inkscape) Color is used for the text in Inkscape, but the package 'color.sty' is not loaded}%
    \renewcommand\color[2][]{}%
  }%
  \providecommand\transparent[1]{%
    \errmessage{(Inkscape) Transparency is used (non-zero) for the text in Inkscape, but the package 'transparent.sty' is not loaded}%
    \renewcommand\transparent[1]{}%
  }%
  \providecommand\rotatebox[2]{#2}%
  \newcommand*\fsize{\dimexpr\f@size pt\relax}%
  \newcommand*\lineheight[1]{\fontsize{\fsize}{#1\fsize}\selectfont}%
  \ifx\svgwidth\undefined%
    \setlength{\unitlength}{600bp}%
    \ifx\svgscale\undefined%
      \relax%
    \else%
      \setlength{\unitlength}{\unitlength * \real{\svgscale}}%
    \fi%
  \else%
    \setlength{\unitlength}{\svgwidth}%
  \fi%
  \global\let\svgwidth\undefined%
  \global\let\svgscale\undefined%
  \makeatother%
  \begin{picture}(1,1.16666664)%
    \lineheight{1}%
    \setlength\tabcolsep{0pt}%
    \put(0,0){\includegraphics[width=\unitlength,page=1]{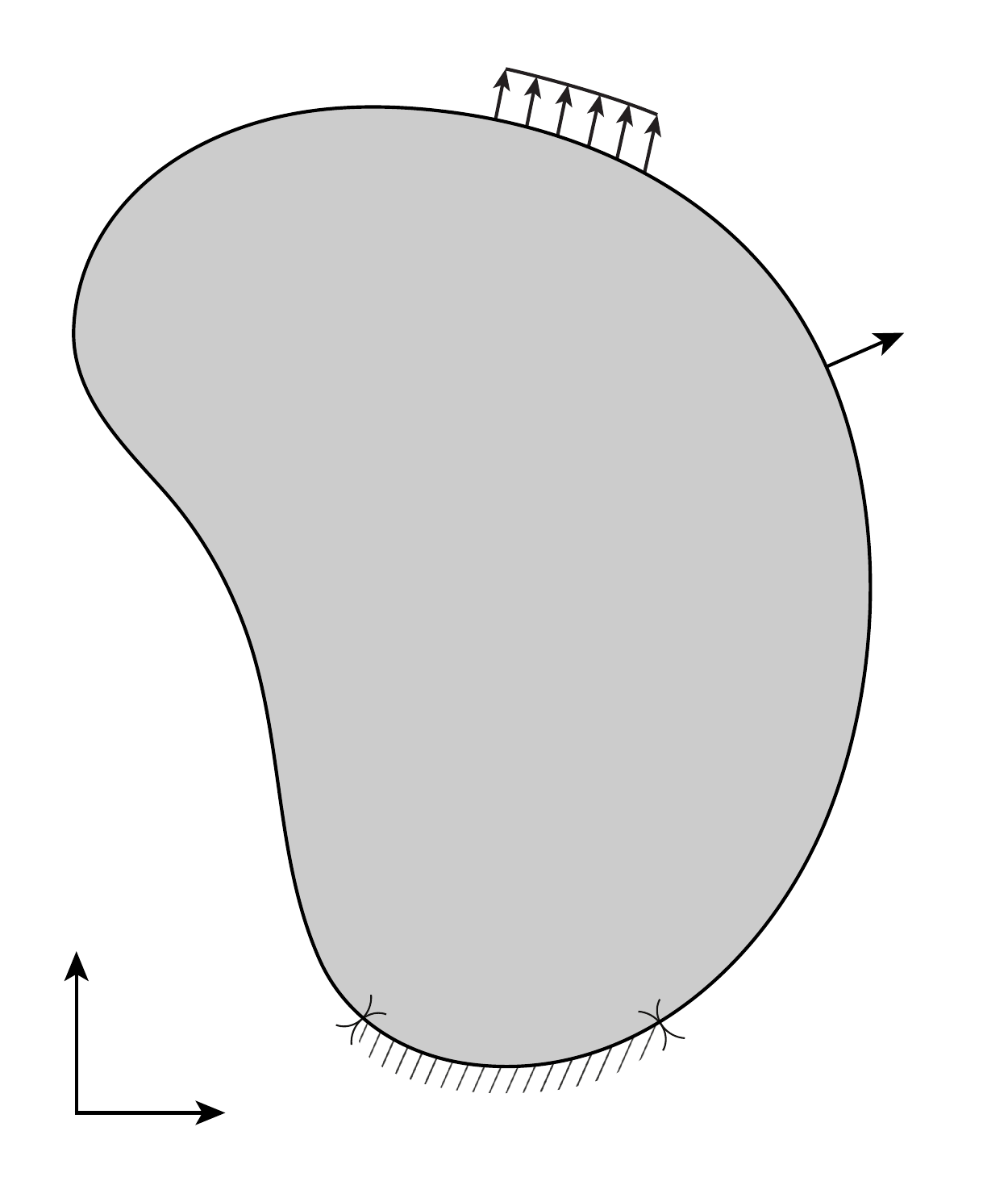}}%
    \put(0.23,0.05){\color[rgb]{0,0,0}\makebox(0,0)[lt]{\lineheight{1.25}\smash{\begin{tabular}[t]{l}$x$\end{tabular}}}}%
    \put(0.06,0.24){\color[rgb]{0,0,0}\makebox(0,0)[lt]{\lineheight{1.25}\smash{\begin{tabular}[t]{l}$y$\end{tabular}}}}%
    \put(0.50,0.13){\color[rgb]{0,0,0}\makebox(0,0)[lt]{\lineheight{1.25}\smash{\begin{tabular}[t]{l}$\Omega$\end{tabular}}}}%
    \put(0.49,0.01){\color[rgb]{0,0,0}\makebox(0,0)[lt]{\lineheight{1.25}\smash{\begin{tabular}[t]{l}$\bar{\boldsymbol{u}}$\end{tabular}}}}%
    \put(0.68,0.09){\color[rgb]{0,0,0}\makebox(0,0)[lt]{\lineheight{1.25}\smash{\begin{tabular}[t]{l}$\Gamma_\mathrm{D}$\end{tabular}}}}%
    \put(0.88,0.77){\color[rgb]{0,0,0}\makebox(0,0)[lt]{\lineheight{1.25}\smash{\begin{tabular}[t]{l}$\boldsymbol{n}$\end{tabular}}}}%
    \put(0.60,1.08){\color[rgb]{0,0,0}\makebox(0,0)[lt]{\lineheight{1.25}\smash{\begin{tabular}[t]{l}$\bar{\boldsymbol{t}}$\end{tabular}}}}%
    \put(0.36,1.08){\color[rgb]{0,0,0}\makebox(0,0)[lt]{\lineheight{1.25}\smash{\begin{tabular}[t]{l}$\Gamma_\mathrm{N}$\end{tabular}}}}%
  \end{picture}%
\endgroup%
	
        \label{fig:Original_model}
        }
\hspace*{\fill}
     \subfloat[]{
        \centering
        \def\svgwidth{0.3\textwidth}
        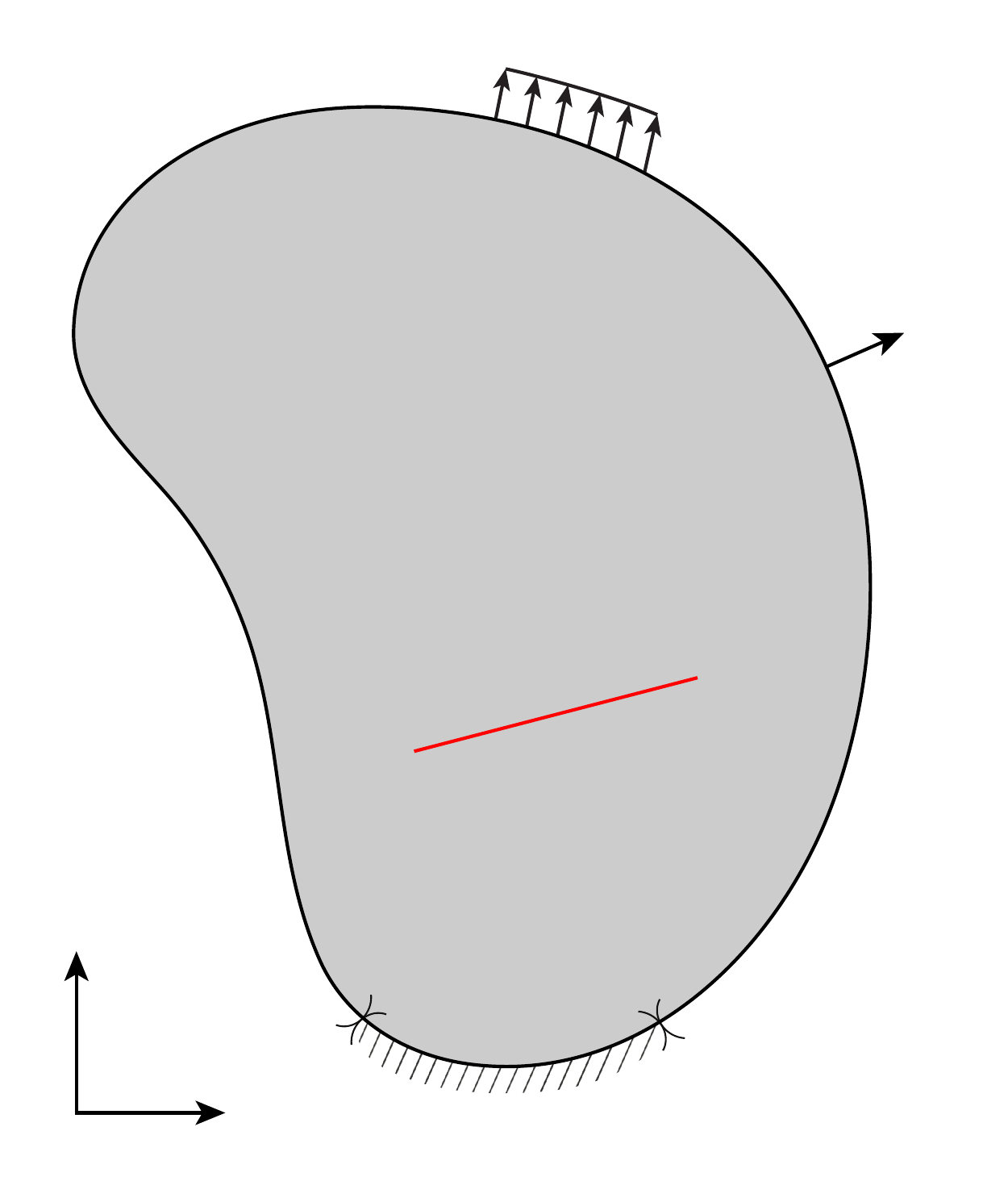	
        \label{fig:No_Phase_field}
        }
\hspace*{\fill}
     \subfloat[]{
        \centering
        \def\svgwidth{0.3\textwidth}
        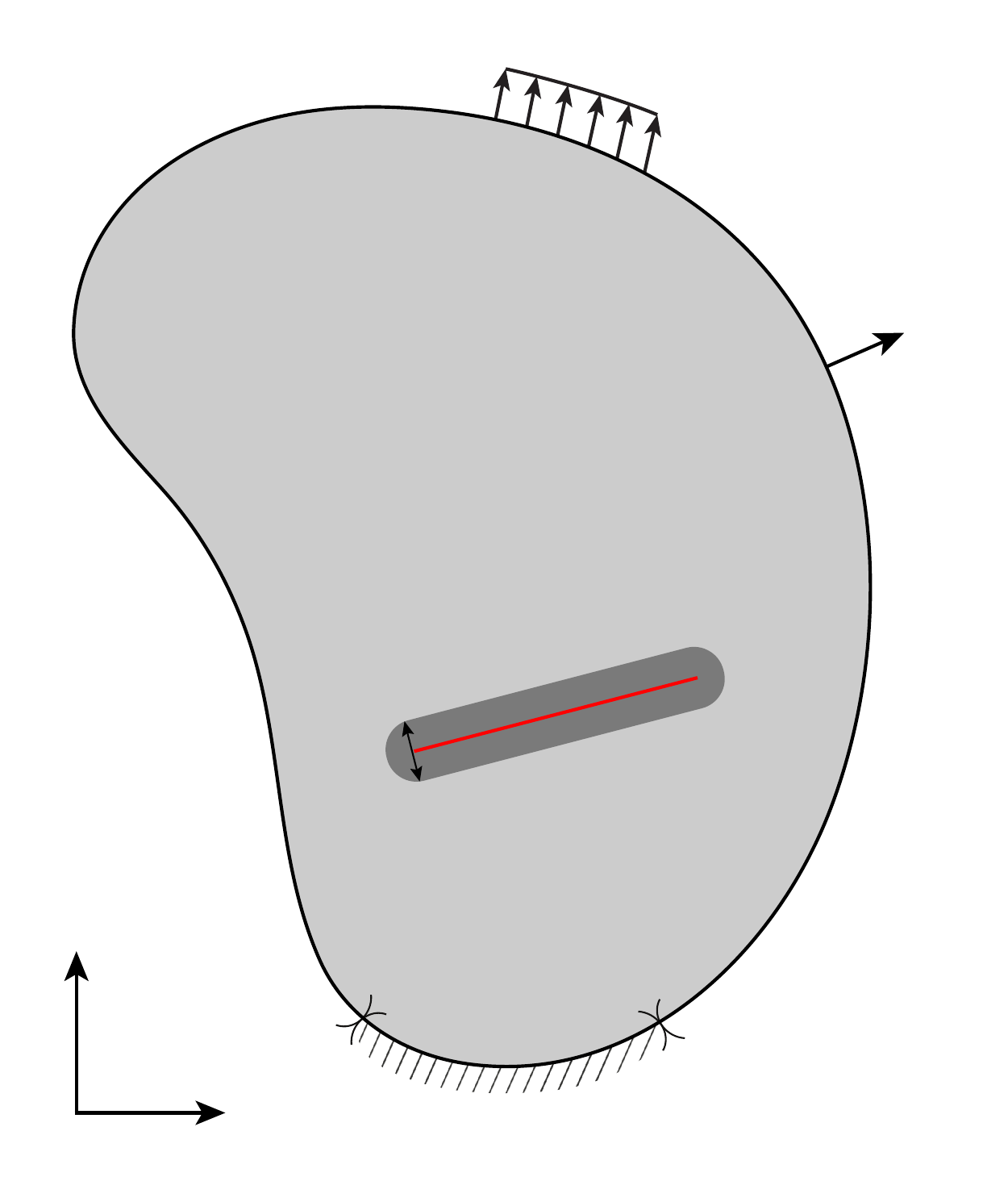	
        \label{fig:Phase_field}
        }
\hspace*{\fill}
       \caption{(a) Solid domain $\Omega$ with smooth boundary $\partial \Omega = \Gamma_\mathrm{D} \cup \Gamma_\mathrm{N}$, where Dirichlet boundary conditions $\bar{\boldsymbol{u}}$ are prescribed on $\Gamma_\mathrm{D}$ and surface tractions $\bar{\boldsymbol{t}}$ are prescribed on $\Gamma_\mathrm{N}$, respectively. (b) A traction-free crack $\Gamma_\mathrm{c}$ is introduced into the domain $\Omega$. (c) The sharp crack $\Gamma_\text{c}$ is regularised as a smooth area $\Omega_{\text{c}}$ and scaled by the length scale parameter $l_0$.}
\end{figure}

We consider the elastostatics boundary value problem for a solid domain $\Omega \subset \mathbb{R}^2$ with closure $\overline{\Omega}$ and boundary $\partial \Omega\equiv\Gamma=\overline{\Omega} \setminus \Omega$, as shown in Figure~\ref{fig:Original_model}.~Two kinds of boundary conditions along the normal vector $\boldsymbol n$ are considered, such that $\Gamma_{\mathrm{D}} \cap \Gamma_{\mathrm{N}} =  \emptyset$, where subscripts D and N denote Dirichlet and Neumann boundary conditions, respectively. In this framework, a non-homogeneous essential boundary condition (BC) $\bar{\boldsymbol u}$ and a traction BC $\bar{\boldsymbol t}$ are imposed on $\Gamma_{\mathrm{D}}$ and $\Gamma_{\mathrm{N}}$, respectively. 

\subsection{Crack Propagation}\label{sec:Crack_propagation}

We adopt the PF method to handle the solid model $\Omega$, which includes a traction-free sharp crack $\Gamma_\text{c}$, as shown in Figure~\ref{fig:No_Phase_field}.
The total potential energy $\Pi$ consists of the elastic strain energy $\Pi_\text{el}$, the fracture energy $\Pi_\text{frac}$, minus the external work ${W}_{\text{ext}}$:
\begin{equation}\label{eq:fracture}
\Pi(\mathbf{u}, \Gamma)=\Pi_\text{el}+\Pi_\text{frac} - W_{\text{ext}}=\int_{\Omega} \psi_\mathrm{e}\left(\boldsymbol{\varepsilon}(\boldsymbol{u})\right) \mathrm{d} \Omega+\int_{\Gamma} G_\text{c} \mathrm{d} \Gamma - W_{\text{ext}},
\end{equation}
with elastic energy density $\psi_\mathrm{e}\left(\boldsymbol{\varepsilon}(\boldsymbol{u})\right)$ and the critical energy release rate ${G}_\text{c}$, where $\boldsymbol{u}: \Omega \rightarrow \mathbb{R}^2$ is the displacement field, and $\boldsymbol{\varepsilon}(\boldsymbol{u}) = \frac{1}{2}\left(\nabla \boldsymbol{u} + \nabla \boldsymbol{u}^{\top} \right)$ is the linearised strain. 

To enable an efficient numerical implementation of Equation~\eqref{eq:fracture}, the fracture energy $\Pi_\text{frac}$ defined on the sharp crack $\Gamma_{\text{c}}$ is regularised by a domain integral defined over $\Omega$ as
\begin{equation}
\int_{\Gamma}{G}_\text{c} \mathrm{d} \Gamma \approx \int_{\Omega} {G}_\text{c} \gamma(\phi, \nabla \phi) \mathrm{d} \Omega,
\end{equation}
where $\phi = \phi(\mathbf{x}) \in[0,1], \forall \mathbf{x} \in \Omega$ is the scalar phase field to represent the material state that continuously transits between $\phi = 0$ (intact material) and $\phi = 1$ (completely damaged material), and $\gamma(\phi, \nabla \phi)$ is the crack surface density functional given by
\begin{equation}
\gamma(\phi, \nabla \phi) = \frac{1}{c_\alpha}\left(\frac{\alpha(\phi)}{l_0}+l_0|\nabla \phi|^2\right),
\end{equation}
where $\alpha(\phi)$ is the geometric function determining the ultimate distribution of the crack field, and $c_\alpha = 4 \int_0^1 \sqrt{\alpha(\phi)} \mathrm{d} \phi$.~Different expressions for $\alpha(\phi)$ and $c_\alpha$ result in variants of the fracture surface energy approximation~\cite{Mandal2020}.~With the introduction of the crack surface density function $\gamma(\phi, \nabla \phi)$, the discrete representation of the sharp crack $\Gamma_\text{c}$ is changed into a diffused crack description, where $l_0$ is the length scale parameter to control the smeared width of the crack zone $\Omega_{\text{c}}$ (Figure~\ref{fig:Phase_field}).
It should be noted that the regularised formulation results into the original variational model of the brittle fracture when $l_0 \rightarrow 0$ in the sense of $\Gamma$-convergence~\cite{Braides2006}.

For the elastic strain energy $\Pi_\text{el}$, a degradation function $g(\phi)$ is introduced to consider the effect of material damage. $\Pi_\text{el}$ is then rewritten as
\begin{equation}\label{eq:pie_elasticenergy}
\Pi_\text{el} = \int_{\Omega} g(\phi) \psi_\text{e} \left(\boldsymbol{\varepsilon}(\boldsymbol{u})\right) \mathrm{d} \Omega,
\end{equation}
where $g(\phi)$ should satisfy the following conditions $g(0)=1, g(1)=0, g^{\prime}(\phi) \leq 0 \text { for } 0 \leq \phi \leq 1$.

The external work of volume and surface forces is given by
\begin{equation}
W_{\text{ext}} = \int_{\Omega}  \boldsymbol{b} \boldsymbol{u} \mathrm{d} \Omega+\int_{\Gamma_{\text{N}}}  \bar{\boldsymbol{t}} \boldsymbol{u} \mathrm{d} \Gamma_{\text{N}},
\end{equation}
where $ \boldsymbol{b}$ is the body force.

According to the above equations, the expression of the total potential energy $\Pi$ is rewritten in the following form:
\begin{equation}
\Pi \approx \int_{\Omega} g(\phi) \psi_\text{e} \left(\boldsymbol{\varepsilon}(\boldsymbol{u})\right) \mathrm{d} \Omega + \int_{\Omega} \frac{{G}_\text{c} }{c_\alpha}\left(\frac{\alpha(\phi)}{l_0}+l_0|\nabla \phi|^2\right) \mathrm{d} \Omega-\left(\int_{\Omega}  \boldsymbol{b} \boldsymbol{u} \mathrm{d} \Omega+\int_{\Gamma_{\text{N}}}  \bar{\boldsymbol{t}} \boldsymbol{u} \mathrm{d} \Gamma_{\text{N}}\right).
\end{equation}
Its variation with respect to the displacement field $\boldsymbol{u}$ and the crack field $\phi$ provides the fundamental equations of the phase field fracture model
\begin{align}
& g(\phi) \nabla \cdot \boldsymbol{\sigma}+\boldsymbol{b}=0,&  &  \text{ in } \Omega \\
& g(\phi) \boldsymbol{\sigma} \cdot \boldsymbol{n}=\boldsymbol{\bar{t}},& & \text{ on } \partial \Gamma_{\text{N}} \\
& \boldsymbol{u} =\boldsymbol{\bar{u}},& & \text{ on } \partial \Gamma_{\text{D}} \\ \label{eq:phasefieldequation} 
& g^{\prime}(\phi) \psi_\text{e} + \frac{G_\text{c}}{c_\alpha}\left(\frac{\alpha^{\prime}(\phi)}{l_0}-2 l_0 \nabla \cdot \nabla \phi \right)=0,& &\text{ in } \Omega \\
& \nabla \phi \cdot \boldsymbol{n}=0,& & \text{ on } \partial \Omega.
\end{align}
In this work, we set $\alpha(\phi) = \phi^2$ with $c_\alpha = 4\int_{0}^{1}\sqrt{\alpha(\phi)} \mathrm{d} \phi = 2$, and adopt the quadratic degradation function $g(\phi) = (1-k)(1-\phi)^2 + k$ with a small value of $k = 10^{-6}$, which can avoid singularity issues within the finite element discretization.
The governing formulation of the crack propagation in Equation~\eqref{eq:phasefieldequation} can be expressed as 
\begin{equation}
\phi - l_0^2 \nabla \cdot \nabla \phi = (1-k)(1-\phi)D_\mathrm{d},
\end{equation}
where $D_\mathrm{d} = \frac{2 \psi_\mathrm{e}}{G_\text{c}/l_0}$ is defined as the crack driving state function that depends on different fracture failure criteria. For instance, instead of the critical energy release rate $G_\text{c}$, the critical fracture energy density $\psi_\text{c}$ formulated as
\begin{equation}
\psi_\mathrm{c}=\frac{\sigma_\mathrm{c}^2}{2 E}
\end{equation}
can be used to define the crack driving state function
\begin{equation}
D_\mathrm{d} = \left\langle \frac{\psi_\mathrm{e}}{\psi_\mathrm{c}} - 1 \right\rangle,
\end{equation} 
where $\sigma_\mathrm{c}$ is the critical fracture stress or material strength, and $E$ denotes the material's Young's modulus. The ramp function $\langle x\rangle:=(x+|x|)/2$ is determined by the Macaulay bracket.~The criteria above do not differentiate between tension and compression modes, and a formulation based on the decomposition of the free energy into tensile and compressive parts was considered in the work of Miehe et al.~\cite{Miehe2010}.

As an alternative to the energy-based phase field model, the stress-based one enables us to capture purely stress-driven fractures, such as the principal tensile stress criterion with 
\begin{equation}
D_\mathrm{d} = \left\langle\sum_{i=1}^3\left(\frac{\left\langle{\sigma}_i\right\rangle}{\sigma_\mathrm{c}}\right)^2-1\right\rangle,
\end{equation}
where ${\sigma}_i$ is the $i$th principal stress.~It can be seen that the stress-based $D_\mathrm{d}$ is independent of the length scale parameter, which reduces the sensitivity of the results with respect to $l_0$.~We adopt this stress-based phase field fracture model in this work.

\subsection{Phase Transformation}\label{sec:Phase_transformation}
The evolution of the T$\rightarrow$M phase transformation in zirconia described by the time-dependent Ginzburg-Landau kinetic equation~\cite{Landau1965}, defined as
\begin{equation}\label{eq:phasetransformation}
\frac{\partial \eta_p}{\partial t}=-L \frac{\delta F(\boldsymbol{u}, \eta_p)}{\delta \eta_p} +\varsigma, \quad p = 1, ..., n ,
\end{equation}
where $\eta_p \in [0, 1]$ is the {order parameter} that characterizes the $p$th variant of the monoclinic phase, which takes 0 in the tetragonal phase and 1 in the monoclinic phase, respectively. $L$ is the kinetic coefficient controlling the speed of the transformation, and $F$ is the total free energy consisting of {chemical} free energy $F_\mathrm{ch}$, {elastic} strain energy $F_\mathrm{el}$, and surface energy $F_\mathrm{surf}$~\cite{Becher1992}. The functional derivative $\frac{\delta F(\boldsymbol{u}, \eta_p)}{\delta \eta_p }$ is the thermodynamic driving force for the spatial and temporal evolution of $\eta_p$, and $\varsigma$ is the Langevin noise describing the thermal fluctuation.

The chemical free energy $F_\mathrm{ch}$ primarily originates from the difference of Gibbs free energy between tetragonal and monoclinic phases.~Considering the interfacial energies between the coexisting phases~\cite{Wang1997}, $F_\mathrm{ch}$ is expressed as
\begin{equation}\label{eq:chemicalenergy}
F_\mathrm{ch}=\int_{\Omega}\left[\Delta G f\left(\eta_1, ..., \eta_n\right)+\frac{1}{2} \sum_{p=1}^{n} \beta_{i j}(p) \nabla_{i} \eta_p \nabla_{j} \eta_p \right] \mathrm{d}{\Omega},
\end{equation}
where $\Delta G f\left(\eta_1, ..., \eta_n\right)$ is the local specific free energy describing the basic bulk thermodynamic properties of the system, and $\beta_{ij}(p)$ is a positive coefficient determining the contribution of gradient energy. $\Delta G$ represents the difference in the specific chemical free energy between the tetragonal and monoclinic phases, and $f\left(\eta_1, ..., \eta_n\right)$ can be approximated by the forth-order Landau polynomial~\cite{Levitas2002}, given by
\begin{equation}\label{eq:landaupolynomial}
f\left(\eta_1, ..., \eta_n\right) = a\sum_{p=1}^{n} \eta_p^{2} + b \sum_{p=1}^{n} \eta_p^{3} + c \sum_{p=1}^{n} \eta_p^{4},
\end{equation}
where $a$, $b$, and $c$ are the expansion coefficients at a fixed temperature.

The elastic strain energy $F_\mathrm{el}$ is defined as
\begin{equation}\label{eq:elasticenergy}
F_\mathrm{el} = \frac{1}{2} \int_{\Omega} \mathrm{C}_{ijkl}(\eta_p) \varepsilon_{kl}^\mathrm{el}(\boldsymbol{u})  \varepsilon_{ij}^\mathrm{el}(\boldsymbol{u}) \mathrm{d} \Omega,
\end{equation}
where the elastic strain $\varepsilon_{ij}^\mathrm{el}(\boldsymbol{u})$ is the difference between the total strain $\varepsilon_{ij}(\boldsymbol{u})$ and the stress free strain $\varepsilon_{ij}^{0}$~\cite{Mamivand2014} defined as
\begin{equation}\label{eq:elasticstrain}
\varepsilon_{ij}^\mathrm{el}(\boldsymbol{u}) =\varepsilon_{i j}(\boldsymbol{u}) - \varepsilon_{i j}^{0} = \frac{1}{2}\left(\frac{\partial u_{i}}{\partial x_{j}}+\frac{\partial u_{j}}{\partial x_{i}}\right) - \sum_{p=1}^n \eta_p^{2} \varepsilon_{ij}^{00}(p).
\end{equation}
Stress-free strain $\varepsilon_{ij}^{0}$ characterizes the degree of lattice mismatch between the tetragonal and monoclinic phases, where $\varepsilon_{i j}^{00}(p)$ is the transformation-induced strain of the $p$th variant of the monoclinic phase.~Elastic constants $\mathrm{C}_{ijkl}(\eta_p)$ are given by
\begin{equation}\label{eq:elasticconstants}
\mathrm{C}_{ijkl}(\eta_p) = P\left(\sum_{p=1}^n \eta_p\right) \mathrm{C}_{ijkl}^{\mathrm{Mon}}+\left(1-P\left(\sum_{p=1}^n \eta_p\right)\right) \mathrm{C}_{ijkl}^\mathrm{Tet},
\end{equation}
where $\mathrm{C}_{ijkl}^\mathrm{Mon}$ and $\mathrm{C}_{ijkl}^{\mathrm{Tet}}$ are monoclinic and tetragonal elastic constants, respectively, and we choose the polynomial $P(\eta)$ with $P(0) = 0$ and $P(1) = 1$ in line with~\cite{Mahmood2014A} as
\begin{equation}\label{eq:peta}
P(\eta)=\eta^{3}\left(6 \eta^{2}-15 \eta+10\right).
\end{equation}

It is well-known that zirconia is usually in the monoclinic phase at room temperature. However, by decreasing the grain sizes to a certain value or adding stabilizers like ceria or yttria, the surface energy $F_\text{surf}$ increases to stabilize zirconia in the tetragonal phase at room temperature, which can be defined as
\begin{equation}\label{eq:surfaceEnergy}
F_\mathrm{surf} = \frac{6 \Delta S}{d} f\left(\eta_1, ..., \eta_n\right),
\end{equation}
where $\Delta S$ includes the surface free energy and twinning energy, and $d$ is the diameter of individual grains~\cite{Garvie1985}.

According to Equations~\eqref{eq:chemicalenergy}-\eqref{eq:surfaceEnergy}, the thermodynamic driving force for evolving the phase variable $\eta_p$ can be derived as
\begin{equation}\label{eq:finalphasetransformation}
\begin{aligned}
\frac{\delta F}{\delta \eta_p} & =  \left(\Delta G + \frac{6 \Delta S}{d} \right) \frac{\partial f(\eta_1, ..., \eta_n)}{\partial \eta_p} - \beta (p) \nabla \cdot \nabla \eta_p\\
& - \eta_p \mathrm{C}_{ijkl} \left( \varepsilon_{kl}^{00}(p) \varepsilon_{ij}^\mathrm{el} + \varepsilon_{kl}^\mathrm{el} \varepsilon_{ij}^{00}(p)\right) + \frac{1}{2} \frac{\partial P\left(\sum_{p=1}^{n}\eta_p\right)}{ \partial \eta_p} \left(\mathrm{C}_{ijkl}^{\mathrm{Mon}}-\mathrm{C}_{ijkl}^\mathrm{Tet}\right)\varepsilon_{kl}^\mathrm{el} \varepsilon_{ij}^\mathrm{el}.
\end{aligned}
\end{equation}

In this work, we set the Langevin noise $\varsigma$ to 0, and only the volumetric part of transformation-induced strain $(\varepsilon_{11}^{00} \neq 0, \varepsilon_{22}^{00} \neq 0, \varepsilon_{12}^{00} =  \varepsilon_{21}^{00} = 0)$ is taken into account, which results in the single variant of the monoclinic phase. It has indeed been shown experimentally that the shear component of the transformation is often compensated for by twinning ~\cite{Chevalier2007}. Then, the governing equations of the displacement field $\boldsymbol{u}$ and the phase field $\eta$ are given by

\begin{align}
& \nabla \cdot \boldsymbol{\sigma} + \boldsymbol{b}=0,  &   \text{ in } &\Omega \\
&  \boldsymbol{\sigma} \cdot \boldsymbol{n}=\boldsymbol{\bar{t}},  & \text{ on } &\partial \Gamma_{\text{N}} \\
& \boldsymbol{u} =\boldsymbol{\bar{u}},  & \text{ on }& \partial \Gamma_{\text{D}} \\ 
& \begin{aligned}
\frac{\partial \eta}{\partial t} + L \beta \nabla \cdot \nabla \eta  & =   L \eta \mathrm{C}_{ijkl} \left( \varepsilon_{kl}^{00} \varepsilon_{ij}^\mathrm{el}
+ \varepsilon_{kl}^\mathrm{el} \varepsilon_{ij}^{00}\right) - 15 L \eta^2 (\eta-1)^2 \left(\mathrm{C}_{ijkl}^{\mathrm{Mon}}-\mathrm{C}_{ijkl}^\mathrm{Tet}\right)\varepsilon_{kl}^\mathrm{el} \varepsilon_{ij}^\mathrm{el} \\
& -L \left(\Delta G + \frac{6 \Delta S}{d} \right) \left( 2a \eta + 3b\eta^2 + 4c\eta^3 \right)   \\
\end{aligned}, & \text{ in }& \Omega  \label{eq1} \\ 
& \nabla \eta \cdot \boldsymbol{n}=0, & \text{ on }& \partial \Omega.
\end{align}

\section{Results}\label{sec:Numericalexamples}
We start at the nanoscale, where we assess the external stress required to trigger the T$\rightarrow$M phase transformation in the zirconia mortar and the resulting transformation-induced strain at room temperature. The stress-strain curve derived from this analysis is then employed as the constitutive law for the mortar material in the microscale model, where we investigate crack propagation in the brick-and-mortar structure and examine the influence of geometric parameters and material properties on fracture performance. Subsequently, a gradient-free optimization algorithm is applied to refine the brick-and-mortar structure layout, aiming to maximize fracture toughness. The governing equations for phase transformation and crack propagation are implemented in the commercial software COMSOL Multiphysics.

\subsection{Nanoscale}

\begin{figure}[!ht]
\centering
\hspace*{\fill}
     \subfloat[]{
        \centering
        \def\svgwidth{0.4\textwidth}
\begingroup%
  \makeatletter%
  \providecommand\color[2][]{%
    \errmessage{(Inkscape) Color is used for the text in Inkscape, but the package 'color.sty' is not loaded}%
    \renewcommand\color[2][]{}%
  }%
  \providecommand\transparent[1]{%
    \errmessage{(Inkscape) Transparency is used (non-zero) for the text in Inkscape, but the package 'transparent.sty' is not loaded}%
    \renewcommand\transparent[1]{}%
  }%
  \providecommand\rotatebox[2]{#2}%
  \newcommand*\fsize{\dimexpr\f@size pt\relax}%
  \newcommand*\lineheight[1]{\fontsize{\fsize}{#1\fsize}\selectfont}%
  \ifx\svgwidth\undefined%
    \setlength{\unitlength}{309.19100189bp}%
    \ifx\svgscale\undefined%
      \relax%
    \else%
      \setlength{\unitlength}{\unitlength * \real{\svgscale}}%
    \fi%
  \else%
    \setlength{\unitlength}{\svgwidth}%
  \fi%
  \global\let\svgwidth\undefined%
  \global\let\svgscale\undefined%
  \makeatother%
  \begin{picture}(1,0.7233716)%
    \lineheight{1}%
    \setlength\tabcolsep{0pt}%
    \put(0,0){\includegraphics[width=\unitlength,page=1]{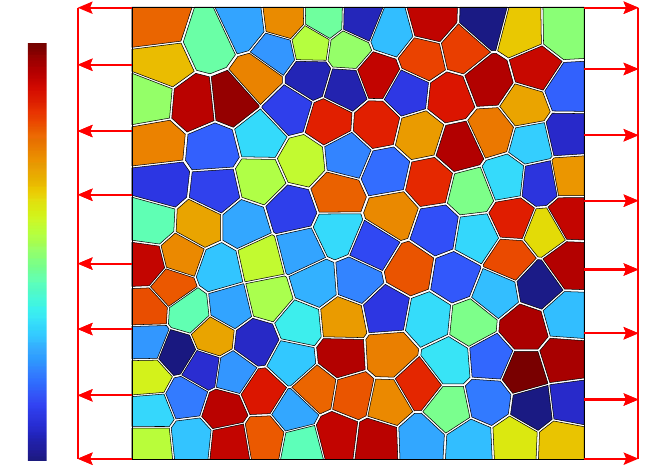}}%
    \put(0.04,0.70){\color[rgb]{0,0,0}\makebox(0,0)[lt]{\lineheight{1.25}\smash{\begin{tabular}[t]{l}$\zeta$\end{tabular}}}}%
    \put(-0.09,0.63){\color[rgb]{0,0,0}\makebox(0,0)[lt]{\lineheight{1.25}\smash{\begin{tabular}[t]{l}180$^\circ$\end{tabular}}}}%
    \put(-0.06,0.32){\color[rgb]{0,0,0}\makebox(0,0)[lt]{\lineheight{1.25}\smash{\begin{tabular}[t]{l}90$^{\circ}$\end{tabular}}}}%
    \put(-0.02,-0.01){\color[rgb]{0,0,0}\makebox(0,0)[lt]{\lineheight{1.25}\smash{\begin{tabular}[t]{l}0\end{tabular}}}}%
    \put(0.95,0.34){\color[rgb]{0,0,0}\makebox(0,0)[lt]{\lineheight{1.25}\smash{\begin{tabular}[t]{l}$\boldsymbol{t}$\end{tabular}}}}%
    \put(0.15,0.35){\color[rgb]{0,0,0}\makebox(0,0)[lt]{\lineheight{1.25}\smash{\begin{tabular}[t]{l}$\boldsymbol{t}$\end{tabular}}}}%
  \end{picture}%
\endgroup%
	
        \label{fig:Grain_orientations}
        }
\hspace*{\fill}
\hspace{0.5cm}
     \subfloat[]{
        \centering
        \def\svgwidth{0.355\textwidth}
\begingroup%
  \makeatletter%
  \providecommand\color[2][]{%
    \errmessage{(Inkscape) Color is used for the text in Inkscape, but the package 'color.sty' is not loaded}%
    \renewcommand\color[2][]{}%
  }%
  \providecommand\transparent[1]{%
    \errmessage{(Inkscape) Transparency is used (non-zero) for the text in Inkscape, but the package 'transparent.sty' is not loaded}%
    \renewcommand\transparent[1]{}%
  }%
  \providecommand\rotatebox[2]{#2}%
  \newcommand*\fsize{\dimexpr\f@size pt\relax}%
  \newcommand*\lineheight[1]{\fontsize{\fsize}{#1\fsize}\selectfont}%
  \ifx\svgwidth\undefined%
    \setlength{\unitlength}{270bp}%
    \ifx\svgscale\undefined%
      \relax%
    \else%
      \setlength{\unitlength}{\unitlength * \real{\svgscale}}%
    \fi%
  \else%
    \setlength{\unitlength}{\svgwidth}%
  \fi%
  \global\let\svgwidth\undefined%
  \global\let\svgscale\undefined%
  \makeatother%
  \begin{picture}(1,0.81481484)%
    \lineheight{1}%
    \setlength\tabcolsep{0pt}%
    \put(0,0){\includegraphics[width=\unitlength,page=1]{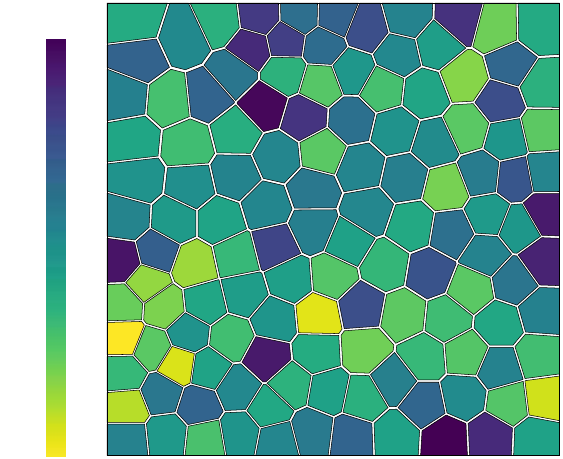}}%
    \put( 0.08,0.78){\color[rgb]{0,0,0}\makebox(0,0)[lt]{\lineheight{1.25}\smash{\begin{tabular}[t]{l}$\eta$\end{tabular}}}}%
    \put(-0.18,0.70){\color[rgb]{0,0,0}\makebox(0,0)[lt]{\lineheight{1.25}\smash{\begin{tabular}[t]{l}1.14e-4\end{tabular}}}}%
    \put(-0.18,-0.02){\color[rgb]{0,0,0}\makebox(0,0)[lt]{\lineheight{1.25}\smash{\begin{tabular}[t]{l}9.88e-5\end{tabular}}}}%
  \end{picture}%
\endgroup%
	
        \label{fig:Initial_eta}
        }
\hspace*{\fill}
       \caption{The nanoscale model, $\SI{400}{\nano\metre} \times \SI{400}{\nano\metre}$, includes 128 grains with an average grain size of $\SI{40}{\nano\metre}$ and 8.93\% of the area occupied by grain boundaries. Traction boundary conditions $\boldsymbol{t}$ are applied in the horizontal direction. The model is illustrated with (a) the distribution of grain orientations $\zeta$ relative to the horizontal axis and (b) the initial phase state $\eta$ (0 indicating tetragonal and 1 monoclinic), which follows a random distribution with a mean value of $1\times10^{-4}$ and a standard deviation of $1\times10^{-5}$.}
\label{fig:Polycrystalline_zirconia}
\end{figure}
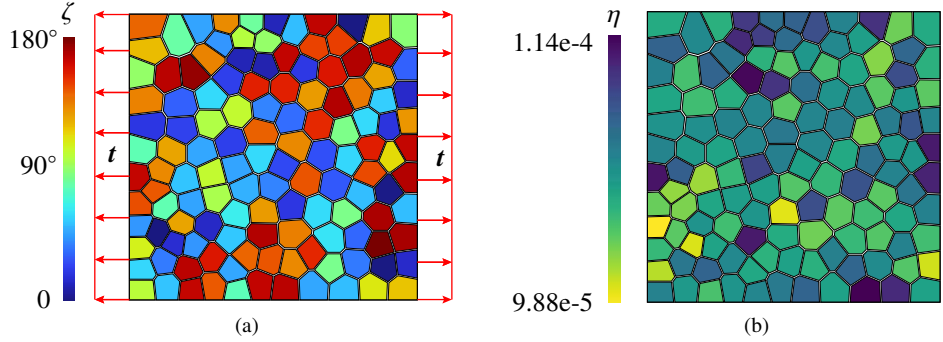

The nanoscale model is defined as a square domain with edge length $\SI{400}{\nano\metre}$, as shown schematically in Figure~\ref{fig:Grain_orientations}. The model contains $n = 128$ grains and 8.93\% grain boundary (GB) area. The corresponding grain size $d$ is calculated as $2\sqrt{length^2/n/\pi}$, yielding a grain size of $\SI{40}{\nano\metre}$. This grain size corresponds to the experimental one, where it has been selected to have an appropriate thickness of the mortar interfacing the given 300-400 $\si{\nano\metre}$ thick bricks (alumina platelets). This 2D polycrystalline model is discretized by linear triangular elements with a maximum mesh size $h = \SI{2}{\nano\metre}$. A time-dependent traction $\boldsymbol{t}$, incremented at a constant rate of $\Delta \boldsymbol{t} = \SI{0.2}{\mega\pascal/\second}$, is applied in the horizontal direction. 
Although a larger model with more grains of the same average size would provide more accurate predictions for the T$\rightarrow$M phase transformation behaviour, we selected the model size, number of grains, and mesh size to strike a balance between accuracy and computational cost. In this model, ceria-stabilized zirconia is considered an anisotropic material, with the elastic constants of both the tetragonal and monoclinic phases provided in Table~\ref{tab:Tetragonal_Monoclinic}. Random orientations $\zeta$ are assigned to all grains, as shown in Figure~\ref{fig:Grain_orientations}. The initial phase variable $\eta$, which characterizes the monoclinic/tetragonal phases, is randomly distributed with a mean value of $1\times10^{-4}$ and a standard deviation of $1\times10^{-5}$, as illustrated in Figure~\ref{fig:Initial_eta}. All other parameters used in Equation~\eqref{eq1} are listed in Table~\ref{tab:parameters}. 

\begin{table}[!ht]
\renewcommand{\arraystretch}{1.2}
\centering
\begin{tabular}{p{0.05\textwidth}p{0.05\textwidth}p{0.05\textwidth}p{0.05\textwidth}p{0.05\textwidth}p{0.05\textwidth}}
\hline $\mathrm{C}_{11}^{\text{M}}$ & $\mathrm{C}_{33}^{\text{M}}$ & $\mathrm{C}_{44}^{\text{M}}$ & $\mathrm{C}_{66}^{\text{M}}$ & $\mathrm{C}_{12}^{\text{M}}$ & $\mathrm{C}_{13}^{\text{M}}$ \\
\hline 393 & 344 & 44 & 74 & 148 & 72 \\
\hline
\end{tabular}
\begin{tabular}{ccccccccccccc}
\hline $\mathrm{C}_{11}^{\text{T}}$ & $\mathrm{C}_{22}^{\text{T}}$ & $\mathrm{C}_{33}^{\text{T}}$ & $\mathrm{C}_{44}^{\text{T}}$ & $\mathrm{C}_{55}^{\text{T}}$ & $\mathrm{C}_{66}^{\text{T}}$ & $\mathrm{C}_{12}^{\text{T}}$ & $\mathrm{C}_{13}^{\text{T}}$ & $\mathrm{C}_{16}^{\text{T}}$ & $\mathrm{C}_{23}^{\text{T}}$ & $\mathrm{C}_{26}^{\text{T}}$ & $\mathrm{C}_{36}^{\text{T}}$ & $\mathrm{C}_{45}^{\text{T}}$ \\
\hline 361 & 408 & 258 & 100 & 81 & 126 & 142 & 55 & -21 & 196 & 31 & -18 & -23 \\
\hline
\end{tabular}
\caption{Elastic constants ($\si{\giga\pascal}$) of ceria-stabilized zirconia, for monoclinic (M) and tetragonal (T) phases~\cite{Zhao2011}.}
\label{tab:Tetragonal_Monoclinic}
\end{table}

\begin{table}[!ht]
\centering
\begin{tabular}{lc}
\hline Temperature $T \left( \si{\kelvin} \right)$ & $293$ \\
Chemical driving force $\Delta G \left(\si{\joule}/\si{\metre}^3\right)$~\cite{Zhang2003} & $-215.34 \times 10^6$ \\
Surface energy density $\Delta S \left(\si{\joule}/\si{\metre}^2\right)$ & $1.4784$\\
Kinetic coefficient $L\left(\si{\meter}^3/\si{\joule}/\si{\second}\right)$ & $2$ \\
Gradient energy coefficient $\beta\left(\si{\joule}/\si{\metre}\right)$ & $1 \times 10^{-8}$ \\
Coefficient $a$ & 0.02 \\
Coefficient $b$ & 11.94 \\
Coefficient $c$ & -11.96 \\
The Langevin noise $\varsigma$ & 0\\
Transformation strain $\varepsilon_{ij}^{00}$ & $\begin{bmatrix}
0.0049 & 0 \\
0 & 0.0180 
\end{bmatrix}$ \\
\hline
\end{tabular}
\caption{Input parameters used in the governing equation of the T$\rightarrow$M phase transformation (Equation \eqref{eq1})}
\label{tab:parameters}
\end{table}

\subsubsection{Grain boundary properties}

\begin{figure}[htbp]
\centering
\hspace*{\fill}
     \subfloat[]{
        \centering
        \def\svgwidth{0.85\textwidth}
        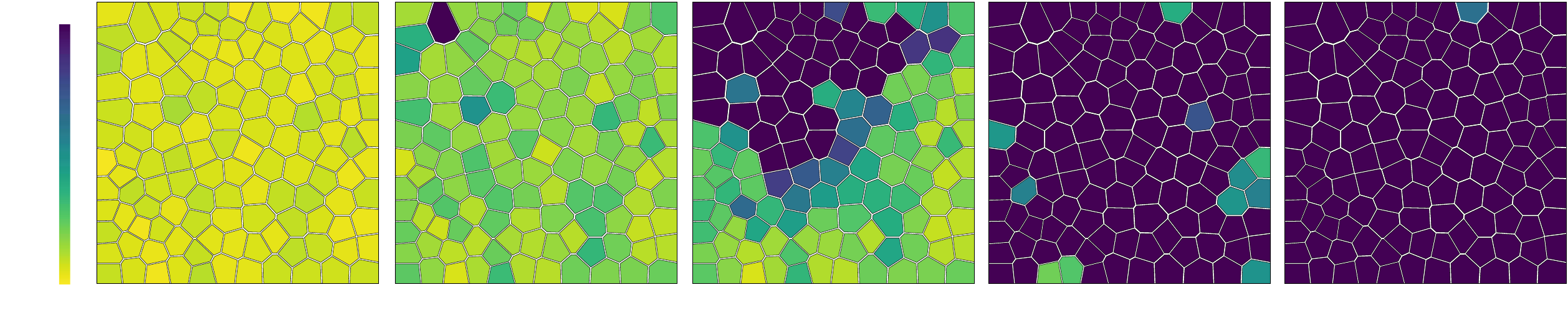	
        \label{fig:Orientation5_005}
        }
\hspace*{\fill}
\\
\hspace*{\fill}
     \subfloat[]{
        \centering
        \def\svgwidth{0.85\textwidth}
        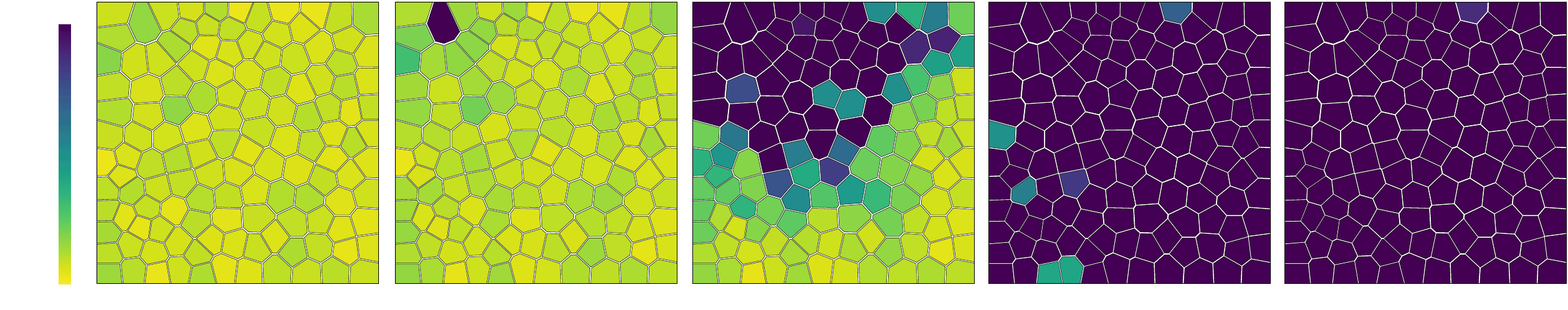	
        \label{fig:Orientation5_025}
        }
\hspace*{\fill}
\\
\hspace*{\fill}
     \subfloat[]{
        \centering
        \def\svgwidth{0.85\textwidth}
        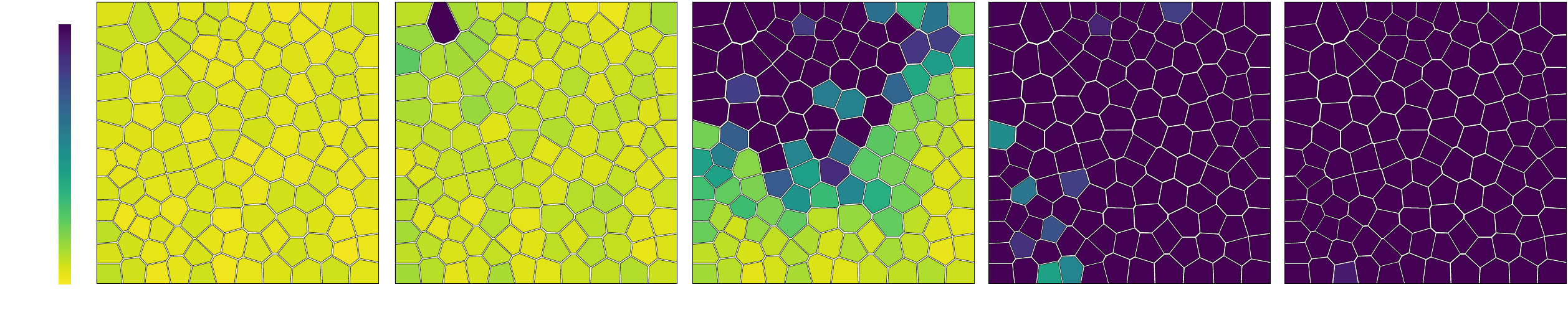	
        \label{fig:Orientation5_050}
        }
\hspace*{\fill}
\\
\hspace*{\fill}
     \subfloat[]{
        \centering
        \def\svgwidth{0.85\textwidth}
        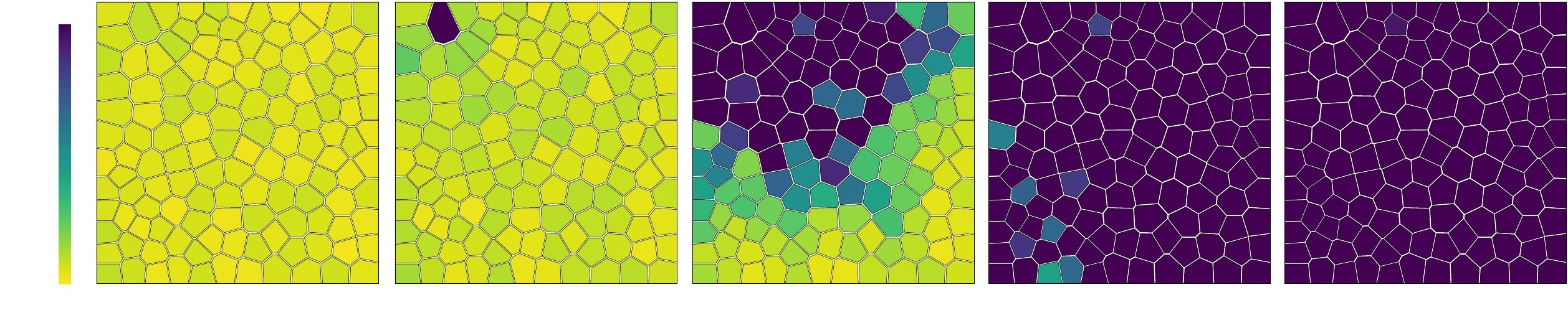	
        \label{fig:Orientation5_100}
        }
\hspace*{\fill}
\\
\hspace*{\fill}
     \subfloat[]{
        \centering
        \includegraphics[width=0.80\textwidth]{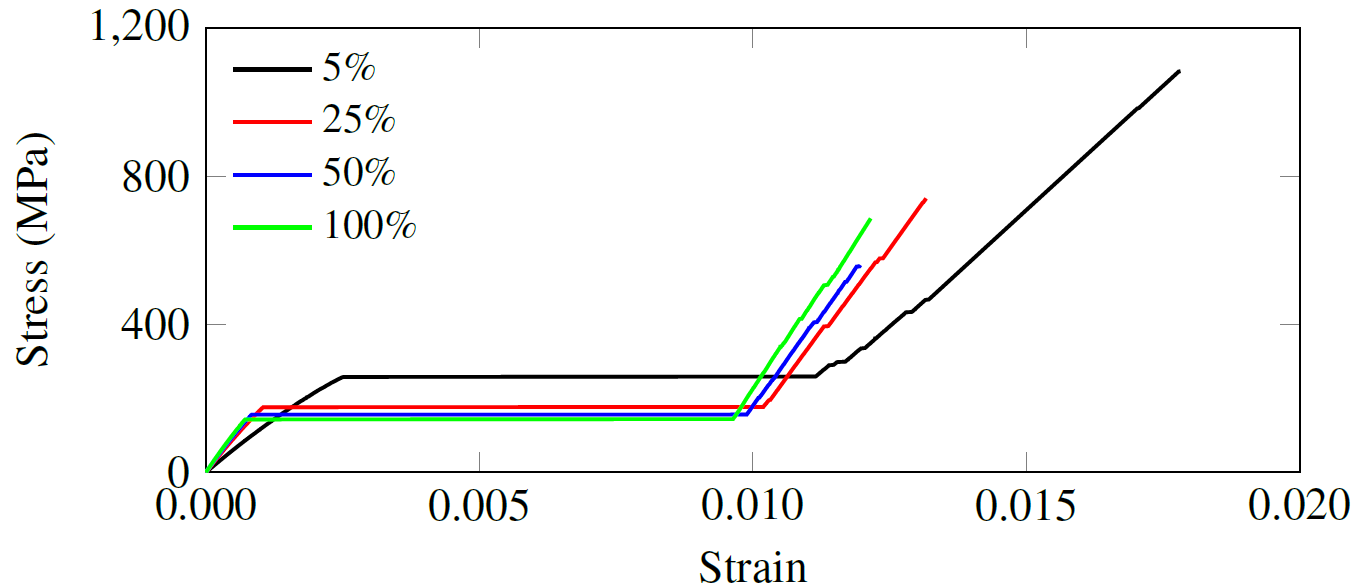}
\label{fig:stress_strain_curves_grain_boundary}
}
\hspace*{\fill}
\caption{The evolution of the phase variable $\eta$ (0 indicating tetragonal and 1 monoclinic) under different grain boundary properties scaled to (a) 5\%, (b) 25\%, (c) 50\%, and (d) 100\% of the bulk grain properties. Phase patterns obtained under $\SI{100}{\mega\pascal}$ are shown in the $1^{\text{st}}$ column. $\SI{259}{\mega\pascal}$, $\SI{176}{\mega\pascal}$, $\SI{156}{\mega\pascal}$, and $\SI{143}{\mega\pascal}$ are the stresses needed to trigger the simultaneous T$\rightarrow$M phase transformation in most of grains, where the corresponding phase patterns are shown in the $2^{\text{nd}} - 4^{\text{th}}$ columns. The stresses shown in the last column - $\SI{1084}{\mega\pascal}$, $\SI{729}{\mega\pascal}$, $\SI{557}{\mega\pascal}$, and $\SI{685}{\mega\pascal}$ - are those needed to fully accomplish the phase transformation. The plots demonstrate that the grain boundary properties only affect the stress required to trigger and accomplish the phase transformation, not the phase evolution patterns, and that softer GBs require higher external stresses to initiate the phase transformation. Figure (e) shows the corresponding stress-strain curves obtained for the 5\%, 25\%, 50\%, and 100\% cases.}
 \label{fig:evolution_eta_grain_boundary}
\end{figure}

Due to the challenges in measuring grain boundary (GB) properties, finding reliable experimental data for proper references is difficult. In this work, we first assign different GB properties and investigate their effect on phase transformation behaviour, setting the elastic constants of GBs to 5\%, 25\%, 50\%, and 100\% of those of the tetragonal phase. 
The corresponding evolution of the phase variable $\eta$ is shown in Figures~\ref{fig:Orientation5_005}--\ref{fig:Orientation5_100}, where all models exhibit similar patterns, with only slight differences in the model with the lowest GB properties.
Since grain properties depend on their orientation, the stresses required to initiate the phase transformation vary. Thus, when most grains fully transform from the tetragonal to the monoclinic phase, a few grains may still be undergoing transformation.
We therefore define two stresses: the stress required to trigger the phase transformation and the stress needed to complete the transformation in all grains.
The stresses required to trigger the transformation under different GB properties are $\SI{259}{\mega\pascal}$, $\SI{176}{\mega\pascal}$, $\SI{156}{\mega\pascal}$, and $\SI{143}{\mega\pascal}$, for the 5\%, 25\%, 50\%, and 100\%, respectively. These values are inversely proportional to the GB elastic properties, as softer GBs (with lower elastic constants) undergo more deformation, requiring higher external stress to transfer the amount of energy required to initiate the phase transformation in the grains.
Accordingly, the stresses needed to fully accomplish the phase transformation are $\SI{1084}{\mega\pascal}$, $\SI{729}{\mega\pascal}$, $\SI{557}{\mega\pascal}$, and $\SI{685}{\mega\pascal}$, respectively. The corresponding stress-strain curves are shown in Figure~\ref{fig:stress_strain_curves_grain_boundary}, where the horizontal regions indicate phase transformation accompanied by inelastic strain. It emerges that GB properties affect only the stresses required to trigger and complete the phase transformation, not the evolution patterns of the phase. Even though the elastic properties of the grain boundaries are not immediately controllable when experimentally processing the material, significantly different sintering heating rates can influence them \cite{Ji2020}. This result is thus relevant for future material developments, and here for a better understanding of the effect of each model parameter. In the following phase transformation analyses, we set the GB properties to 25\% of those of the grains.

\subsubsection{Grain orientations}

\begin{figure}[htbp]
\centering
\hspace*{\fill}
     \subfloat[]{
        \centering
        \def\svgwidth{0.95\textwidth}
        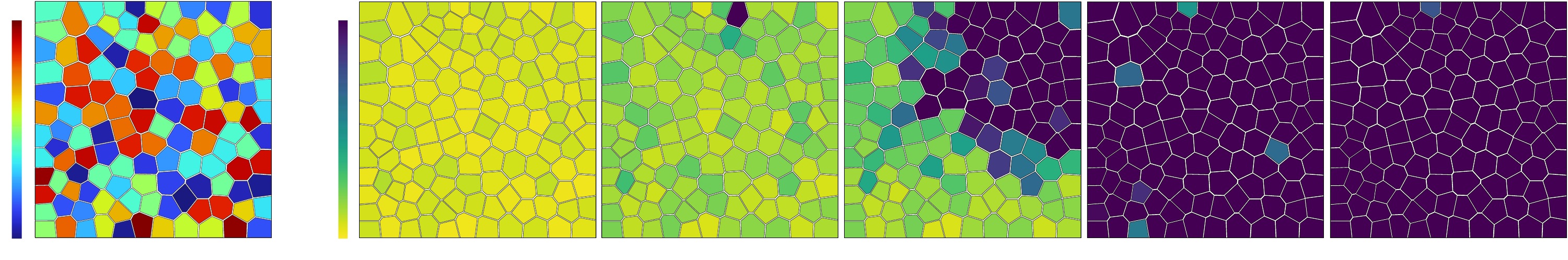	
        \label{fig:Orientation1_Eta}
        }
\hspace*{\fill}
\\
\hspace*{\fill}
     \subfloat[]{
        \centering
        \def\svgwidth{0.95\textwidth}
        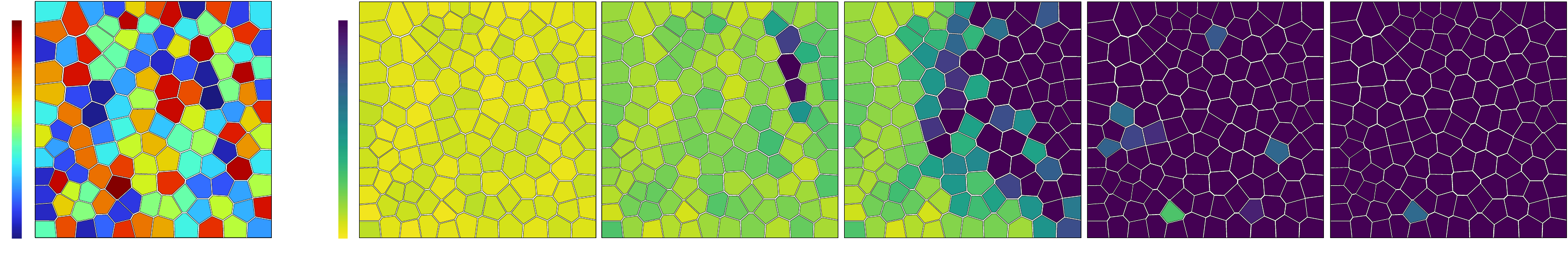	
        \label{fig:Orientation2_Eta}
        }
\hspace*{\fill}
\\
\hspace*{\fill}
     \subfloat[]{
        \centering
        \def\svgwidth{0.95\textwidth}
        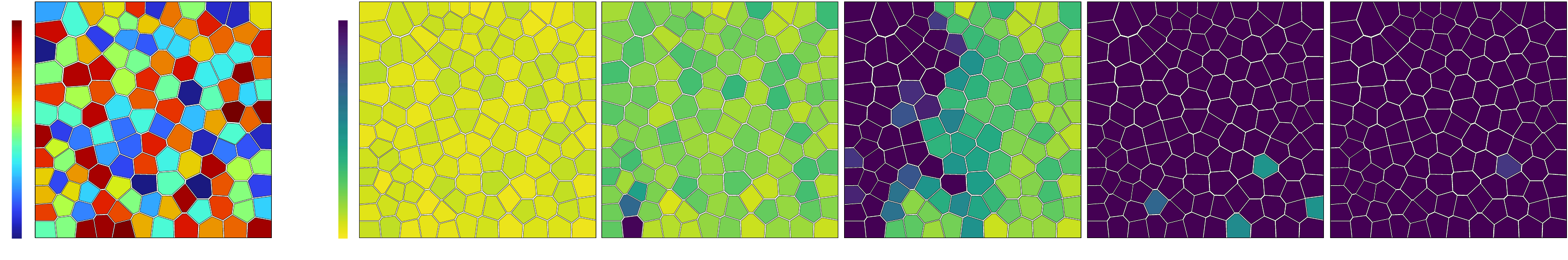	
        \label{fig:Orientation3_Eta}
        }
\hspace*{\fill}
\\
\hspace*{\fill}
     \subfloat[]{
        \centering
        \def\svgwidth{0.95\textwidth}
        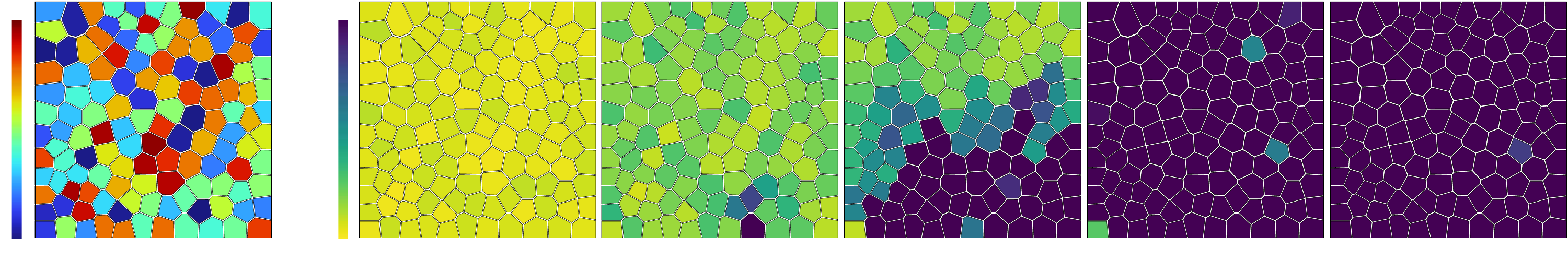	
        \label{fig:Orientation4_Eta}
        }
\hspace*{\fill}
\\
\hspace*{\fill}
     \subfloat[]{
        \centering
        \includegraphics[width=0.8\textwidth]{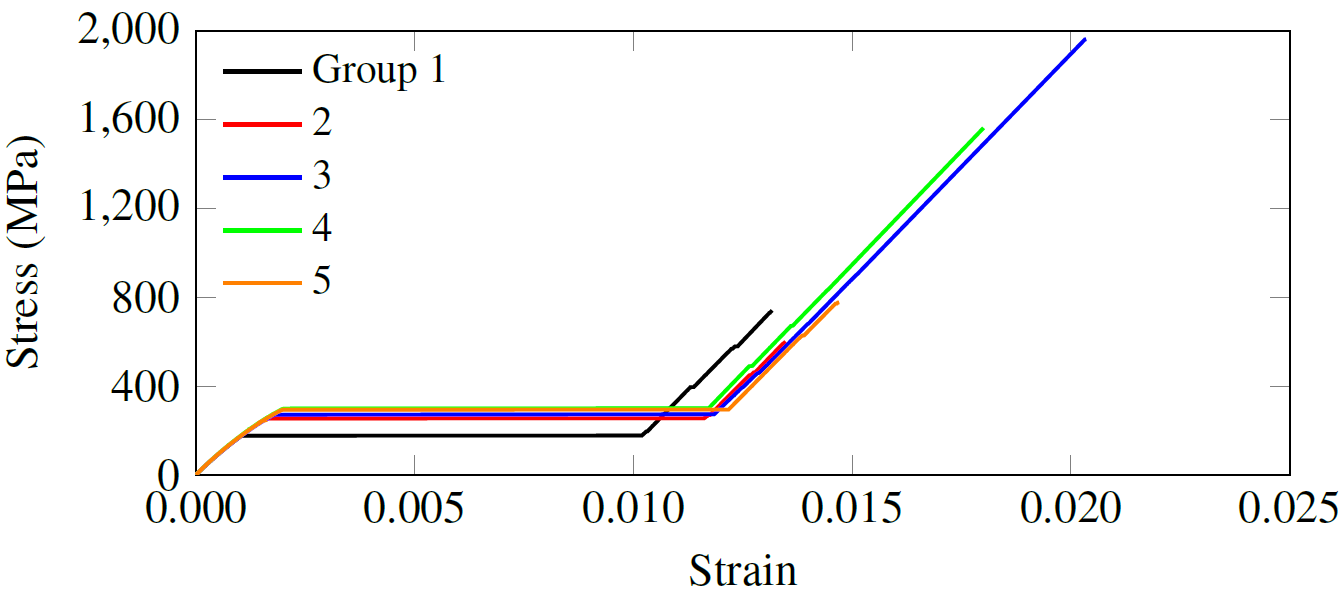}
\label{fig:stress_strain_curves_grain_orientation}
        }
\hspace*{\fill}
\caption{The evolution of the phase variable $\eta$ (0 indicating tetragonal and 1 monolinic) under different groups of grain orientations (a) Group 2, (b) Group 3, (c) Group 4, and (d) Group 5. The $1^{\text{st}}$ column of the phase patterns are obtained under $\SI{100}{\mega\pascal}$. $\SI{254}{\mega\pascal}$, $\SI{271}{\mega\pascal}$, $\SI{299}{\mega\pascal}$, and $\SI{294}{\mega\pascal}$ are the stresses needed to trigger the simultaneous T$\rightarrow$M phase transformation in most grains, where the corresponding phase patterns are shown in the $2^{\text{nd}} - 4^{\text{th}}$ columns. The values $\SI{590}{\mega\pascal}$, $\SI{1961}{\mega\pascal}$, $\SI{1562}{\mega\pascal}$, and $\SI{776}{\mega\pascal}$ shown in the last column are the stresses needed to fully accomplish the phase transformation. It can be seen that grain orientations can have a strong influence on the phase transformation behavior. Figure (e) shows the corresponding stress-strain curves obtained under different grain orientations (Group 1-5).}
 \label{fig:evolution_eta_grain_orientation}
\end{figure}

We also investigate the influence of grain orientations on the phase transformation, by assigning four additional groups of grain orientations to the polycrystalline model. The corresponding evolution of the phase field $\eta$ during the transformation is illustrated in Figures~\ref{fig:Orientation1_Eta}--\ref{fig:Orientation4_Eta}. The location of the first grain to complete the phase transformation varies across the different orientation groups. Furthermore, the direction of transformation propagation is influenced by the grain orientations. The resulting stress-strain curves, presented in Figure~\ref{fig:stress_strain_curves_grain_orientation}, indicate that the stresses required to trigger the phase transformation vary between $\SI{176}{\mega\pascal}$, $\SI{254}{\mega\pascal}$, $\SI{271}{\mega\pascal}$, $\SI{299}{\mega\pascal}$, and $\SI{294}{\mega\pascal}$. The corresponding transformation-induced strains are 0.0091, 0.0100, 0.0100, 0.0097, and 0.0102, respectively. As mentioned in \S\ref{sec:Phase_transformation}, the elastic energy contributes to the driving force of the transformation. Consequently, since orientations influence material properties, they can have a significant impact on the phase transformation. This finding further elucidates the phase transformation behavior in nanocrystalline materials, and becomes relevant for cases of textured microstructures. In the following, we proceed with the grain orientation distribution shown in Figure \ref{fig:Grain_orientations} (Group 1).

\subsubsection{Kinetic coefficient}

\begin{figure}[htbp]
        \centering
        \includegraphics[width=0.8\textwidth]{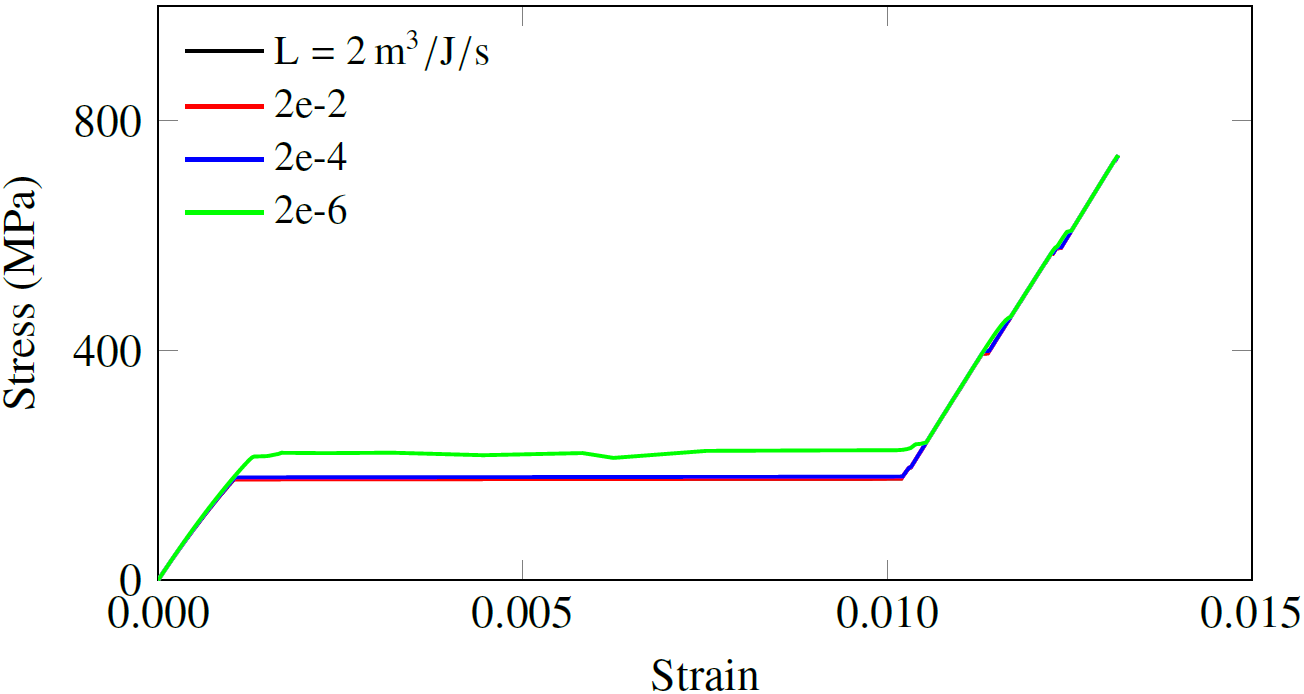}
\caption{Stress-strain curves obtained under different kinetic coefficients $L \left(\si{\meter}^3/\si{\joule}/\si{\second}\right)$ = 2, 2e-2, 2e-4, 2e-6. The corresponding stresses required to trigger the phase transformation are $\SI{176}{\mega\pascal}$, $\SI{176}{\mega\pascal}$, $\SI{180}{\mega\pascal}$, and $\SI{216}{\mega\pascal}$. It emerges that the kinetic coefficient has a minor influence on the phase transformation process.}
\label{fig:Kinetic_coefficients_stress_strain_curves}
\end{figure}

According to the governing equation for the T$\rightarrow$M phase transformation, Equation \eqref{eq1}, the phase evolution is controlled by the kinetic coefficient $L$, which describes the transformation speed. Since the phase transformation occurs almost instantaneously, there is currently no experimental data specifying the exact value of the kinetic coefficient. To investigate the effect of the kinetic coefficient on transformation behaviour, we set $L \left(\si{\meter}^3/\si{\joule}/\si{\second}\right)$ to 2, 2e-2, 2e-4, and 2e-6. The corresponding stress-strain curves are shown in Figure~\ref{fig:Kinetic_coefficients_stress_strain_curves}. The stresses required to trigger the phase transformation are measured at $\SI{176}{\mega\pascal}, \SI{176}{\mega\pascal}, \SI{180}{\mega\pascal}$, and $\SI{216}{\mega\pascal}$, respectively, demonstrating an inverse relationship with the kinetic coefficients. Notably, the triggering stress converges to  \SI{176}{\mega\pascal} as the transformation speed increases. Additionally, the phase evolution patterns remain consistent across all kinetic coefficients. We can thus conclude that the kinetic coefficient has a minimal impact on the phase transformation process.

\subsection{Microscale}

The microscale model consists of a brick-and-mortar structure with dimensions of $\SI{30}{\micro\meter} \times \SI{5}{\micro\meter}$. A predefined notch, $\SI{1.5}{\micro\meter}$ in length, is located in the center of the bottom boundary. Displacement boundary conditions are applied on both the left and right sides in the horizontal direction with an incremental rate of $\SI{10}{\nano\meter/\second}$, implementing mode I fracture to capture the stress state around the crack tip under three-point bending. The bricks have uniform dimensions, length $l = \SI{10}{\micro\meter}$ and width $w = \SI{0.33}{\micro\meter}$, corresponding to those used in the corresponding experimental work~\cite{Francesco2024}, and they are separated by a mortar layer of thickness $t = \SI{0.04}{\micro\meter}$. Under these conditions, the resulting zirconia content is 11.5\%.
Isotropic material properties are assigned to both bricks and mortar, with alumina used for the bricks and ceria-stabilized zirconia for the mortar. The material properties are provided in Table~\ref{tab:Alumina_Zirconia}. The T$\rightarrow$M phase transformation behaviour in the mortar, characterized by a $\SI{259}{\mega\pascal}$ activation stress and 0.0098 induced inelastic strain, is captured by a pseudoelastic constitutive law derived from the nanoscale model with orientation Group 1 (shown in Figure \ref{fig:Grain_orientations}) and grain boundary properties set as 5\% of the bulk ones. 
The domain is discretized using bilinear quadrilateral elements with a mesh size $h = \SI{10}{\nano\meter}$. For the fracture analysis, the phase-field method is applied with a length scale of $l_0 = \SI{20}{\nano\meter}$. Unless otherwise specified, the default units for brick dimensions and mortar thickness are $\si{\micro\meter}$.

\begin{table}[!ht]
\newcolumntype{G}{>{\centering\arraybackslash}m{4.00em}}
 \newcolumntype{E}{>{\centering\arraybackslash}m{9.00em}}
\newcolumntype{P}{>{\centering\arraybackslash}m{8.00em}}
 \newcolumntype{X}{>{\centering\arraybackslash}m{9.00em}}
 \newcolumntype{T}{>{\centering\arraybackslash}m{14.00em}}
\begin{center}
\begin{tabular}{G*1{E}@{}*1{P}@{}*1{X}@{}}
\hline                      & Elastic modulus $\mathrm{E}$ &  Poisson's ratio $\nu$      & Fracture strength $\sigma_f$ \\
\hline  Alumina        &       $\SI{431}{\giga\pascal}$                   &          0.29                   &             $\SI{20}{\giga\pascal}$                      \\
          Zirconia         &       $\SI{195}{\giga\pascal}$                    &           0.32                  &           $\SI{0.5}{\giga\pascal}$                  \\
\hline
\end{tabular}
\end{center}
\caption{Material properties of alumina and zirconia~\cite{Attaoui2007, Konstantiniuk2021}}
\label{tab:Alumina_Zirconia}
\end{table}

\subsubsection{Toughening mechanisms}

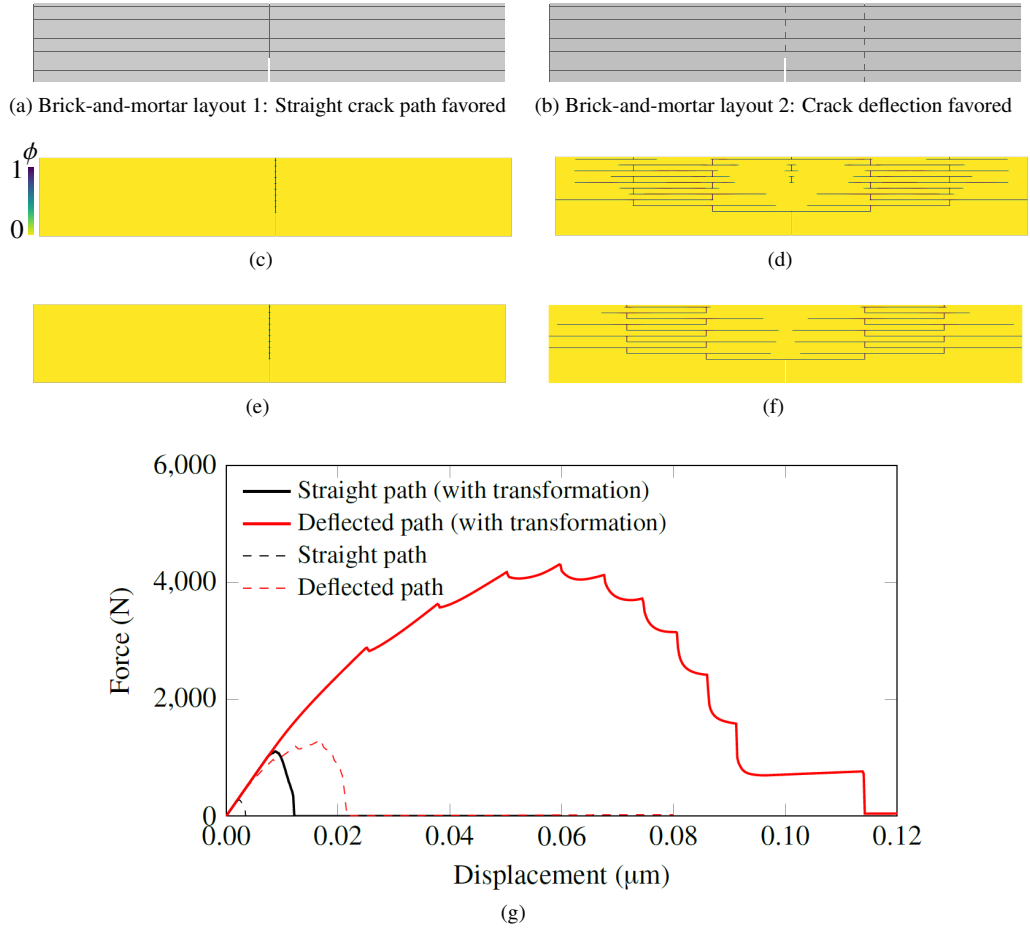
\begin{figure}[!ht]
\centering
\hspace*{\fill}
     \subfloat[Brick-and-mortar layout 1: Straight crack path favored]{
        \centering
        \includegraphics[width=0.485\textwidth]{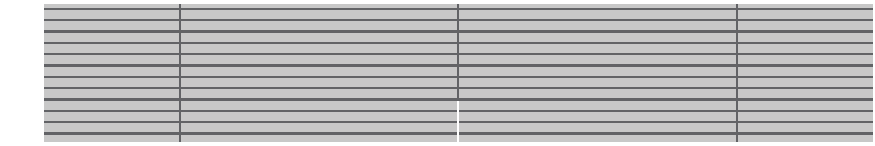}
        \label{fig:Brick_10_033_Mortar_004_Straight_1}
        }
\hspace*{\fill}
     \subfloat[Brick-and-mortar layout 2: Crack deflection favored]{
        \centering
        \includegraphics[width=0.485\textwidth]{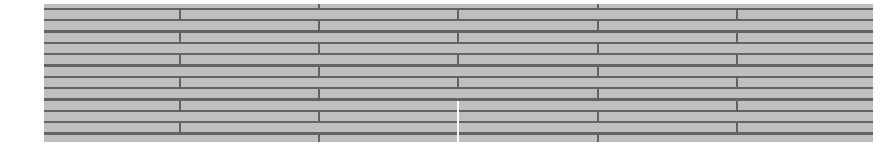}
        \label{fig:Brick_10_033_Mortar_004_Transformation_1}
        }
\hspace*{\fill}
\\
\hspace*{\fill}
     \subfloat[]{
        \centering
        \def\svgwidth{0.485\textwidth}
\begingroup%
  \makeatletter%
  \providecommand\color[2][]{%
    \errmessage{(Inkscape) Color is used for the text in Inkscape, but the package 'color.sty' is not loaded}%
    \renewcommand\color[2][]{}%
  }%
  \providecommand\transparent[1]{%
    \errmessage{(Inkscape) Transparency is used (non-zero) for the text in Inkscape, but the package 'transparent.sty' is not loaded}%
    \renewcommand\transparent[1]{}%
  }%
  \providecommand\rotatebox[2]{#2}%
  \newcommand*\fsize{\dimexpr\f@size pt\relax}%
  \newcommand*\lineheight[1]{\fontsize{\fsize}{#1\fsize}\selectfont}%
  \ifx\svgwidth\undefined%
    \setlength{\unitlength}{420bp}%
    \ifx\svgscale\undefined%
      \relax%
    \else%
      \setlength{\unitlength}{\unitlength * \real{\svgscale}}%
    \fi%
  \else%
    \setlength{\unitlength}{\svgwidth}%
  \fi%
  \global\let\svgwidth\undefined%
  \global\let\svgscale\undefined%
  \makeatother%
  \begin{picture}(1,0.18)%
    \lineheight{1}%
    \setlength\tabcolsep{0pt}%
    \put(0,0){\includegraphics[width=\unitlength,page=1]{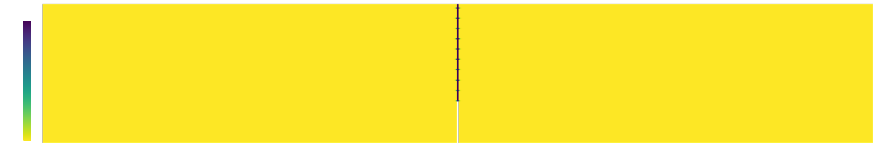}}%
    \put(0.015,0.16){\color[rgb]{0,0,0}\makebox(0,0)[lt]{\lineheight{1.25}\smash{\begin{tabular}[t]{l}$\phi$\end{tabular}}}}%
    \put(-0.01,0.12){\color[rgb]{0,0,0}\makebox(0,0)[lt]{\lineheight{1.25}\smash{\begin{tabular}[t]{l}1\end{tabular}}}}%
    \put(-0.01,0.00){\color[rgb]{0,0,0}\makebox(0,0)[lt]{\lineheight{1.25}\smash{\begin{tabular}[t]{l}0\end{tabular}}}}%
  \end{picture}%
\endgroup%
	
        \label{fig:Brick_10_033_Mortar_004_Straight_2}
        }
\hspace*{\fill}
             \subfloat[]{
        \centering
        \includegraphics[width=0.485\textwidth]{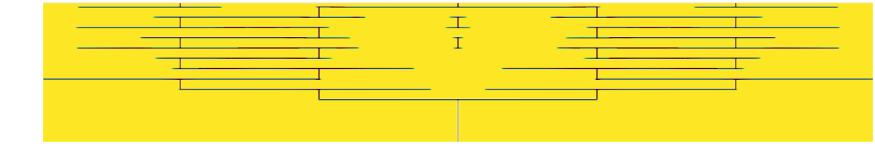}
        \label{fig:Brick_10_033_Mortar_004_Transformation_2}
        }
\hspace*{\fill}
\\
\hspace*{\fill}
     \subfloat[]{
        \centering
        \includegraphics[width=0.485\textwidth]{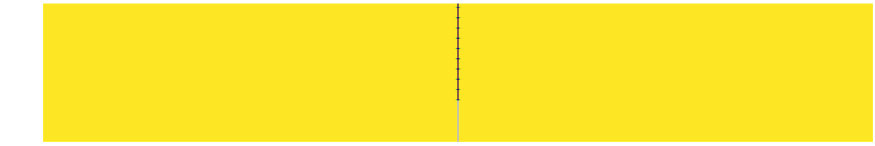}
        \label{fig:Brick_10_033_Mortar_004_Straight_3}
        }
\hspace*{\fill}
     \subfloat[]{
        \centering
        \includegraphics[width=0.485\textwidth]{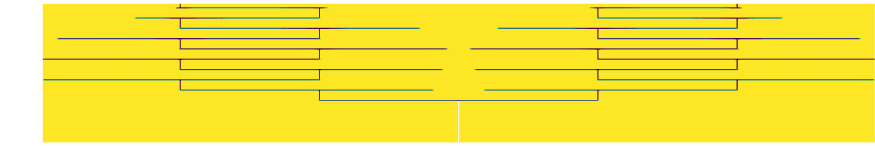}
        \label{fig:Brick_10_033_Mortar_004_Transformation_3}
        }
\hspace*{\fill}
\\
\hspace*{\fill}
     \subfloat[]{
        \centering
        \includegraphics[width=0.8\textwidth]{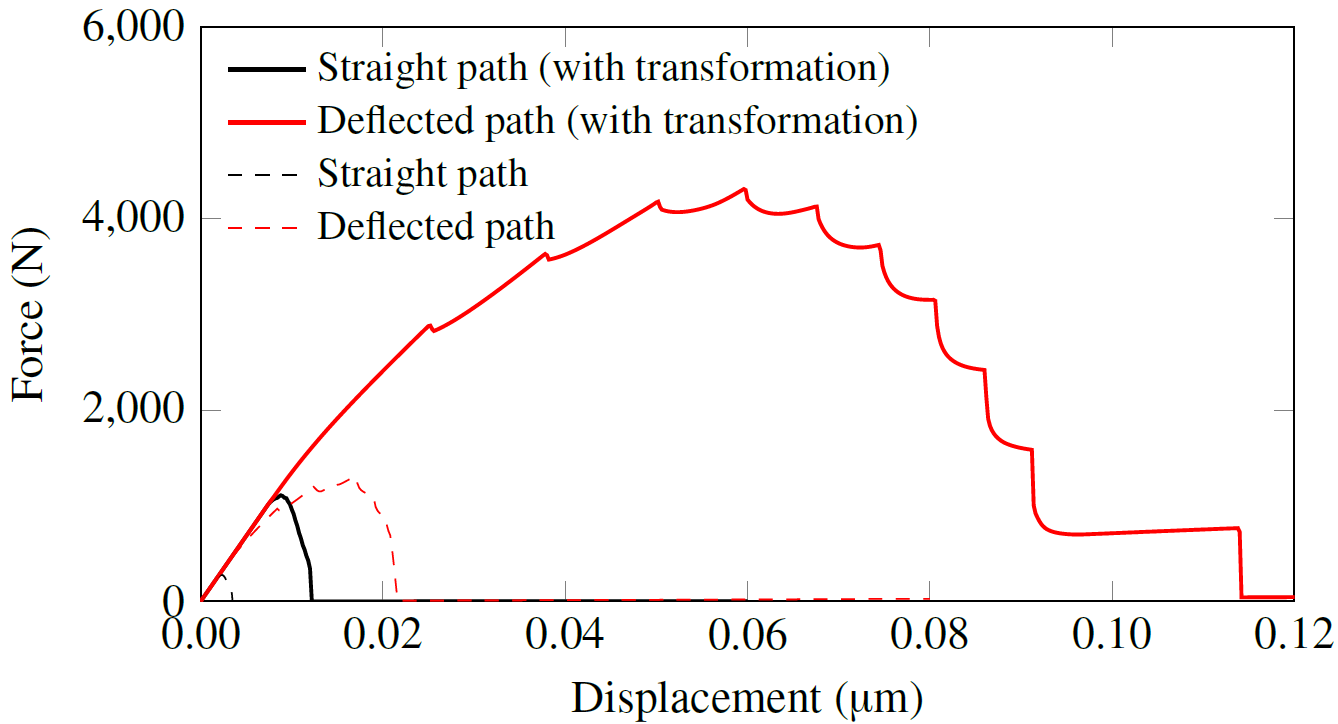}
\label{fig:force_displacement_curves_straight_deflect}
        }
\hspace*{\fill}

\caption{Microscale models with dimensions $\SI{30}{\micro\meter} \times \SI{5}{\micro\meter}$ consisting of brick-and-mortar structures with brick size $\SI{10}{\micro\meter} \times \SI{0.33}{\micro\meter}$ and mortar thickness $\SI{0.04}{\micro\meter}$, where layouts 1 and 2 lead to (a) straight and (b) deflected crack propagation paths (0 indicating intact material and 1 crack), respectively. Figures (c, d) show the corresponding final crack paths obtained with layouts 1 and 2, without phase transformation. It can be seen that the brick-and-mortar structure leads to crack deflection, with significantly longer crack paths. Figures (e, f) are the corresponding final crack fields obtained with the phase transformation. Figure (g) shows force-displacement curves obtained under different brick-and-mortar layouts with/without the phase transformation, where both brick-and-mortar structures leading to crack deflections and the T$\rightarrow$M phase transformation in the zirconia mortar are proven to be effective toughening mechanisms.}
\label{fig:crack_fields_straight_deflect}
\end{figure}

We first demonstrate the crack deflection behaviour in the brick-and-mortar structure without considering the phase transformation. For comparison, two microscale models with different brick-and-mortar arrangements are constructed, as shown in Figures~\ref{fig:Brick_10_033_Mortar_004_Straight_1} and~\ref{fig:Brick_10_033_Mortar_004_Transformation_1}. The crack fields for both models are presented in Figures~\ref{fig:Brick_10_033_Mortar_004_Straight_2} and~\ref{fig:Brick_10_033_Mortar_004_Transformation_2}, and their corresponding force-displacement curves are depicted in Figure~\ref{fig:force_displacement_curves_straight_deflect}. Although both models have the same number of bricks, the second brick-and-mortar arrangement, which promotes crack deflection, dissipates more energy compared to the first model, where the crack can easily propagate along a straight path. This confirms that designing brick-and-mortar structures to encourage crack deflection is an effective strategy for enhancing fracture toughness.

We then add the phase transformation to both models. While the transformation does not alter the final crack path in the straight propagation case, as shown in Figure~\ref{fig:Brick_10_033_Mortar_004_Straight_3}, it increases the resistance force, and thus leads to greater energy dissipation before the final fracture, as indicated by the black curve in Figure~\ref{fig:force_displacement_curves_straight_deflect}. This effect is even more remarkable for the deflected crack path case (Figure~\ref{fig:Brick_10_033_Mortar_004_Transformation_3} and red solid line in Figure~\ref{fig:force_displacement_curves_straight_deflect}). The force-displacement curves demonstrate that coupling both toughening mechanisms brings on a boosted fracture performance, yielding a fracture toughness $K_\text{I} = \SI{6.00}{\mega\pascal\sqrt\meter}$. This value is calculated based on the dissipation energy, crack length, and elastic properties. Although our experimental work showed R-curve behaviour~\cite{Francesco2024}, the fracture toughness $K_\text{I}$ obtained here refers to the crack initiation toughness, representing the starting point of the R-curve.

\subsubsection{Brick-and-mortar arrangements}

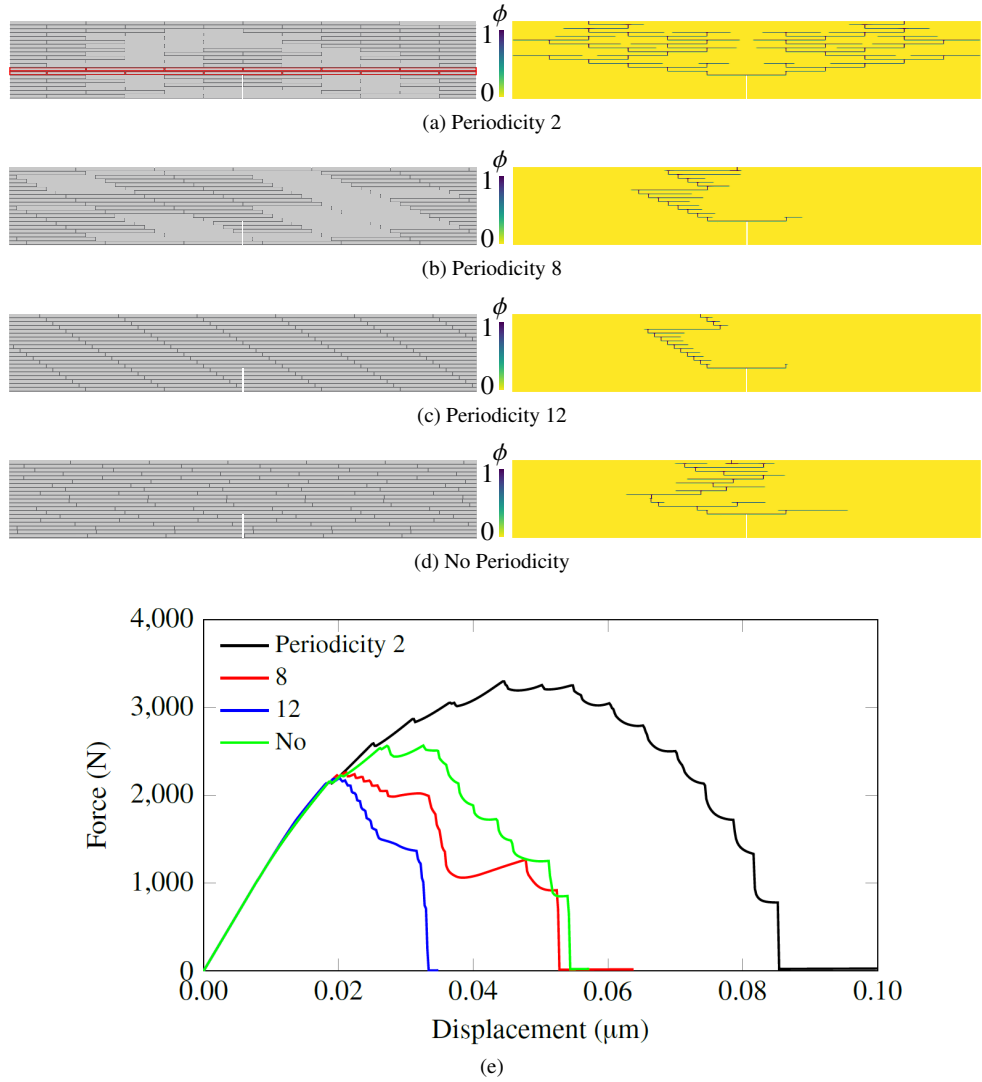
\begin{figure}[!ht]
\centering
     \subfloat[Periodicity 2]{
        \centering
        \def\svgwidth{0.95\textwidth}
\begingroup%
  \makeatletter%
  \providecommand\color[2][]{%
    \errmessage{(Inkscape) Color is used for the text in Inkscape, but the package 'color.sty' is not loaded}%
    \renewcommand\color[2][]{}%
  }%
  \providecommand\transparent[1]{%
    \errmessage{(Inkscape) Transparency is used (non-zero) for the text in Inkscape, but the package 'transparent.sty' is not loaded}%
    \renewcommand\transparent[1]{}%
  }%
  \providecommand\rotatebox[2]{#2}%
  \newcommand*\fsize{\dimexpr\f@size pt\relax}%
  \newcommand*\lineheight[1]{\fontsize{\fsize}{#1\fsize}\selectfont}%
  \ifx\svgwidth\undefined%
    \setlength{\unitlength}{830.00006104bp}%
    \ifx\svgscale\undefined%
      \relax%
    \else%
      \setlength{\unitlength}{\unitlength * \real{\svgscale}}%
    \fi%
  \else%
    \setlength{\unitlength}{\svgwidth}%
  \fi%
  \global\let\svgwidth\undefined%
  \global\let\svgscale\undefined%
  \makeatother%
  \begin{picture}(1,0.08433735)%
    \lineheight{1}%
    \setlength\tabcolsep{0pt}%
    \put(0,0){\includegraphics[width=\unitlength,page=1]{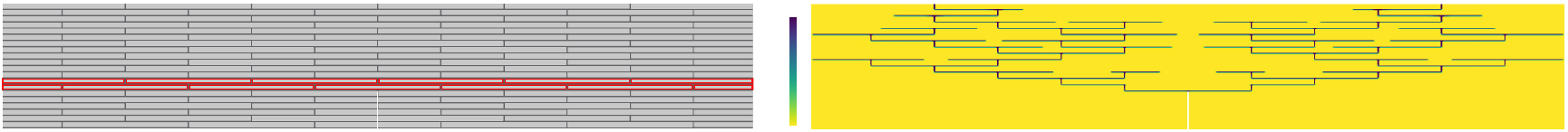}}%
    \put(0.4975,0.0825){\color[rgb]{0,0,0}\makebox(0,0)[lt]{\lineheight{1.25}\smash{\begin{tabular}[t]{l}$\phi$\end{tabular}}}}%
    \put(0.485,0.06){\color[rgb]{0,0,0}\makebox(0,0)[lt]{\lineheight{1.25}\smash{\begin{tabular}[t]{l}1\end{tabular}}}}%
    \put(0.485,0.00){\color[rgb]{0,0,0}\makebox(0,0)[lt]{\lineheight{1.25}\smash{\begin{tabular}[t]{l}0\end{tabular}}}}%
  \end{picture}%
\endgroup%
	
        \label{fig:Periodicity_2_New}
        }
\\
     \subfloat[Periodicity 8]{
        \centering
        \def\svgwidth{0.95\textwidth}
\begingroup%
  \makeatletter%
  \providecommand\color[2][]{%
    \errmessage{(Inkscape) Color is used for the text in Inkscape, but the package 'color.sty' is not loaded}%
    \renewcommand\color[2][]{}%
  }%
  \providecommand\transparent[1]{%
    \errmessage{(Inkscape) Transparency is used (non-zero) for the text in Inkscape, but the package 'transparent.sty' is not loaded}%
    \renewcommand\transparent[1]{}%
  }%
  \providecommand\rotatebox[2]{#2}%
  \newcommand*\fsize{\dimexpr\f@size pt\relax}%
  \newcommand*\lineheight[1]{\fontsize{\fsize}{#1\fsize}\selectfont}%
  \ifx\svgwidth\undefined%
    \setlength{\unitlength}{830.00006104bp}%
    \ifx\svgscale\undefined%
      \relax%
    \else%
      \setlength{\unitlength}{\unitlength * \real{\svgscale}}%
    \fi%
  \else%
    \setlength{\unitlength}{\svgwidth}%
  \fi%
  \global\let\svgwidth\undefined%
  \global\let\svgscale\undefined%
  \makeatother%
  \begin{picture}(1,0.08433735)%
    \lineheight{1}%
    \setlength\tabcolsep{0pt}%
    \put(0,0){\includegraphics[width=\unitlength,page=1]{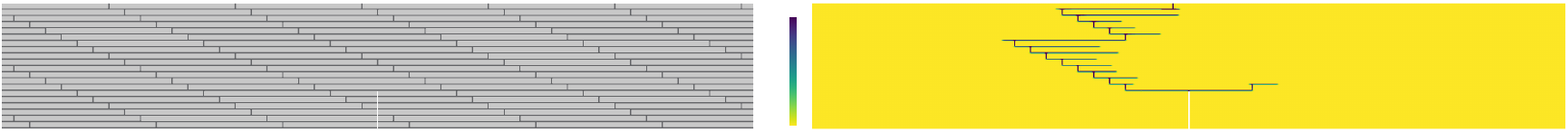}}%
    \put(0.4975,0.0825){\color[rgb]{0,0,0}\makebox(0,0)[lt]{\lineheight{1.25}\smash{\begin{tabular}[t]{l}$\phi$\end{tabular}}}}%
    \put(0.485,0.06){\color[rgb]{0,0,0}\makebox(0,0)[lt]{\lineheight{1.25}\smash{\begin{tabular}[t]{l}1\end{tabular}}}}%
    \put(0.485,0.00){\color[rgb]{0,0,0}\makebox(0,0)[lt]{\lineheight{1.25}\smash{\begin{tabular}[t]{l}0\end{tabular}}}}%
  \end{picture}%
\endgroup%
	
        \label{fig:Periodicity_8_New}
        }
\\
     \subfloat[Periodicity 12]{
        \centering
        \def\svgwidth{0.95\textwidth}
\begingroup%
  \makeatletter%
  \providecommand\color[2][]{%
    \errmessage{(Inkscape) Color is used for the text in Inkscape, but the package 'color.sty' is not loaded}%
    \renewcommand\color[2][]{}%
  }%
  \providecommand\transparent[1]{%
    \errmessage{(Inkscape) Transparency is used (non-zero) for the text in Inkscape, but the package 'transparent.sty' is not loaded}%
    \renewcommand\transparent[1]{}%
  }%
  \providecommand\rotatebox[2]{#2}%
  \newcommand*\fsize{\dimexpr\f@size pt\relax}%
  \newcommand*\lineheight[1]{\fontsize{\fsize}{#1\fsize}\selectfont}%
  \ifx\svgwidth\undefined%
    \setlength{\unitlength}{830.00006104bp}%
    \ifx\svgscale\undefined%
      \relax%
    \else%
      \setlength{\unitlength}{\unitlength * \real{\svgscale}}%
    \fi%
  \else%
    \setlength{\unitlength}{\svgwidth}%
  \fi%
  \global\let\svgwidth\undefined%
  \global\let\svgscale\undefined%
  \makeatother%
  \begin{picture}(1,0.08433735)%
    \lineheight{1}%
    \setlength\tabcolsep{0pt}%
    \put(0,0){\includegraphics[width=\unitlength,page=1]{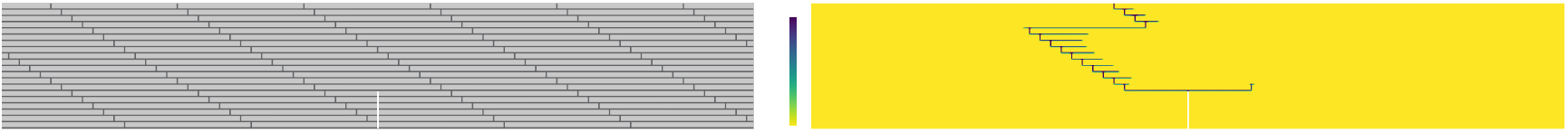}}%
    \put(0.4975,0.0825){\color[rgb]{0,0,0}\makebox(0,0)[lt]{\lineheight{1.25}\smash{\begin{tabular}[t]{l}$\phi$\end{tabular}}}}%
    \put(0.485,0.06){\color[rgb]{0,0,0}\makebox(0,0)[lt]{\lineheight{1.25}\smash{\begin{tabular}[t]{l}1\end{tabular}}}}%
    \put(0.485,0.00){\color[rgb]{0,0,0}\makebox(0,0)[lt]{\lineheight{1.25}\smash{\begin{tabular}[t]{l}0\end{tabular}}}}%
  \end{picture}%
\endgroup%
	
        \label{fig:Periodicity_12_New}
        }
\\
     \subfloat[No Periodicity]{
        \centering
        \def\svgwidth{0.95\textwidth}
\begingroup%
  \makeatletter%
  \providecommand\color[2][]{%
    \errmessage{(Inkscape) Color is used for the text in Inkscape, but the package 'color.sty' is not loaded}%
    \renewcommand\color[2][]{}%
  }%
  \providecommand\transparent[1]{%
    \errmessage{(Inkscape) Transparency is used (non-zero) for the text in Inkscape, but the package 'transparent.sty' is not loaded}%
    \renewcommand\transparent[1]{}%
  }%
  \providecommand\rotatebox[2]{#2}%
  \newcommand*\fsize{\dimexpr\f@size pt\relax}%
  \newcommand*\lineheight[1]{\fontsize{\fsize}{#1\fsize}\selectfont}%
  \ifx\svgwidth\undefined%
    \setlength{\unitlength}{830.00006104bp}%
    \ifx\svgscale\undefined%
      \relax%
    \else%
      \setlength{\unitlength}{\unitlength * \real{\svgscale}}%
    \fi%
  \else%
    \setlength{\unitlength}{\svgwidth}%
  \fi%
  \global\let\svgwidth\undefined%
  \global\let\svgscale\undefined%
  \makeatother%
  \begin{picture}(1,0.08433735)%
    \lineheight{1}%
    \setlength\tabcolsep{0pt}%
    \put(0,0){\includegraphics[width=\unitlength,page=1]{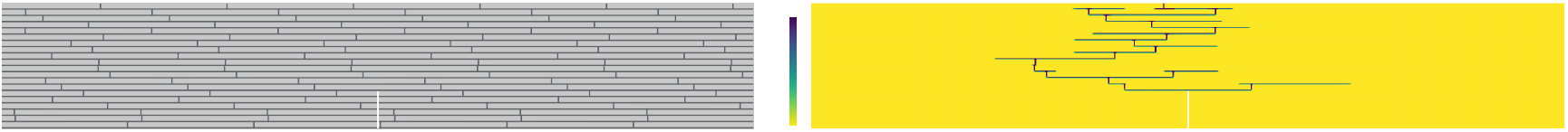}}%
    \put(0.4975,0.0825){\color[rgb]{0,0,0}\makebox(0,0)[lt]{\lineheight{1.25}\smash{\begin{tabular}[t]{l}$\phi$\end{tabular}}}}%
    \put(0.485,0.06){\color[rgb]{0,0,0}\makebox(0,0)[lt]{\lineheight{1.25}\smash{\begin{tabular}[t]{l}1\end{tabular}}}}%
    \put(0.485,0.00){\color[rgb]{0,0,0}\makebox(0,0)[lt]{\lineheight{1.25}\smash{\begin{tabular}[t]{l}0\end{tabular}}}}%
  \end{picture}%
\endgroup%
	
        \label{fig:Periodicity_NoP_New}
        }
\\
     \subfloat[]{
        \centering
        \includegraphics[width=0.8\textwidth]{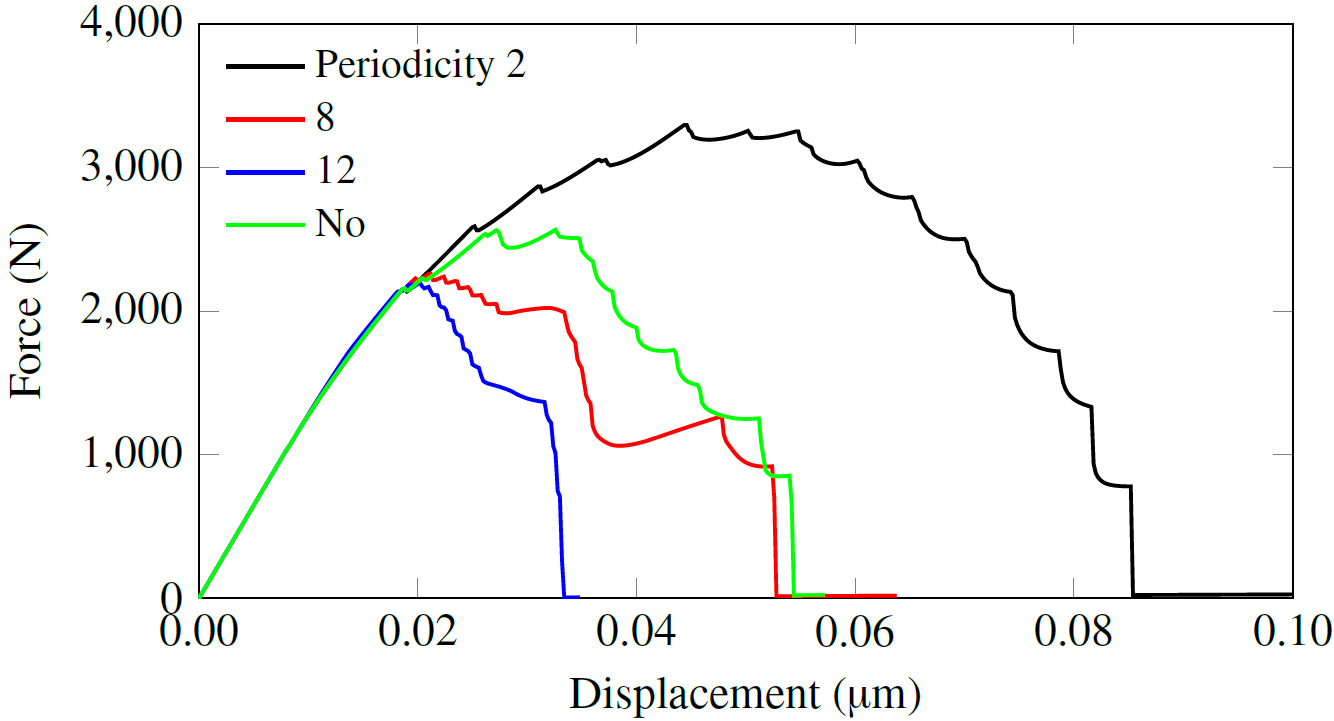}
\label{fig:force_displacement_curves_periodicity}
        }
\caption{Microscale brick-and-mortar models consisting of brick length $l = 5$, brick width $w = 0.20$, and mortar thickness $t = 0.05$  with different layouts and the corresponding crack fields (0 indicating intact material and 1 crack), where the patterns are repeated in the vertical direction: (a) 2 layers, (b) 8 layers, (c) 12 layers, and (d) No periodicity. The brick-and-mortar layout with Periodicity 2 has the largest damaged area. The one without periodicity outperforms Periodicity 8 and 12, as the fractured area reveals. Figure (e) shows the corresponding force-displacement curves obtained by setting different periodicities of the brick-and-mortar structures, where the fracture performance and the number of layers in the periodic pattern are in inverse proportion. Again, we see that the model without periodicity outperforms those with Periodicity 8 and 12.}
\label{fig:crack_fields_periodicity}
\end{figure}

We further explore the impact of periodicity in brick-and-mortar arrangements. Periodicity refers to the number of repeating layers in the vertical direction, as highlighted in Figure~\ref{fig:Periodicity_2_New}. Microscale models are constructed using bricks and mortar with dimensions $(l, w, t) = (5, 0.20, 0.05)$, and periodicities of 2, 8, and 12 are considered. Additionally, a non-periodic model is created, where each layer has a different brick-and-mortar pattern. The corresponding force-displacement curves are presented in Figure~\ref{fig:force_displacement_curves_periodicity}. The model with 2 periodicities shows the highest resistance force and dissipates the most energy during crack propagation. These findings suggest an inverse relationship between fracture performance and the number of repeated layers in the periodic pattern. Interestingly, the non-periodic model performs better than those with 8 and 12 periodicities, as indicated by the larger damaged area and more distributed crack paths shown in Figures~\ref{fig:Periodicity_8_New}--\ref{fig:Periodicity_NoP_New}. These findings are relevant for thin films or microscale components that can leverage fine-tuned self-assembly or high-resolution 3D printing techniques for such fine control of the material structure.

\subsubsection{Brick size and mortar thickness}

\begin{figure}[htbp]
\centering
\hspace*{\fill}
    \subfloat[($l, w, t = \SI{2}{\micro\meter}, \SI{0.04}{\micro\meter}, \SI{0.01}{\micro\meter}$)]{
        \centering
        \def\svgwidth{0.495\textwidth}
\begingroup%
  \makeatletter%
  \providecommand\color[2][]{%
    \errmessage{(Inkscape) Color is used for the text in Inkscape, but the package 'color.sty' is not loaded}%
    \renewcommand\color[2][]{}%
  }%
  \providecommand\transparent[1]{%
    \errmessage{(Inkscape) Transparency is used (non-zero) for the text in Inkscape, but the package 'transparent.sty' is not loaded}%
    \renewcommand\transparent[1]{}%
  }%
  \providecommand\rotatebox[2]{#2}%
  \newcommand*\fsize{\dimexpr\f@size pt\relax}%
  \newcommand*\lineheight[1]{\fontsize{\fsize}{#1\fsize}\selectfont}%
  \ifx\svgwidth\undefined%
    \setlength{\unitlength}{420bp}%
    \ifx\svgscale\undefined%
      \relax%
    \else%
      \setlength{\unitlength}{\unitlength * \real{\svgscale}}%
    \fi%
  \else%
    \setlength{\unitlength}{\svgwidth}%
  \fi%
  \global\let\svgwidth\undefined%
  \global\let\svgscale\undefined%
  \makeatother%
  \begin{picture}(1,0.16666667)%
    \lineheight{1}%
    \setlength\tabcolsep{0pt}%
    \put(0,0){\includegraphics[width=\unitlength,page=1]{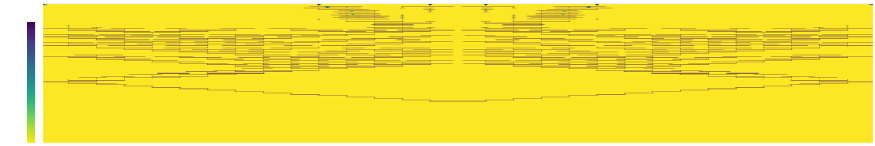}}%
    \put(-0.005,-0.01){\color[rgb]{0,0,0}\makebox(0,0)[lt]{\lineheight{1.25}\smash{\begin{tabular}[t]{l}0\end{tabular}}}}%
    \put(-0.005,0.13){\color[rgb]{0,0,0}\makebox(0,0)[lt]{\lineheight{1.25}\smash{\begin{tabular}[t]{l}1\end{tabular}}}}%
    \put(0.02,0.165){\color[rgb]{0,0,0}\makebox(0,0)[lt]{\lineheight{1.25}\smash{\begin{tabular}[t]{l}$\phi$\end{tabular}}}}%
  \end{picture}%
\endgroup%

        \label{fig:Brick_2_004_Mortar_001}
        }
\hspace*{\fill}
     \subfloat[(2, 0.20, 0.05)]{
        \centering
        \includegraphics[width=0.495\textwidth]{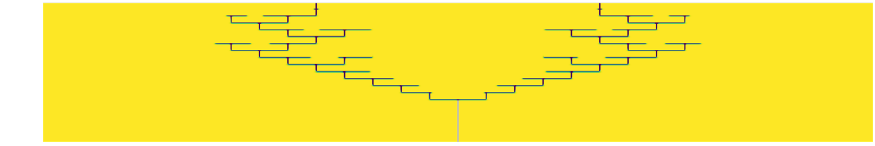}
        \label{fig:Brick_2_020_Mortar_005}
        }
\hspace*{\fill}
\\
\hspace*{\fill}
     \subfloat[(5, 007, 0.018)]{
        \centering
        \includegraphics[width=0.495\textwidth]{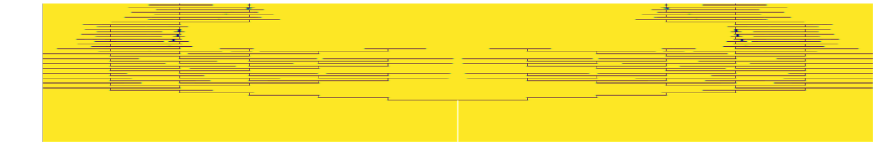}
        \label{fig:Brick_5_007_Mortar_0018}
        }
\hspace*{\fill}
     \subfloat[(5, 0.20, 0.05)]{
        \centering
        \includegraphics[width=0.495\textwidth]{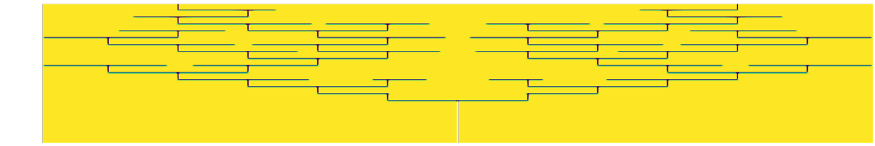}
        \label{fig:Brick_5_020_Mortar_005}
        }
\hspace*{\fill}
\\
\hspace*{\fill}
     \subfloat[(5, 0.50, 0.12)]{
        \centering
        \includegraphics[width=0.495\textwidth]{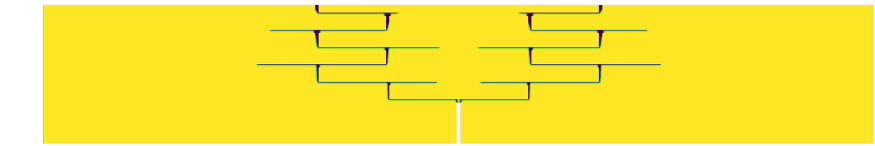}
        \label{fig:Brick_5_050_Mortar_012}
        }
\hspace*{\fill}
     \subfloat[(7, 0.10, 0.025)]{
        \centering
        \includegraphics[width=0.495\textwidth]{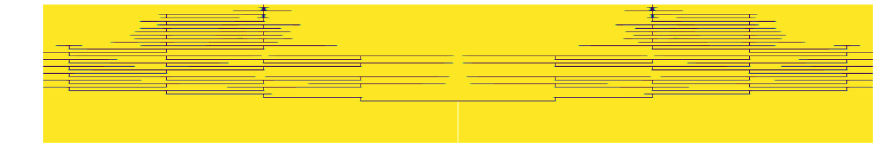}
        \label{fig:Brick_7_010_Mortar_0025}
        }
\hspace*{\fill}
\\
\hspace*{\fill}
     \subfloat[(10, 0.20, 0.05)]{
        \centering
        \includegraphics[width=0.495\textwidth]{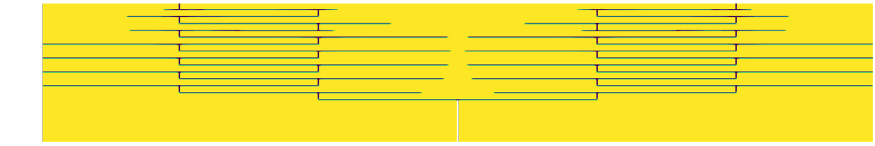}
        \label{fig:Brick_10_020_Mortar_005}
        }
\hspace*{\fill}
     \subfloat[(10, 0.33, 0.04)]{
        \centering
        \includegraphics[width=0.495\textwidth]{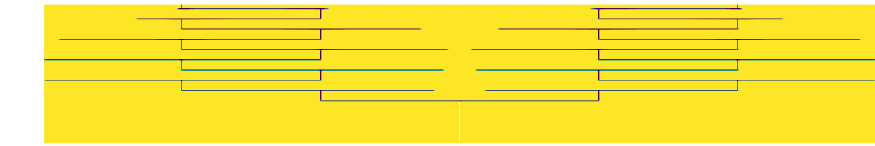}
        \label{fig:Brick_10_033_Mortar_004}
        }
\hspace*{\fill}
\\
\hspace*{\fill}
     \subfloat[(10, 0.33, 0.08)]{
        \centering
        \includegraphics[width=0.495\textwidth]{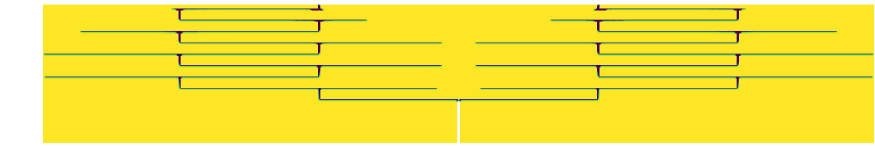}
        \label{fig:Brick_10_033_Mortar_008}
        }
\hspace*{\fill}
     \subfloat[(10, 0.33, 0.13)]{
        \centering
        \includegraphics[width=0.495\textwidth]{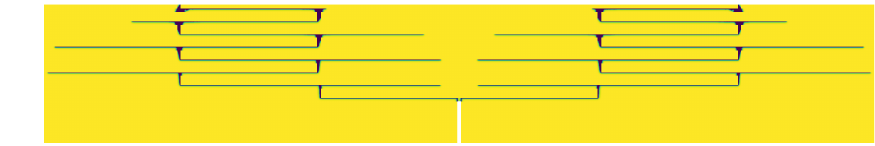}
        \label{fig:Brick_10_033_Mortar_013}
        }
\hspace*{\fill}
\caption{Final crack fields $\phi$ (0 indicating intact material and 1 crack) obtained with different brick length $l$, brick width $w$, and mortar thickness $t$, where $(l, w, t) = $ (a) (2, 0.04, 0.01), (b) (2, 0.20, 0.05), (c) (5, 007, 0.018), (d) (5, 0.20, 0.05), (e) (5, 0.50, 0.12), (f) (7, 0.10, 0.025), (g) (10, 0.20, 0.05), (h) (10, 0.33, 0.04), (i) (10, 0.33, 0.08), and (j) (10, 0.33, 0.13). Different combinations of brick sizes and mortar thicknesses can significantly alter crack propagation paths.}
\label{fig:crack_path_total1}
\end{figure}

\begin{figure}[htbp]
\centering
\hspace*{\fill}
     \subfloat[]{
        \centering
        \includegraphics[width=0.875\textwidth]{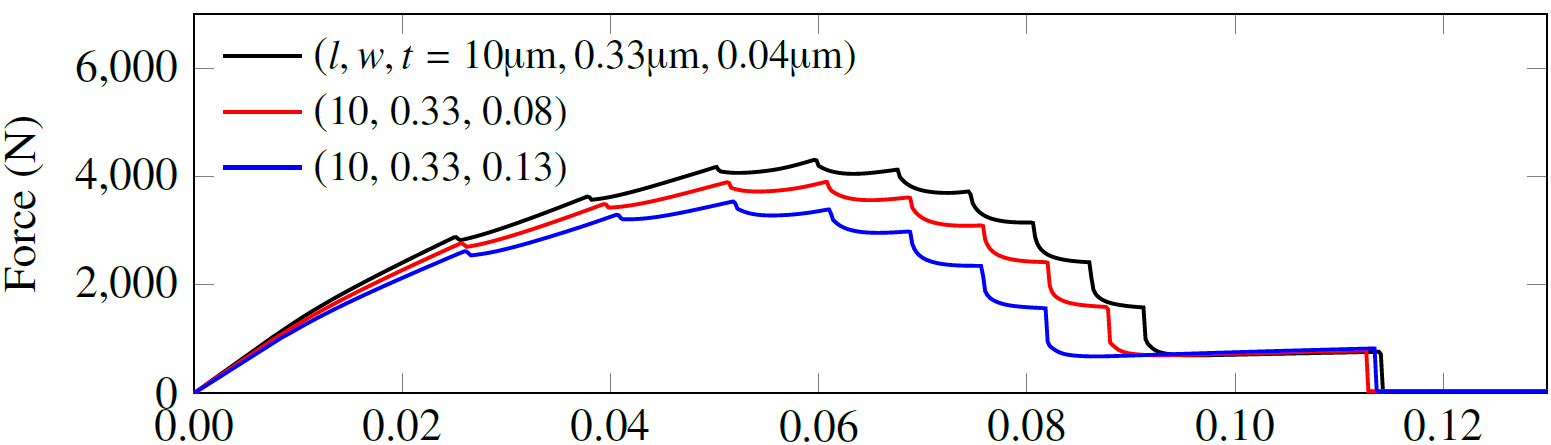}
\label{fig:force_displacement_curves_zirconia}
        }
\hspace*{\fill}
\\
\hspace*{\fill}
     \subfloat[]{
         \centering
        \includegraphics[width=0.875\textwidth]{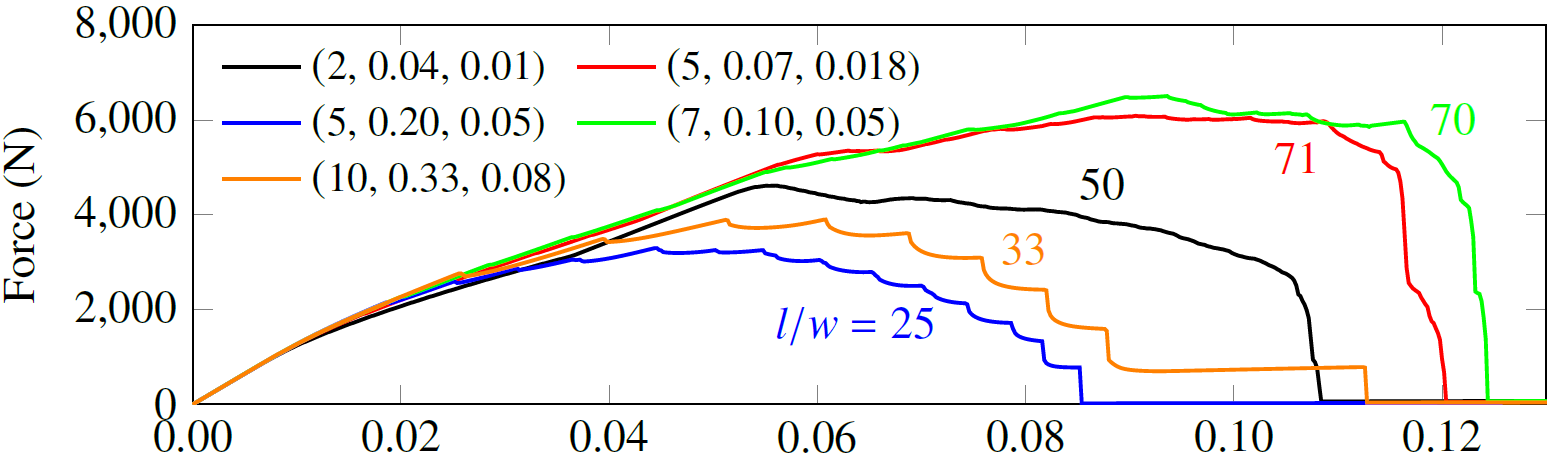}
\label{fig:force_displacement_curves_brick_mortar}
}
\hspace*{\fill}
\\
\hspace*{\fill}
     \subfloat[]{
        \centering
        \includegraphics[width=0.875\textwidth]{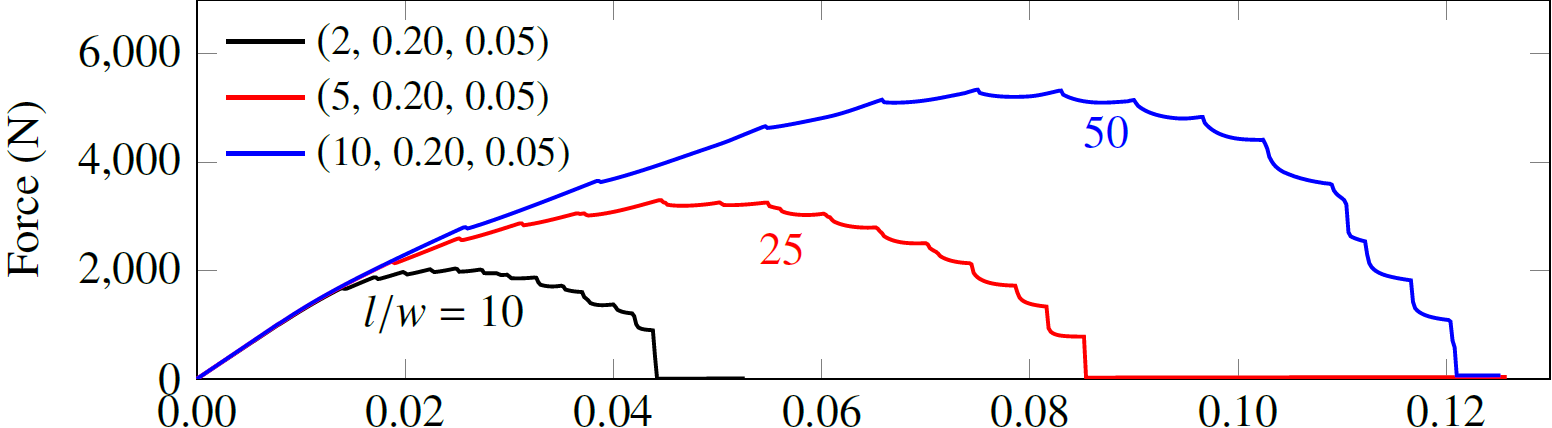}
\label{fig:force_displacement_curves_brick_length}
        }
\hspace*{\fill}
\\
\hspace*{\fill}
     \subfloat[]{
        \centering
        \includegraphics[width=0.875\textwidth]{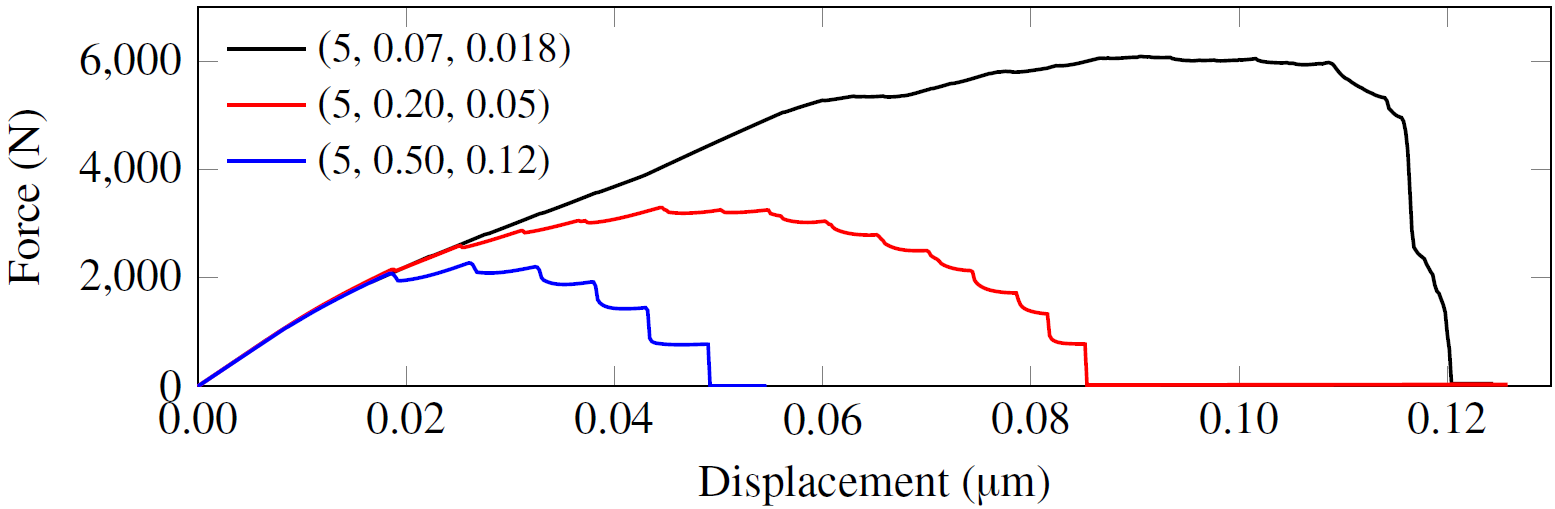}
\label{fig:force_displacement_curves_brick_width}
        }
\hspace*{\fill}
\caption{Force-displacement curves obtained with different brick length $l$, brick width $w$, and mortar thickness $t$ (a) $(l, w, t) = $ (10, 0.33, 0.04), (10, 0.33, 0.08), and (10, 0.33, 0.13), resulting in amounts of zirconia 11.5\%, 19.7\%, and 29.4\%. In this context, thinner mortars improve the fracture toughness. (b) $(l, w, t) = $ (2, 0.04, 0.01),  (5, 0.07, 0.018), (5, 0.20, 0.05), (7, 0.10, 0.05), (10, 0.33, 0.08), with the resulting amounts of zirconia being 21.2\%, 20.8\%, 20.7\%, 20.3\%, and 19.7\%. Here, the bricks with higher aspect ratios $l/w$ (indicated next to each curve) lead to improved fracture behaviour. (c) $(l, w, t) = $ (2, 0.20, 0.05), (5, 0.20, 0.05), and (10, 0.20, 0.05), resulting in amounts of zirconia of 21.9\%, 20.7\%, and 21.3\%, and in aspect ratios $l/w$ (shown next to each curve) of 10, 25, and 50, respectively. (d) $(l, w, t) = $ (5, 0.07, 0.018), (5, 0.20, 0.05), and (5, 0.50, 0.12), with the resulting amount of zirconia being 20.8\%, 20.7\%, and 21\%. The brick width is inversely proportional to the fracture performance. As the length of the brick increases, the fracture toughness also increases.}
\label{fig:crack_path_total2}
\end{figure}

Key parameters to investigate in terms of influence on fracture toughness are the brick size and mortar thickness, since these can be easily manipulated experimentally. We adopt bricks with dimensions $(l, w)$ = (10, 0.33) and vary the mortar thickness $t$ between 0.04, 0.08, and 0.13, corresponding to zirconia percentages of 11.5\%, 19.7\%, and 29.4\%, respectively. The final crack fields are displayed in Figures~\ref{fig:Brick_10_033_Mortar_004}-\ref{fig:Brick_10_033_Mortar_013}. As the mortar thickness decreases, the number of bricks in the microscale domain increases, while the overall domain size remains constant. The model with $(l, w, t)$ = (10, 0.33, 0.04) then exhibits the highest level of crack deflections and the largest crack areas. The force-displacement curves in Figure~\ref{fig:force_displacement_curves_zirconia} show that both dissipated energy and maximum resistance force are inversely proportional to the mortar thickness. Thus, thinner mortar layers - corresponding to higher number of bricks and thus more chances for crack deflection - result in improved fracture performance.

It is, however, interesting to look into the effect of microstructure geometry (brick dimensions and mortar thickness) when the zirconia content remains constant. We thus construct microscale models using five different combinations of brick dimensions and mortar thicknesses: $(l, w, t)$ = (2, 0.04, 0.01), (5, 0.07, 0.018), (5, 0.20, 0.05), (7, 0.10, 0.05), (10, 0.33, 0.08). The zirconia content in these models is approximately 20\% (21.2\%, 20.8\%, 20.7\%, 20.3\%, and 19.7\%, respectively). The force-displacement curves corresponding to these models are shown in Figure~\ref{fig:force_displacement_curves_brick_mortar}.
Significant drops in the curves are observed when approaching fracture for the models with $(l, w, t)$ = (2, 0.04, 0.01), (5, 0.07, 0.018), and (7, 0.10, 0.05), indicating brittle failure once the fracture strength of alumina is reached. In these cases, the cracks crosses the bricks instead of propagating around them, as shown in Figures~\ref{fig:Brick_2_004_Mortar_001},~\ref{fig:Brick_5_007_Mortar_0018}, and~\ref{fig:Brick_7_010_Mortar_0025}. 
These results suggest that an optimal balance between the strengths of alumina and zirconia needs to be identified to maximize the toughening effect for certain brick sizes and mortar thicknesses. Among the models, the one with $(l, w, t) = (7, 0.10, 0.05)$ demonstrates the highest resistance force and greatest energy dissipation, with a resulting fracture toughness of $K_\text{I} = \SI{8.12}{\mega\pascal\sqrt\meter}$. Therefore, a high aspect ratio $l/w$ and higher brick lengths are beneficial for enhancing fracture performance.

To further validate these findings, we modify the brick length while keeping the brick width and mortar thickness constant, with $(l, w, t)$ set to (2, 0.20, 0.05), (5, 0.20, 0.05), and (10, 0.20, 0.05). The corresponding aspect ratios $l/w$ are 10, 25, and 50, with zirconia contents of 21.9\%, 20.7\%, and 21.3\%, respectively. The force-displacement curves, presented in Figure~\ref{fig:force_displacement_curves_brick_length}, confirm that the dissipated energy during crack propagation is directly proportional to the aspect ratio and the brick length. The final crack fields for these configurations are shown in Figures~\ref{fig:Brick_2_020_Mortar_005},~\ref{fig:Brick_5_020_Mortar_005}, and~\ref{fig:Brick_10_020_Mortar_005}, further illustrating the trend that longer bricks and higher aspect ratios contribute to enhanced fracture resistance.

Finally, we examine the effect of brick width on fracture behaviour. Since the overall model dimensions remain fixed, both brick width and mortar thickness are adjusted to maintain the same zirconia content across the models. We construct three models with dimensions $(l, w, t)$ = (5, 0.07, 0.018), (5, 0.20, 0.05), and (5, 0.50, 0.12), yielding zirconia contents of 20.8\%, 20.7\%, and 21\%, respectively. The force-displacement curves are presented in Figure~\ref{fig:force_displacement_curves_brick_width}. 
As the brick width decreases, the number of bricks increases within the fixed model dimensions. The model with the smallest brick width, $w = 0.018$, shows the most complex crack path and the highest energy dissipation, as shown in Figure~\ref{fig:Brick_5_007_Mortar_0018}. This confirms that longer and thinner bricks, characterized by a high aspect ratio, lead to higher fracture toughness. Table~\ref{tab:brick_and_mortar_data} presents the resulting zirconia percentage and fracture toughness obtained for various combinations of brick length $l$, brick width $w$, and mortar thickness $t$.

\begin{table}[htbp]
\newcolumntype{G}{>{\centering\arraybackslash}m{3.0em}}
 \newcolumntype{E}{>{\centering\arraybackslash}m{3.50em}}
\newcolumntype{P}{>{\centering\arraybackslash}m{5.00em}}
 \newcolumntype{X}{>{\centering\arraybackslash}m{6.00em}}
 \newcolumntype{T}{>{\centering\arraybackslash}m{14.00em}}
\begin{center}
\begin{tabular}{G*3{E}@{}*1{X}@{}*1{X}@{}}
\hline Model &  $l$ ($\si{\micro\meter}$)   &  $w$ ($\si{\micro\meter}$)  & $t$ ($\si{\micro\meter}$)   & $\mathrm{ZrO_2}$ (\%) & $K_\text{I}$ ($\si{\mega\pascal}\sqrt{\si{\meter}}$)\\
\hline    1     &        2                         &          0.04                   &             0.010            &       21.18             &    5.41   \\
            2     &        2                         &          0.20                   &             0.050            &        21.93             &   2.76   \\
            3      &       5                         &          0.07                   &             0.018          &           20.78            &   7.79 \\
            4      &       5                         &          0.20                   &             0.050            &          20.73           &   4.85   \\
            5      &       5                         &          0.50                   &             0.120            &          20.98           &   2.96   \\
            6      &       7                         &          0.10                   &             0.025          &            20.30           &   8.12 \\
            7      &      10                       &          0.20                   &             0.050            &           21.33          &    7.21  \\
            8      &      10                       &          0.33                   &             0.040            &            11.49         &   6.00   \\
            9      &      10                       &          0.33                   &             0.080            &            19.74         &   5.57 \\
            10    &      10                       &          0.33                   &             0.130            &             29.37        &   5.04   \\
\hline
\end{tabular}
\end{center}
\caption{Fracture toughness $K_\text{I}$ evaluated under different brick length $l$, brick width $w$, and mortar thickness $t$, where the highest value $\SI{8.12}{\mega\pascal}\sqrt{\si{\meter}}$ and lowest value $\SI{2.76}{\mega\pascal}\sqrt{\si{\meter}}$ of fracture toughness are found with $(l, w, t) = (7, 0.10, 0.025)$ and $(2, 0.20, 0.05)$.}
\label{tab:brick_and_mortar_data}
\end{table}

\subsubsection{Material strength}

\begin{figure}[htbp]
\centering
\hspace*{\fill}
     \subfloat[$(\sigma_{f\mathrm{Z}}, \sigma_{f\mathrm{A}} = \SI{0.5}{\giga\pascal}, \SI{5}{\giga\pascal})$]{
        \centering
        \def\svgwidth{0.475\textwidth} 
\begingroup%
  \makeatletter%
  \providecommand\color[2][]{%
    \errmessage{(Inkscape) Color is used for the text in Inkscape, but the package 'color.sty' is not loaded}%
    \renewcommand\color[2][]{}%
  }%
  \providecommand\transparent[1]{%
    \errmessage{(Inkscape) Transparency is used (non-zero) for the text in Inkscape, but the package 'transparent.sty' is not loaded}%
    \renewcommand\transparent[1]{}%
  }%
  \providecommand\rotatebox[2]{#2}%
  \newcommand*\fsize{\dimexpr\f@size pt\relax}%
  \newcommand*\lineheight[1]{\fontsize{\fsize}{#1\fsize}\selectfont}%
  \ifx\svgwidth\undefined%
    \setlength{\unitlength}{430.90402222bp}%
    \ifx\svgscale\undefined%
      \relax%
    \else%
      \setlength{\unitlength}{\unitlength * \real{\svgscale}}%
    \fi%
  \else%
    \setlength{\unitlength}{\svgwidth}%
  \fi%
  \global\let\svgwidth\undefined%
  \global\let\svgscale\undefined%
  \makeatother%
  \begin{picture}(1,0.16244917)%
    \lineheight{1}%
    \setlength\tabcolsep{0pt}%
    \put(0,0){\includegraphics[width=\unitlength,page=1]{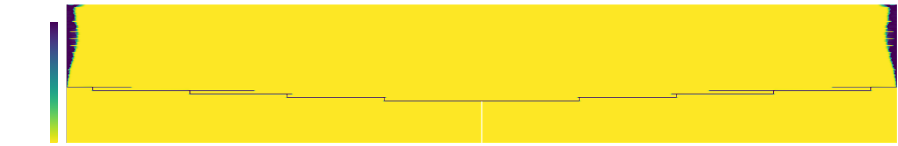}}%
    \put(0.02,-0.01){\color[rgb]{0,0,0}\makebox(0,0)[lt]{\lineheight{1.25}\smash{\begin{tabular}[t]{l}0\end{tabular}}}}%
    \put(0.02,0.12){\color[rgb]{0,0,0}\makebox(0,0)[lt]{\lineheight{1.25}\smash{\begin{tabular}[t]{l}1\end{tabular}}}}%
    \put(0.0475,0.16){\color[rgb]{0,0,0}\makebox(0,0)[lt]{\lineheight{1.25}\smash{\begin{tabular}[t]{l}$\phi$\end{tabular}}}}%
  \end{picture}%
\endgroup%

        \label{fig:Brick_7_010_Mortar_0025_5GPa}
        }
\hspace*{\fill}
      \subfloat[$(0.5, 10)$]{
        \centering
        \includegraphics[width=0.475\textwidth]{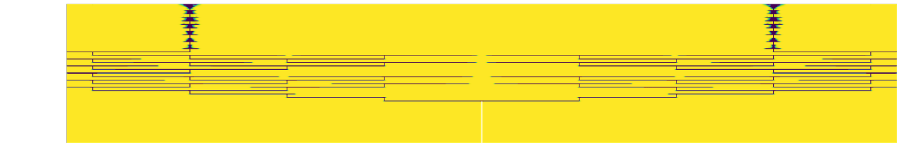}
        \label{fig:Brick_7_010_Mortar_0025_10GPa}
        }
\hspace*{\fill}
\\
\hspace*{\fill}
      \subfloat[$(0.5, 20)$]{
      \centering
        \includegraphics[width=0.475\textwidth]{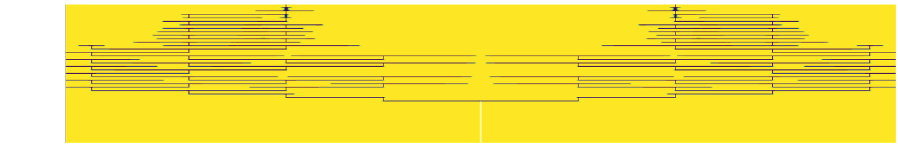}
        \label{fig:Brick_7_010_Mortar_0025_20GPa}
        }
\hspace*{\fill}
      \subfloat[$(0.5, 30)$]{
      \centering
        \includegraphics[width=0.475\textwidth]{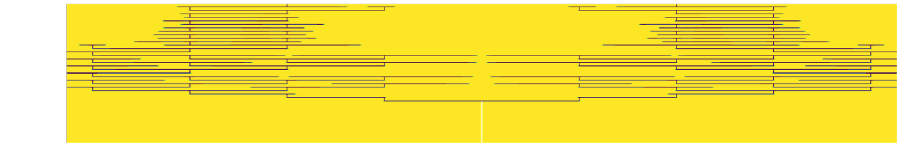}
        \label{fig:Brick_7_010_Mortar_0025_30GPa}
        }
\hspace*{\fill}
\\
\hspace*{\fill}
      \subfloat[$(0.8, 20)$]{
      \centering
        \includegraphics[width=0.475\textwidth]{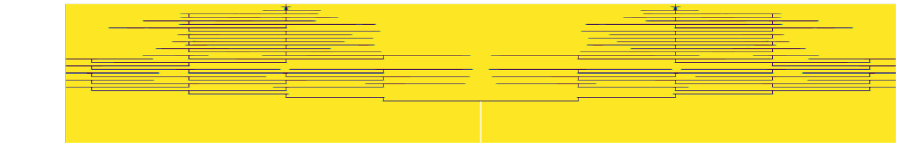}
        \label{fig:Brick_7_010_Mortar_0025_800MPa}
        }
\hspace*{\fill}
      \subfloat[$(1.0, 20)$]{
      \centering
        \includegraphics[width=0.475\textwidth]{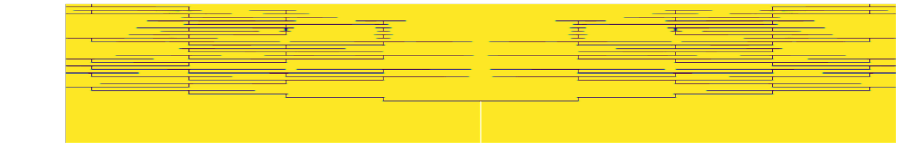}
        \label{fig:Brick_7_010_Mortar_0025_1GPa}
        }
\hspace*{\fill}
\\
\hspace*{\fill}
     \subfloat[]{
        \centering
        \includegraphics[width=0.80\textwidth]{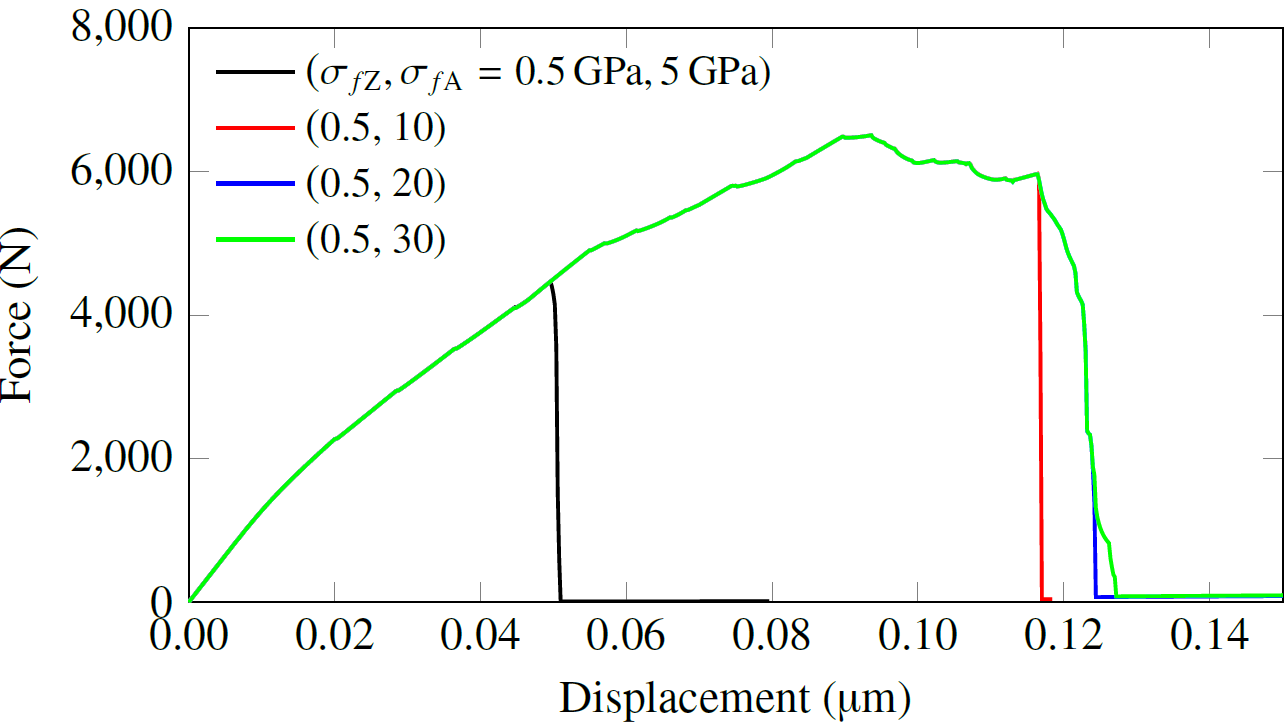}
\label{fig:force_displacement_curves_alumina_strength}
        }
\hspace*{\fill}
\\
\hspace*{\fill}
     \subfloat[]{
        \centering
        \includegraphics[width=0.80\textwidth]{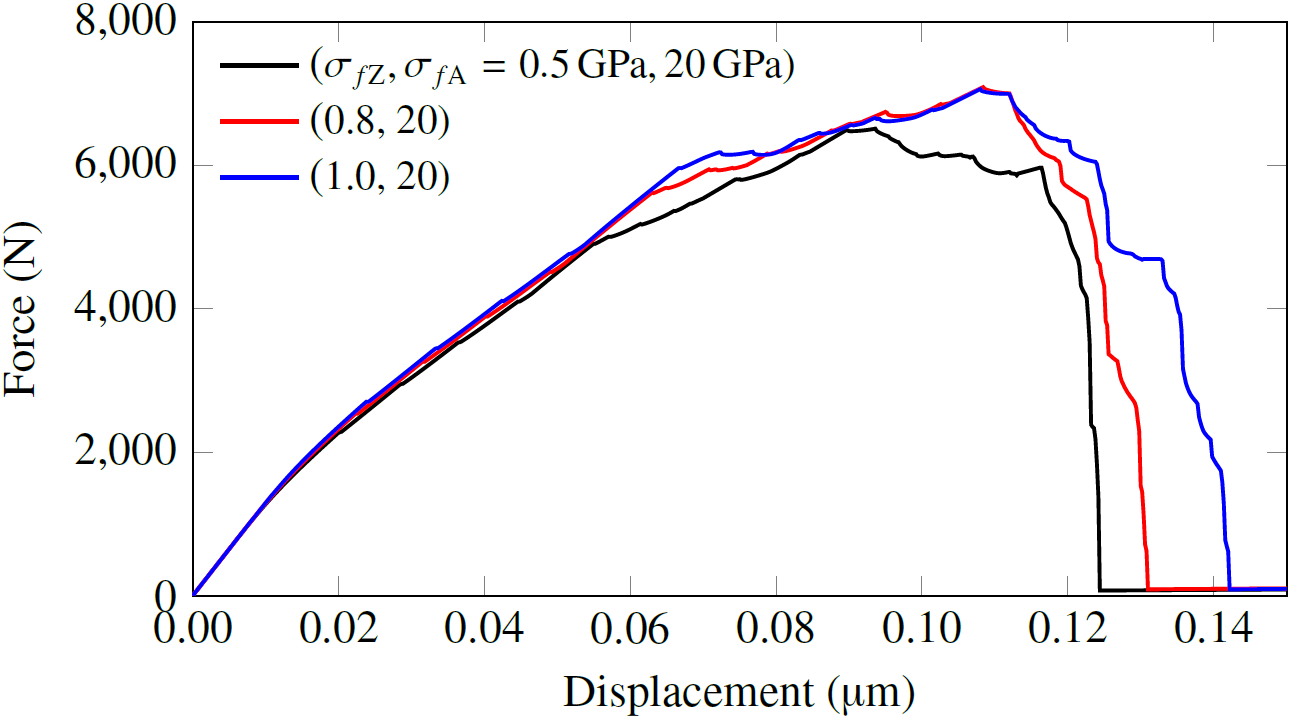}
\label{fig:force_displacement_curves_zirconia_strength}
        }
\hspace*{\fill}
\caption{Final crack fields $\phi$ (0 indicating intact material and 1 crack) obtained with different fracture strengths ($\si{\giga\pascal}$) of zirconia $\sigma_{f\mathrm{Z}}$ and alumina $\sigma_{f\mathrm{A}}$ (a) (0.5, 5), (b) (0.5, 10), (c) (0.5, 20), (d) (0.5, 30), (e) (0.8, 20), and (f) (1, 20), where the higher material strength results in more complex crack fields. Figure (g) shows force-displacement curves obtained with different alumina (brick) fracture strengths $\sigma_{f\mathrm{A}}$, where high alumina strength benefits fracture performance. Figure (h) shows force-displacement curves obtained with different zirconia (mortar) fracture strengths, where high zirconia strength benefits fracture performance.}
 \label{fig:crack_path_strength}
\end{figure}
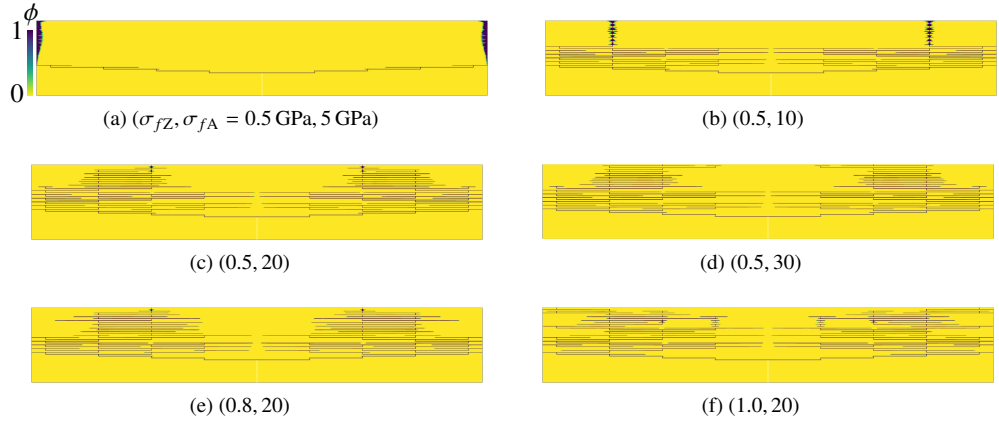
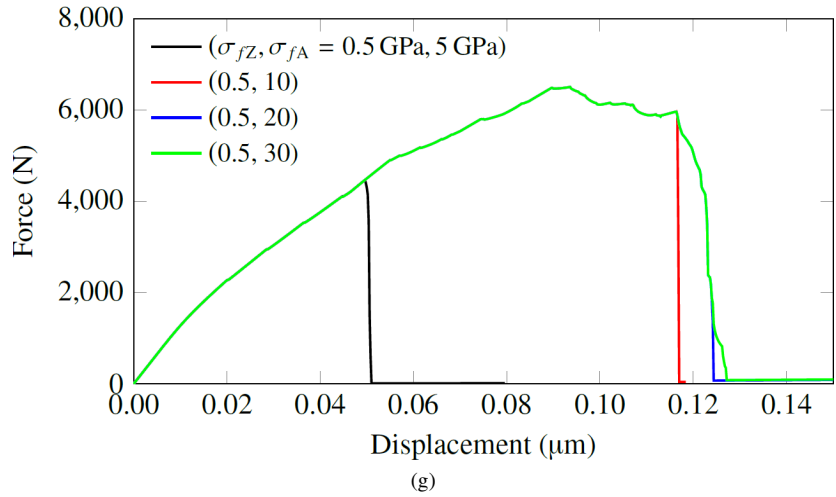
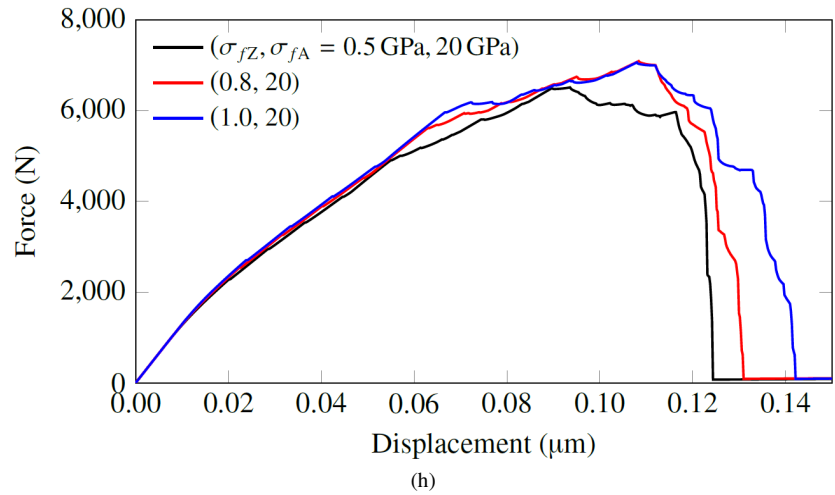

As hinted in the previous section when detecting fracture crossing alumina bricks, the strengths of alumina and zirconia are expected to also play a role in the composites' fracture behaviour. We investigate this effect in the microscale model with brick and mortar dimensions $(l, w, t) = (7, 0.10, 0.05)$. With the zirconia strength set at $\SI{0.5}{\giga\pascal}$, we first vary the alumina strength to $\SI{5}{\giga\pascal}$, $\SI{10}{\giga\pascal}$, $\SI{20}{\giga\pascal}$, and $\SI{30}{\giga\pascal}$. The resulting crack fields are presented in Figures~\ref{fig:Brick_7_010_Mortar_0025_5GPa}-\ref{fig:Brick_7_010_Mortar_0025_30GPa}. Under the lowest alumina strength of $\SI{5}{\giga\pascal}$, cracks propagate through the bricks with sudden fracture reaching until the model boundaries. As alumina strength increases, more crack deflections occur during propagation. 
The force-displacement curves in Figure~\ref{fig:force_displacement_curves_alumina_strength} reveal that the curves for alumina strengths of $\SI{20}{\giga\pascal}$ and $\SI{30}{\giga\pascal}$ nearly overlap, except for the final part. 
In the case of $\SI{20}{\giga\pascal}$, cracks cross several bricks at the end of the fracture process (Figure~\ref{fig:Brick_7_010_Mortar_0025_20GPa}). 
For $\SI{30}{\giga\pascal}$, the fracture occurs exclusively along the mortar (Figure~\ref{fig:Brick_7_010_Mortar_0025_30GPa}), as the stress does not reach the alumina’s strength.
Additionally, we increase the zirconia strength to $\SI{0.8}{\giga\pascal}$ and $\SI{1}{\giga\pascal}$ while maintaining an alumina strength of $\SI{20}{\giga\pascal}$. The final crack fields and the corresponding force-displacement curves are shown in Figures~\ref{fig:Brick_7_010_Mortar_0025_800MPa},~\ref{fig:Brick_7_010_Mortar_0025_1GPa}, and~\ref{fig:force_displacement_curves_zirconia_strength}, illustrating cracks propagating through a few bricks.
 With the higher zirconia strengths, the resulting fracture toughness values are $K_\text{I} = \SI{8.52}{\mega\pascal\sqrt\meter}$ and $K_\text{I} = \SI{8.91}{\mega\pascal\sqrt\meter}$, respectively, demonstrating improved fracture resistance.

\subsection{Optimization}
Based on all the examples above, it has become evident that the fracture behaviour of ‘double-tough' ceramics with brick-and-mortar structures and phase transformation-toughened mortar can be tuned by altering several factors. Among the most influential, the microstructure's geometry and the properties of the material constituents stand out as particularly relevant to guide experiments. Brick dimensions and alumina strength can be altered by selecting the material building blocks accordingly, the mortar thickness can be tuned in the early steps of material synthesis via e.g. sol-gel processes\cite{Francesco2024}, and the strength of zirconia can be tailored by controlling its composition and grain size. From our numerical analysis, it has emerged that longer and thinner bricks with a high aspect ratio enhance fracture resistance, whereas a thicker mortar layer tends to reduce fracture performance. Variations in alumina and zirconia strengths across different brick-and-mortar configurations also lead to distinct crack propagation patterns. 

Given the large number of influencing parameters and the complex interplays among them, optimization algorithms become instrumental for an effective and efficient material design. The particle swarm optimization (PSO) method is selected here. It is implemented to maximize fracture resistance, with the objective function defined as the fracture toughness extracted from the force-displacement curves during the fracture process. This approach allows for the determination of the optimal combination of brick sizes, mortar thickness, and material properties to enhance the mechanical performance of the target ‘double-tough' ceramics.

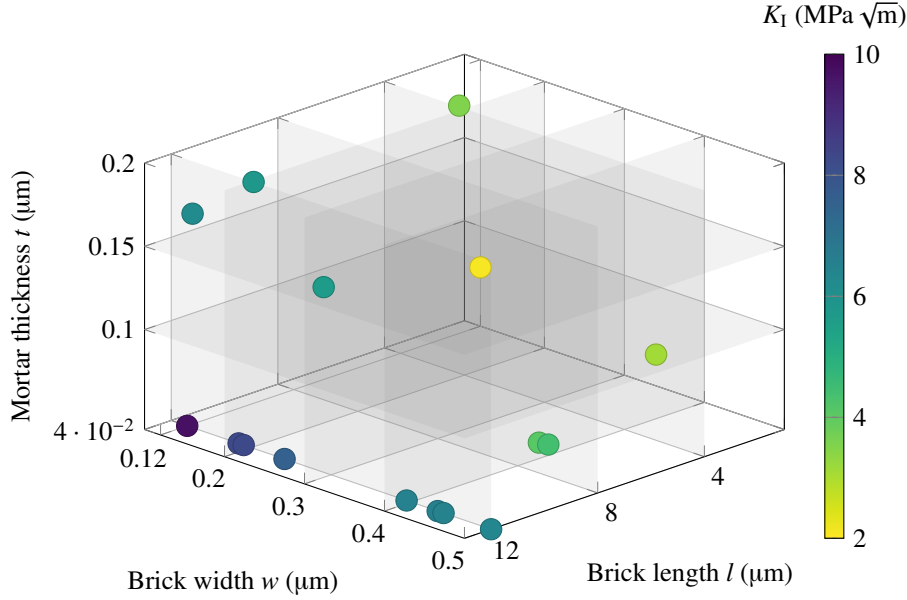
\begin{figure}[htbp]
\centering
   \begin{tikzpicture}[]
        \begin{axis}[
            colorbar,
    colormap={reverse viridis}{
        indices of colormap={
            \pgfplotscolormaplastindexof{viridis},...,0 of viridis}
    },
colorbar/width=2.5mm,
colorbar style={title= $K_\text{I}$ $(\si{\mega\pascal}\sqrt{\si{\meter}})$},
            point meta min=2,
            point meta max=10,
        ymajorgrids,
        xmajorgrids,
        zmajorgrids,
	xmin = 1,
	xmax = 13,
	ymin = 0.10,
	ymax = 0.5,
	zmin = 0.04,
	zmax = 0.2,
        width=10cm,
        height=8cm,
	xlabel={Brick length $l$ ($\si{\micro\meter})$},
        ylabel={Brick width $w$ $(\si{\micro\meter})$},
        zlabel={Mortar thickness $t$ $(\si{\micro\meter})$},
        xtick={4, 8, 12},
        ytick={0.12 ,0.2, 0.3, 0.4, 0.5},
        ztick={0.04 ,0.1, 0.15, 0.2},
        view={135}{30}
        ]
            \addplot3 [
               only marks, scatter, mark size=4,point meta=explicit
            ] table [x=X,y=Y,z=Z, meta=meta] {
            X           Y           Z         meta
         4.8683   0.4685   0.1009     3.11
        10.1391  0.2286   0.1309     5.6322
        7.4379    0.4077   0.0521     4.075
        7.047     0.4064    0.0486     4.3856
        12          0.4332   0.04         6.5477 
        12          0.203     0.2          5.7757
        12          0.12       0.04        9.6849
        1            0.12      0.0753     2.0739
        12          0.3943   0.04        6.4863
        12          0.1267   0.1685     6.1596
        3.9337    0.1912   0.2          3.5119
        12          0.1843   0.04         8.2252
        12          0.2417   0.04         7.5331
        12          0.4405   0.04         6.4485
        12          0.1906   0.04         8.2371
        12          0.5        0.04         6.3335
        };

         \fill[black!50,opacity=0.10] (4, 0.10, 0.04) -- (4, 0.10, 0.20) -- (4, 0.5, 0.20) -- (4, 0.50, 0.04) -- cycle;
         \fill[black!50,opacity=0.10] (8, 0.10, 0.04) -- (8, 0.10, 0.20) -- (8, 0.5, 0.20) -- (8, 0.50, 0.04) -- cycle;
         \fill[black!50,opacity=0.10] (12, 0.10, 0.04) -- (12, 0.10, 0.20) -- (12, 0.5, 0.20) -- (12, 0.50, 0.04) -- cycle;

         \fill[black!50,opacity=0.10] (13, 0.50, 0.10) -- (1, 0.50, 0.10) -- (1, 0.10, 0.10) -- (13, 0.10, 0.10) -- cycle;
         \fill[black!50,opacity=0.10] (13, 0.50, 0.15) -- (1, 0.50, 0.15) -- (1, 0.10, 0.15) -- (13, 0.10, 0.15) -- cycle;

         \fill[black!50,opacity=0.10] (13, 0.40, 0.04) -- (13, 0.40, 0.20) -- (1, 0.40, 0.20) -- (1, 0.40, 0.04) -- cycle;
         \fill[black!50,opacity=0.10] (13, 0.30, 0.04) -- (13, 0.30, 0.20) -- (1, 0.30, 0.20) -- (1, 0.30, 0.04) -- cycle;
         \fill[black!50,opacity=0.10] (13, 0.20, 0.04) -- (13, 0.20, 0.20) -- (1, 0.20, 0.20) -- (1, 0.20, 0.04) -- cycle;
        \end{axis}
    \end{tikzpicture}
\caption{Relationship between the three geometric design variables (brick length $l \in [1, 12]$, brick width $w \in [0.12, 0.50]$ and mortar thickness $t \in [0.04, 0.20]$, in $\si{\micro\meter}$) and the corresponding objective function (the fracture toughness $K_\text{I}$ $(\si{\mega\pascal}\sqrt{\si{\meter}})$), showing how high brick aspect ratios $l/w$ and thin mortar layers benefit the fracture resistance.}
\label{fig:optimization_plot1}
\end{figure}

We begin by focusing on the geometric parameters, where the brick length $l$, brick width $w$, and mortar thickness $t$ are selected as the three design variables. These variables are constrained within the following ranges: $[1, 12]$, $[0.12, 0.50]$, and $[0.04, 0.20]$, again in $\si{\micro\meter}$, resulting in brick aspect ratios within the range of $[2, 100]$. The optimization algorithm is set to run for a maximum of 10 iterations, with four candidates evaluated per iteration. The optimal design variables and corresponding objective values for each iteration are shown in Table~\ref{tab:optimization_data1} in the Appendix. The highest fracture toughness, $K_\text{I} = \SI{9.68}{\mega\pascal}\sqrt{\si{\meter}}$, is achieved with brick sizes and mortar thickness of $(l, w, t) = (12, 0.12, 0.04)$, corresponding to the maximum aspect ratio of 100. Figure~\ref{fig:optimization_plot1} illustrates the relationship between the design variables and the objective function. 

\begin{figure}[htbp]
\captionsetup[subfigure]{justification=centering}
\centering
\hspace*{\fill}
     \subfloat[($l, w, t, \sigma_{f\mathrm{Z}}, \sigma_{f\mathrm{A}}$ = \\ $ 10.96\si{\micro\meter}, 0.33\si{\micro\meter}, 0.12\si{\micro\meter}, 1.94\si{\giga\pascal}, 23.22\si{\giga\pascal}$)]{
        \centering
        \def\svgwidth{0.475\textwidth}
\begingroup%
  \makeatletter%
  \providecommand\color[2][]{%
    \errmessage{(Inkscape) Color is used for the text in Inkscape, but the package 'color.sty' is not loaded}%
    \renewcommand\color[2][]{}%
  }%
  \providecommand\transparent[1]{%
    \errmessage{(Inkscape) Transparency is used (non-zero) for the text in Inkscape, but the package 'transparent.sty' is not loaded}%
    \renewcommand\transparent[1]{}%
  }%
  \providecommand\rotatebox[2]{#2}%
  \newcommand*\fsize{\dimexpr\f@size pt\relax}%
  \newcommand*\lineheight[1]{\fontsize{\fsize}{#1\fsize}\selectfont}%
  \ifx\svgwidth\undefined%
    \setlength{\unitlength}{420bp}%
    \ifx\svgscale\undefined%
      \relax%
    \else%
      \setlength{\unitlength}{\unitlength * \real{\svgscale}}%
    \fi%
  \else%
    \setlength{\unitlength}{\svgwidth}%
  \fi%
  \global\let\svgwidth\undefined%
  \global\let\svgscale\undefined%
  \makeatother%
  \begin{picture}(1,0.16666667)%
    \lineheight{1}%
    \setlength\tabcolsep{0pt}%
    \put(0,0){\includegraphics[width=\unitlength,page=1]{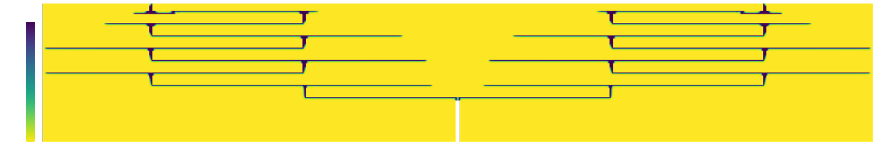}}%
    \put(0.02,0.16){\color[rgb]{0,0,0}\makebox(0,0)[lt]{\lineheight{1.25}\smash{\begin{tabular}[t]{l}$\phi$\end{tabular}}}}%
    \put(0.00,0.12){\color[rgb]{0,0,0}\makebox(0,0)[lt]{\lineheight{1.25}\smash{\begin{tabular}[t]{l}1\end{tabular}}}}%
    \put(0.00,-0.01){\color[rgb]{0,0,0}\makebox(0,0)[lt]{\lineheight{1.25}\smash{\begin{tabular}[t]{l}0\end{tabular}}}}%
  \end{picture}%
\endgroup%
\label{fig:MicroModel_BrickLength_10.9637_BrickHeight_0.3278_MortarThickness_0.1177_StrengthZ_1.9392_StrengthA_23.2167}
        }
        \hspace*{\fill}
     \subfloat[(12, 0.37, 0.20, 2, 2)]{
        \centering
        \includegraphics[width=0.475\textwidth]{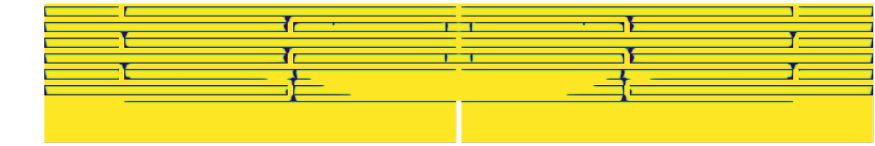}
        \label{fig:MicroModel_BrickLength_12_BrickHeight_0.3674_MortarThickness_0.1986_StrengthZ_2_StrengthA_2}
        }
        \hspace*{\fill}
\\
\hspace*{\fill}
     \subfloat[(12, 0.20, 0.06, 1.95, 23.12)]{
        \centering
        \includegraphics[width=0.475\textwidth]{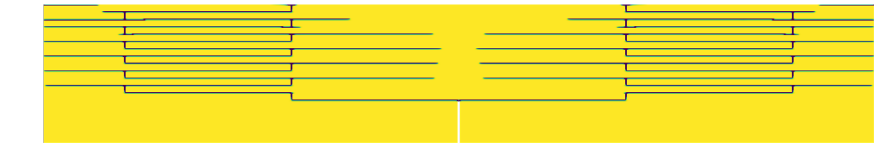}
        \label{fig:MicroModel_BrickLength_12_BrickHeight_0.2014_MortarThickness_0.0641_StrengthZ_1.9459_StrengthA_23.1247}
        }
        \hspace*{\fill}
     \subfloat[(12, 0.23, 0.04, 1.89, 23.13)]{
        \centering
        \includegraphics[width=0.475\textwidth]{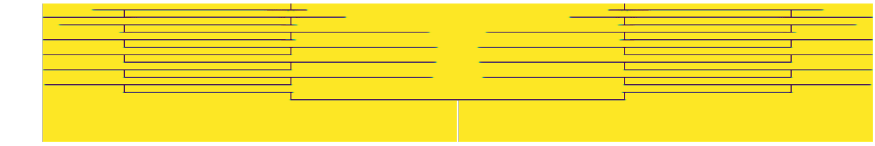}
        \label{fig:MicroModel_BrickLength_12_BrickHeight_0.2306_MortarThickness_0.04_StrengthZ_1.8864_StrengthA_23.1343}
        }
        \hspace*{\fill}
\\
\hspace*{\fill}
     \subfloat[(12, 0.26, 0.04, 1.82, 23.14)]{
        \centering
        \includegraphics[width=0.475\textwidth]{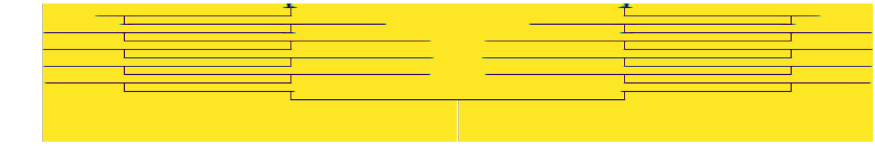}
        \label{fig:MicroModel_BrickLength_12_BrickHeight_0.2627_MortarThickness_0.04_StrengthZ_1.8209_StrengthA_23.1448}
        }
        \hspace*{\fill}
     \subfloat[(12, 0.20, 0.05, 2, 28.34)]{
        \centering
        \includegraphics[width=0.475\textwidth]{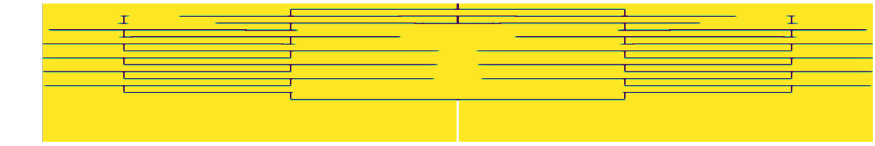}
        \label{fig:MicroModel_BrickLength_12_BrickHeight_0.1966_MortarThickness_0.0542_StrengthZ_2_StrengthA_28.3352}
        }
        \hspace*{\fill}
\\
\hspace*{\fill}
     \subfloat[(12, 0.12, 0.04, 2, 24.58)]{
        \centering
        \includegraphics[width=0.475\textwidth]{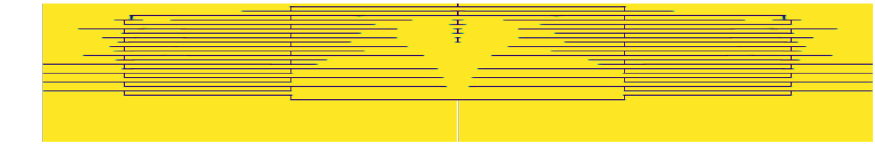}
        \label{fig:MicroModel_BrickLength_12_BrickHeight_0.12_MortarThickness_0.04_StrengthZ_2_StrengthA_24.58}
        }
        \hspace*{\fill}
     \subfloat[(12, 0.12, 0.04, 2, 30)]{
        \centering
        \includegraphics[width=0.475\textwidth]{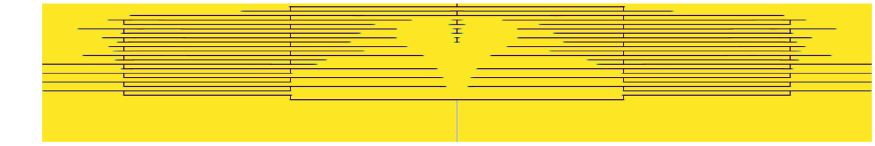}
        \label{fig:MicroModel_BrickLength_12_BrickHeight_0.12_MortarThickness_0.04_StrengthZ_2_StrengthA_30}
        }
        \hspace*{\fill}
\caption{Final crack fields $\phi$ (0 indicating intact material and 1 crack) of the optimal design variables at each iteration are given by brick length $l$ $(\si{\micro\meter})$, brick width $w$ $(\si{\micro\meter})$, mortar thickness $t$ $ (\si{\micro\meter})$, zirconia fracture strength $\sigma_{f\mathrm{Z}}$ $(\si{\giga\pascal})$, and alumina fracture strength $\sigma_{f\mathrm{A}}$ $(\si{\giga\pascal})$, where $(l, w, t, \sigma_{f\mathrm{Z}}, \sigma_{f\mathrm{A}})=$ (a) (10.96, 0.33, 0.12, 1.94, 23.22), (b) (12, 0.37, 0.20, 2, 2), (c) (12, 0.20, 0.06, 1.95, 23.12), (d) (12, 0.23, 0.04, 1.89, 23.13), (e) (12, 0.26, 0.04, 1.82, 23.14), (f) (12, 0.20, 0.05, 2, 28.34), (g) (12, 0.12, 0.04, 2, 24.58), and (h) (12, 0.12, 0.04, 2, 30).}
 \label{fig:crack_path_optimization}
\end{figure}

\begin{figure}[htbp]
\centering
   \begin{tikzpicture}[]
   \begin{pgfonlayer}{background}
        \begin{axis}[
            colorbar,
    colormap={reverse viridis}{
        indices of colormap={
            \pgfplotscolormaplastindexof{viridis},...,0 of viridis}
    },
colorbar/width=2.5mm,
colorbar style={title= $K_\text{I}$ $(\si{\mega\pascal}\sqrt{\si{\meter}})$},
            point meta min=1,
            point meta max=13,
        ymajorgrids,
        xmajorgrids,
        zmajorgrids,
	zmin = 0,
	zmax = 100,
	ymin = -0.1,
	ymax = 2.2,
	xmin = 0,
	xmax = 32,
        width=10cm,
        height=8cm,
	zlabel={Brick aspect ratio $l/w$},
        ylabel={Zirconia fracture strength $\sigma_{f\mathrm{Z}}$ ($\si{\giga\pascal}$)},
        xlabel={Alumina fracture strength $\sigma_{f\mathrm{A}}$ ($\si{\giga\pascal})$},
        ztick={2, 50, 100},
        ytick={0, 1, 2.0},
        xtick={2 ,15, 30},
        view={135}{30}
        ]
            \addplot3 [
               only marks, scatter, mark size=4,point meta=explicit
            ] table [x=X,y=Y,z=Z, meta=meta] {
            Z           Y       X          meta
            33.21     1.94      23.22     8.41 
            32.43     2         2         6.11 
            60        1.95      23.12     10.36
            52.17     1.89      23.13     10.67
            46.15     1.82      23.14     9.82
            60        2         28.34     10.84
            44.12 	  1.6883 	21.00 	  6.1811
            4.95 	  1.4836 	22.81 	  3.2521
            22.7 	  0.5536 	12.98 	  5.3692
            44.22 	  1.7737 	20.35 	  7.1974
            4.24 	  1.901 	6.79      2.756
            5.72 	  2 		23.1159   2.6915
            2 		  0.5 		30 		  2.2797
            8.33 	  2 		11.6608   3.0237
            7.77	  2 		25.6059   2.9697
            100 	  0.7036 	7.3317 	  5.185
            29.74 	  1.7272 	19.5764   7.2241
            31.18 	  2 		18.3664   4.8138
            17.49 	  1.5719 	17.8689   5.8504
            60.94 	  1.4837 	21.6753   8.5004
            39.34 	  0.5 	    29.7792   6.5887
            26.25	  1.427 	30 		  5.9627
            95.40 	  0.5 		20.9543   7.4326
            24 		  0.7851 	30 		  6.8208
            44.12	  0.9158 	17.3516   6.5215
            19.03 	  1.8605 	30 		  7.337
            21.19 	  1.566 	28.7036   6.16
            100 	  0.5 		23.801 	  9.7478
            100 	  0.6134 	22.8278   9.0508
            43.24 	  1.1206 	22.2443   7.6855
            15.31 	  2 		30 		  5.7234
            48.13 	  0.6313    14.9919 	6.8666
            46.55 	  1.7867 	23.1289 	9.8411
            61.41 	  1.0263 	26.2615 	9.409
            45.42 	  0.8495 	14.9414 	7.9038
            61.76 	  1.6866 	26.0433 	7.5286
            80.81 	  0.5 		13.3025 	7.3101
            51.70 	  1.8524 	23.122 		10.5789
            72.46 	  0.5 		24.6432 	8.675
            59.14 	  0.5808 	26.8912 	7.8356
            97.95 	  0.5 		16.93 		9.2358
            79.24 	 0.5 		23.4711 	8.5422
            59.11 	 1.9803 	23.1255 	11.5316
            100 	 0.5 		20.9183 	9.5413
            100 	 0.5 		30 		    8.2546
            100 	 0.5 		15.9584 	9.6173
            63		1.7461 		30 		    10.3013
            100 	1.4321 		20.63 		11.7421
            100 	0.9957 		29.4707 	10.6346
            100 	0.5 		15.4982 	9.8899
            64.45 	2 		    30 		    11.7665
            100 	0.6387 		16.9343 	9.6932
            57.36   2 		    28.9449 	11.0702
            100 	1.4367 		29.0665 	11.7528
            100       2         30        12.99
            100       2         24.58     12.99
		  100       2         26.56     12.99
		  100       2         27.37     12.99
		  100       2         28.74     12.99
		  100       2         29.51     12.99
        };
        \coordinate (inset) at (axis description cs:1,1.1);
    \draw[thick,black] (34,2.2,97) -- (34,2.2,115);
    \draw[thick,black] (34,2.2,97) -- (26,2.2,97);
    \draw[thick,black] (26,2.2,97) -- (26,2.2,115);
    \draw[thick,black] (34,2.2,115) -- (26,2.2,115);
    \draw[thick,black] (34,2.0,97) -- (34,2.0,115);
    \draw[thick,black] (34,2.0,97) -- (26,2.0,97);
    \draw[thick,black] (26,2.0,97) -- (26,2.0,115);
    \draw[thick,black] (34,2.0,115) -- (26,2.0,115);
    \draw[thick,black] (34,2.2,97) -- (34,2.0,97);
    \draw[thick,black] (34,2.0,115) -- (34,2.2,115);
    \draw[thick,black] (26,2.2,97) -- (26,2.0,97);
    \draw[thick,black] (26,2.2,115) -- (26,2.0,115);
    \draw[{Circle[red]}-Latex] (25,2.0,98) -- (22,2.2,125);

         \fill[black!50,opacity=0.10] (2, -0.1, 2) -- (2, -0.1, 100) -- (2, 2.2, 100) -- (2, 2.2, 2) -- cycle;
         \fill[black!50,opacity=0.10] (15, -0.1, 2) -- (15, -0.1, 100) -- (15, 2.2, 100) -- (15, 2.2, 2) -- cycle;
         \fill[black!50,opacity=0.10] (30, -0.1, 2) -- (30, -0.1, 100) -- (30, 2.2, 100) -- (30, 2.2, 2) -- cycle;

         \fill[black!50,opacity=0.10] (32, -0.1, 50) -- (32, 2.2, 50) -- (0, 2.2, 50) -- (0, -0.1, 50) -- cycle;

          \fill[black!50,opacity=0.10] (2, 1, 2) -- (2, 1, 100) -- (32, 1, 100) -- (32, 1, 2) -- cycle;

        \end{axis}
            \end{pgfonlayer}

               \begin{pgfonlayer}{foreground}
        \begin{axis}[
                at={(inset)},
        anchor=north east,
    colormap={reverse viridis}{
        indices of colormap={
            \pgfplotscolormaplastindexof{viridis},...,0 of viridis}
    },
            point meta min=6,
            point meta max=13,
        ymajorgrids,
        xmajorgrids,
        zmajorgrids,
	zmin = 99,
	zmax = 101,
	ymin = 1.5,
	ymax = 2.5,
	xmin = 24,
	xmax = 31,
        width=5cm,
        height=4cm,
        ztick={100},
        ytick={2.0},
        xtick={24 ,30},
        view={135}{30}
        ]
            \addplot3 [
               only marks, scatter, mark size=4,point meta=explicit
            ] table [x=X,y=Y,z=Z, meta=meta] {
            Z           Y       X          meta
            100       2         30        12.99
            100       2         24.58     12.99
		  100       2         26.56     12.99
		  100       2         27.37     12.99
		  100       2         28.74     12.99
		  100       2         29.51     12.99
        };

        \end{axis}
   \end{pgfonlayer}    
    \end{tikzpicture}
\caption{Relationship between brick aspect ratio $l/w\in [2, 100]$, zirconia fracture strength ($\si{\giga\pascal}$) $\sigma_{f\mathrm{Z}} \in [0.5, 2]$, alumina fracture strength ($\si{\giga\pascal}$) $\sigma_{f\mathrm{A}} \in [2, 30]$, and the corresponding objective function (the fracture toughness $K_\text{I}$ ($\si{\mega\pascal}\sqrt{\si{\meter}}$), where high brick aspect ratios $l/w$ and high fracture strength of alumina and zirconia benefit the fracture resistance. Under the brick aspect ratio $l/w = 100$ and zirconia fracture strength $\sigma_{f\mathrm{Z}} = \SI{2}{\giga\pascal}$,
the maximum fracture toughness $K_\text{I}$ is obtained when the alumina fracture strength ($\si{\giga\pascal}$) $\sigma_{f\mathrm{A}}$ is in the range of [24.58, 30]}
\label{fig:optimization_plot2}
\end{figure}

\begin{figure}[htbp]
        \centering
        \includegraphics[width=0.875\textwidth]{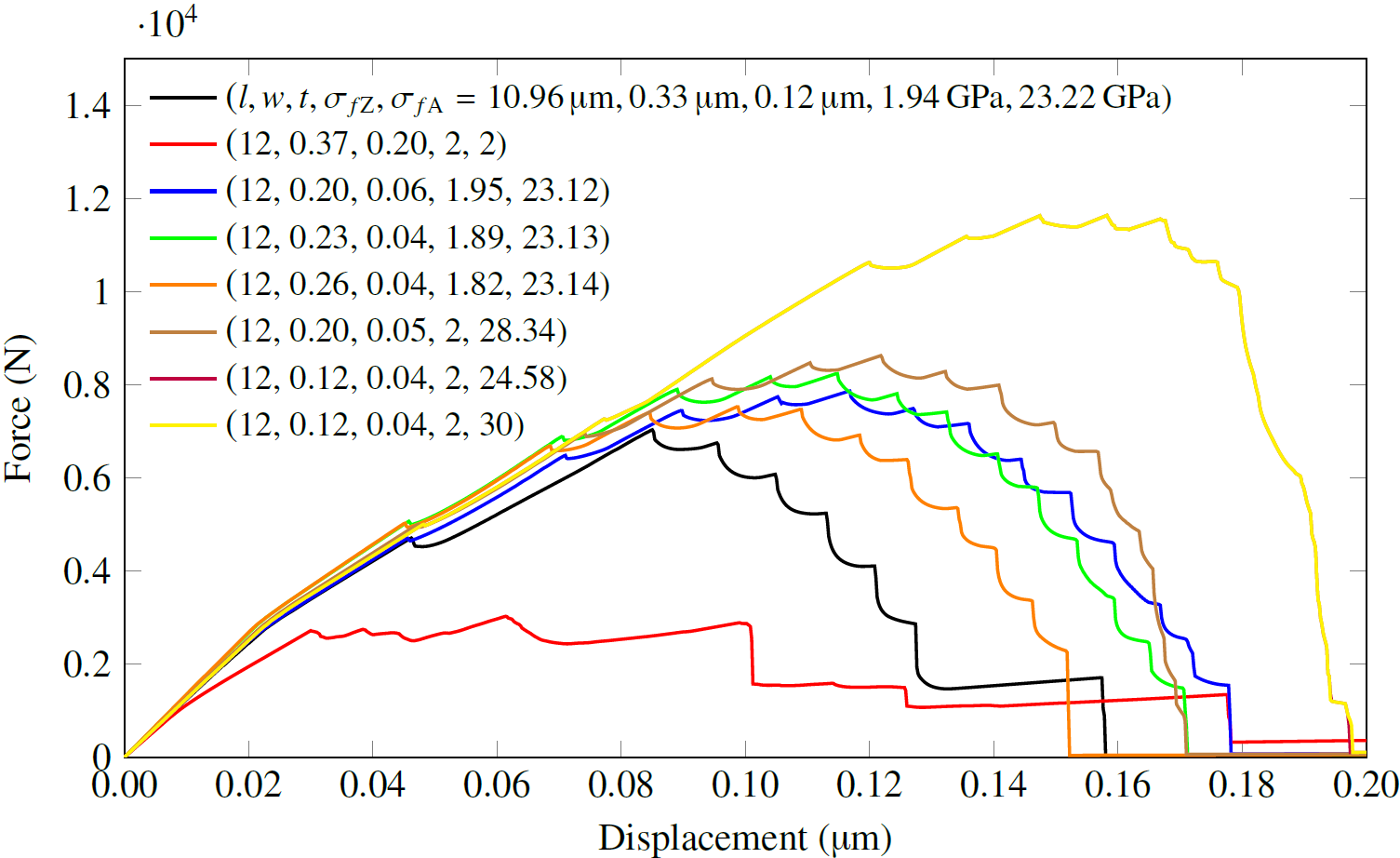}
\caption{Force-displacement curves of each optimal candidate (including five design variables: brick length $l$ $(\si{\micro\meter})$, brick width $w$ $(\si{\micro\meter})$, mortar thickness $t$ $ (\si{\micro\meter})$, zirconia fracture strength $\sigma_{f\mathrm{Z}}$ $(\si{\giga\pascal})$, and alumina fracture strength $\sigma_{f\mathrm{A}}$ $(\si{\giga\pascal})$) per iteration during the optimization. Notably, the same curve is obtained under the geometric parameters $(l, w, t) = (12, 0.12, 0.04)$ with the corresponding aspect ratio $l/w = 100$, $\sigma_{f\mathrm{Z}} = \SI{2}{\giga\pascal}$, and $\sigma_{f\mathrm{A}}$ in the range of [$\SI{24.58}{\giga\pascal}, \SI{30}{\giga\pascal}$].}
\label{fig:force_displacement_optimization2}
\end{figure}

Additionally, we incorporate the fracture strength of zirconia, $\sigma_{f\mathrm{Z}}$, and alumina $\sigma_{f\mathrm{A}}$, as design variables with respective bounds of $[0.5, 2]$ and $[2, 30]$, again in $\si{\giga\pascal}$. Given the increased complexity of the optimization, six candidates are evaluated per iteration, with a maximum of 10 iterations. The optimal design variables and corresponding objective values for each iteration are presented in Table~\ref{tab:optimization_data2} in the Appendix, while the associated crack patterns are shown in Figure~\ref{fig:crack_path_optimization}. For the optimal brick-and-mortar configuration $(l, w, t) = (12, 0.12, 0.04)$, the highest fracture toughness $K_\text{I} = \SI{12.99}{\mega\pascal}\sqrt{\si{\meter}}$ is achieved when the zirconia fracture strength is $\sigma_{f\mathrm{Z}}=\SI{2}{\giga\pascal}$ and the alumina fracture strength $\sigma_{f\mathrm{A}}$ lies in the range of $[24.58, 30]$. It should be noted that this fracture toughness value is relatively high for ceramics. This is due to the assumption in the brick-and-mortar model that the material is flawless and devoid of any initial defects. Figure~\ref{fig:optimization_plot2} illustrates the relationship between the brick aspect ratio $l/w$, fracture strength of zirconia $\sigma_{f\mathrm{Z}}$ and alumina $\sigma_{f\mathrm{A}}$, and fracture toughness $K_\text{I}$. The corresponding force-displacement curves are depicted in Figure~\ref{fig:force_displacement_optimization2}. We see that a minimum value of alumina fracture strength is required to ensure that crack propagation occurs primarily through the zirconia mortar, thereby maximizing energy dissipation during the fracture process.

\section{Summary and conclusions}\label{sec:Summary}

We propose a computational design framework for brick-and-mortar ceramic materials that incorporate stress-induced phase transformation within the mortar. The material system consists of alumina platelets as bricks and ceria-stabilized zirconia as mortar. Our approach integrates three key novelties: (1) Design of ‘double-tough' ceramics: By leveraging the tetragonal-to-monoclinic (T$\rightarrow$M) phase transformation in the mortar of the brick-and-mortar microstructure, the material achieves enhanced toughness at multiple scales. (2) A multiscale computational framework: The toughening mechanisms are captured across scales through a coupled approach. The nanoscale model focuses on the phase transformation behavior of the zirconia mortar, while the microscale model integrates this nanoscale behavior into the overall brick-and-mortar structure. (3) Optimization for fracture performance: The multiscale framework is employed to optimize the material design, highlighting the synergistic effects of nanoscale phase transformations and microscale architecture on fracture toughness. Our results demonstrate that this multiscale design approach enables significant improvements in fracture performance by effectively coupling nanoscale and microscale toughening mechanisms. This study offers a comprehensive pathway for the computational design of advanced ceramic materials with superior toughness.

We began by developing a nanoscale model that focuses exclusively on the volumetric transformation strain to investigate the phase transformation within the mortar. This model allowed us to examine the effects of grain boundary properties and orientations on the stress required to initiate the transformation and the resulting inelastic strain. The outcomes of this analysis, including stress-strain curves, were incorporated into the mortar properties in the microscale model. Subsequently, we analyzed the influence of key geometric and material parameters—such as brick dimensions, mortar thickness, and material properties—on the fracture behavior of the brick-and-mortar structure. Finally, a gradient-free optimization method was used to maximize the fracture toughness of the system. This iterative process involved updating both the layout and material properties of the structure, systematically enhancing its performance.
Our findings demonstrate that the two toughening mechanisms—the phase transformation in the mortar and the brick-and-mortar architecture—can be successfully combined in a single ‘double-tough' ceramic. Leveraging these mechanisms synergistically results in fracture toughness values potentially reaching $\SI{12.99}{\mega\pascal}\sqrt{\si{\meter}}$.
Additionally, this work showcases the potential of multiscale models capable of capturing both phase transformations and fracture processes across two length scales, providing a comprehensive framework for designing high-performance materials.

Here are our final remarks:
\begin{itemize}
\item One of the challenges in modeling the T$\rightarrow$M phase transformation of zirconia is the difficulty in experimentally measuring the transformation's physical speed, making it challenging to determine the appropriate kinetic coefficient for numerical simulations. In this work, we explored a range of kinetic coefficient values and assessed their effects on both the stress required to initiate the transformation and the evolving phase patterns. Our findings indicate that the kinetic coefficient has a minor influence on the stress needed to trigger the transformation and does not significantly impact the transformation patterns.

\item Experimental data on the grain boundary properties of zirconia are limited. To address this, we employed a range of elastic constants for the grain boundaries and found that as the Young's modulus of the grain boundaries decreases, higher stresses are required to activate phase transformation in the bulk grains. This becomes relevant when using different sintering technologies to process the material, which can potentially lead to varying grain boundary structures. Additionally, the distribution of grain orientations in the nanoscale model significantly influence both the transformation patterns and the stresses. This is because the elastic energy associated with grain orientations serves as a key driving force for the phase transformation. This is relevant when textured microstructures are considered.

\item The toughening effect of the brick-and-mortar structure is strongly influenced by its layout - brick dimensions, mortar thickness, and material properties. We demonstrated that reducing periodicity and increasing overlapping areas enhance toughness, as fractures propagate along longer crack paths, dissipating more energy. Longer and thinner bricks were found to significantly improve fracture toughness and strength, primarily due to greater deflections during crack propagation. Additionally, fracture performance was found to be negatively affected by increasing the mortar thickness. 

\item The fracture strengths of zirconia and alumina significantly impact crack propagation patterns and overall structural performance, with higher alumina strength preventing cracks from passing through the bricks and maximizing energy dissipation. By optimizing both geometric and material properties, a substantial increase in fracture toughness was achieved, highlighting the potential of this approach to improve the structural integrity of brick-and-mortar systems.
\end{itemize}

While the proposed multiscale modelling technique effectively simulates the T$\rightarrow$M phase transformation of zirconia, the brick-and-mortar layout, and their toughening effects on fracture behaviour, further refinements are expected to be very beneficial towards applications. The current microscale model assumes a perfect brick-and-mortar structure with uniformly distributed bricks, which may not fully capture the variability present in real materials. Future work should involve a statistical study of brick-and-mortar structures with varying brick sizes and mortar thicknesses, particularly in 3D. Additionally, incorporating built-in defects and random microcracks into the microscale model is recommended. Establishing real-time interactions between the nanoscale and microscale models is also an important next step, with the fracture analysis conducted at both scales to improve the predictive capabilities of the framework.



\section*{Declaration of competing interest}

The authors declare no conflict of interest.

\section*{Acknowledgements}

We gratefully acknowledge the financial support from the Dutch Sectorplan, Zwaartepunt Mechanics of Materials and Zwaartepunt Control Systems Technology. DG gratefully acknowledges the financial support from the Institute of Complex Molecular Systems of TU/e and the Irene Curie Fellowship.

\appendix
\section{Optimization data}
\label{app1}
\begin{table}[H]
\newcolumntype{G}{>{\centering\arraybackslash}m{3.0em}}
 \newcolumntype{E}{>{\centering\arraybackslash}m{3.50em}}
\newcolumntype{P}{>{\centering\arraybackslash}m{5.00em}}
 \newcolumntype{X}{>{\centering\arraybackslash}m{6.00em}}
 \newcolumntype{T}{>{\centering\arraybackslash}m{14.00em}}
\begin{center}
\begin{tabular}{G*3{E}@{}*1{X}@{}*1{X}@{}}
\hline Iteration &  $l$ ($\si{\micro\meter}$)   &  $w$ ($\si{\micro\meter}$)  & $t$ ($\si{\micro\meter}$)   & $\mathrm{ZrO_2}$ (\%) & $K_\text{I}$ ($\si{\mega\pascal}\sqrt{\si{\meter}}$)\\
\hline    1  &   {10.14}   &     {0.23}    &  {0.13}  & 37.34  &  {5.63}   \\
     2  &   12      &     0.12      &  0.04    &    25.05 &  9.68\\
     3  &   12      &     0.12      &  0.04    &    25.05 &  9.68\\
     4  &   12      &     0.12      &  0.04    &    25.05 &  9.68\\
     5  &   12      &     0.12      &  0.04    &    25.05 &  9.68\\
     6      &   12      &     0.12      &  0.04    &    25.05 &  9.68\\
     7      &   12      &     0.12      &  0.04    &    25.05 &  9.68   \\
     8      &   12      &     0.12      &  0.04    &    25.05 &  9.68   \\
     9      &   12      &     0.12      &  0.04    &    25.05 &  9.68  \\
     10    &   12      &     0.12      &  0.04    &    25.05 &  9.68  \\
\hline
\end{tabular}
\end{center}
\caption{Values of the best design variables (brick length $l$ ($\si{\micro\meter}$), brick width $w$ ($\si{\micro\meter}$), and mortar thickness $t$ ($\si{\micro\meter}$)) found per iteration and the corresponding objective function (the fracture toughness $K_\text{I} (\si{\mega\pascal}\sqrt{\si{\meter}})$) during the optimization are presented, where the highest fracture toughness of $K_\text{I} = \SI{9.68}{\mega\pascal}\sqrt{\si{\meter}}$ found with $(l, w, t) = (12, 0.12, 0.04)$. Notably, four candidates with the same combination $(l, w, t) = (12, 0.12, 0.04)$ have been consistently identified since the 6$\text{th}$ iteration.}
\label{tab:optimization_data1}
\end{table}

\begin{table}[H]
\newcolumntype{G}{>{\centering\arraybackslash}m{3.0em}}
 \newcolumntype{E}{>{\centering\arraybackslash}m{3.50em}}
\newcolumntype{P}{>{\centering\arraybackslash}m{4.00em}}
 \newcolumntype{X}{>{\centering\arraybackslash}m{5.00em}}
 \newcolumntype{T}{>{\centering\arraybackslash}m{6.00em}}
\begin{center}
\begin{tabular}{G*3{E}@{}*1{X}@{}*2{P}@{}*1{T}@{}}
\hline Iteration &  $l$ ($\si{\micro\meter}$)   &  $w$ ($\si{\micro\meter}$)  & $t$ ($\si{\micro\meter}$)   & $\mathrm{ZrO_2}$ (\%) & $\sigma_{f\mathrm{Z}}$ ($\si{\giga\pascal}$)  & $\sigma_{f\mathrm{A}}$ ($\si{\giga\pascal}$)  & $K_\text{I}$ ($\si{\mega\pascal}\sqrt{\si{\meter}}$)\\
\hline    1  & 10.96      &  0.33    & 0.12        &   26.62   &1.94      &   23.22   &    8.41\\
            2  & 12             & 0.37       &  0.20      &   34.99   & 2         &     2      &      6.11 \\
            3  & 12              &0.20       &0.06         &  24.45    &  1.95   &   23.12    &   10.36\\
            4  & 12             & 0.23      & 0.04            &   14.68     & 1.89  &    23.13    &   10.67\\
            5  & 12            & 0.26        &  0.04           &   13.62    &1.82   &   23.14    &   9.82\\
            6  &12             & 0.20       &    0.05     &   21.94   &2         &  28.34     &  10.84\\
            7  &12             & 0.12            &   0.04          &  25.05    &2        &      30     &       12.99\\
            8  &12             & 0.12            &   0.04          &  25.05    &2        &      30     &       12.99\\
            9  &12             & 0.12           &   0.04           &  25.05    &2        &      30     &       12.99\\
                &12             & 0.12           &   0.04          &  25.05    &2        &      24.58     &   12.99\\
            10 &12             & 0.12           &   0.04          &   25.05   &2        &      30     &       12.99\\
		 &12             & 0.12           &   0.04          &  25.05    &2        &     26.56     &      12.99\\
		&12             & 0.12           &   0.04          &  25.05    &2        &      27.37     &       12.99\\
            11  &12             & 0.12           &   0.04          &  25.05   &2        &      30     &       12.99\\
		&12               & 0.12           &   0.04          &  25.05   &2        &      28.74     &      12.99\\
		&12               & 0.12           &   0.04          &  25.05   &2        &      29.51     &      12.99\\
\hline
\end{tabular}
\end{center}
\caption{Values of the best design variables (brick length $l$ ($\si{\micro\meter}$), brick width $w$ ($\si{\micro\meter}$), mortar thickness $t$ ($\si{\micro\meter}$), zirconia fracture strength $\sigma_{f\mathrm{Z}}$, and alumina fracture strength $\sigma_{f\mathrm{A}}$) found per iteration and the corresponding objective function (the fracture toughness $K_\text{I}$ ($\si{\mega\pascal}\sqrt{\si{\meter}}$)) during the optimization. The highest fracture toughness, $K_\text{I} = \SI{12.99}{\mega\pascal}\sqrt{\si{\meter}}$, is achieved with the geometric parameters $(l, w, t) = (12, 0.12, 0.04)$ and material properties $\sigma_{f\mathrm{Z}} = \SI{2}{\giga\pascal}$, $\sigma_{f\mathrm{A}} > \SI{24.58}{\giga\pascal}$. It is important to note that a minimum alumina fracture strength is required for certain combinations of brick sizes, mortar thickness and zirconia fracture strength to achieve optimal results.}
\label{tab:optimization_data2}
\end{table}

\section*{}

During the preparation of this work the authors used ChatGPT in order to improve language and readability. After using this tool, the authors reviewed and edited the content as needed and take full responsibility for the content of the publication.


\bibliographystyle{unsrt}
\bibliography{cas-refs}



%
%
%
\end{document}